\documentclass[prd,preprint,nofootinbib]{revtex4}
\usepackage{framed}
\usepackage{bbold}
\usepackage{slashed}
\usepackage{graphicx,hyperref}
\usepackage{amsmath}
\usepackage{amssymb}
\usepackage{epsfig}
\usepackage{hhline}
\usepackage{soul}
\usepackage{tabularx,pifont,multirow}
\usepackage{enumitem}
\usepackage{colordvi}
\usepackage{soul}
\usepackage{xcolor}
\usepackage{yfonts}

\newcommand{\be}{\begin{equation}}
\newcommand{\ee}{\end{equation}}
\newcommand{\bea}{\begin{eqnarray}}
\newcommand{\eea}{\end{eqnarray}}

\newcommand{\nit}[1]{{\textcolor{brown}{#1}}}

\newcommand{\CC}{\Lambda}
\newcommand{\rL}{\rho_{\Lambda}}
\newcommand{\rvm}{\rho^\CC_{\rm RVM}}
\newcommand{\MPl}{M_{\rm Pl}}
\newcommand{\mpl}{m_{\rm Pl}}
\newcommand{\rvo}{\rho_{{\rm vac}\,0}}
\newcommand{\rv}{\rho_{\rm vac}}

\newcommand{\ha}{\hat{a}}
\newcommand{\astar}{a_{*}}
\newcommand{\tHI}{\tilde{H}_I}
\newcommand{\trI}{\tilde{\rho}_I}

\newcommand{\Oro}{\Omega_{r 0}}
\newcommand{\rr}{\rho_{r}}
\newcommand{\rro}{\rho_{r 0}}
\newcommand{\rco}{\rho_{c0}}

\newcommand{\A}{{\cal A}}

\newcommand{\SV}{S_{\cal V}}
\newcommand{\SA}{S_{\cal A}}

\newcommand{\SVp}{S'_{\rm V}}

\newcommand{\Omo}{\Omega_{m 0}}
\newcommand{\OLo}{\Omega_{\CC 0}}

\definecolor{darkgreen}{rgb}{0,0.3,0.05}
\makeatletter                                                         %
\newcommand*\rel@kern[1]{\kern#1\dimexpr\macc@kerna}                  %
\newcommand*\widebar[1]{                                              %
  \begingroup                                                         %
  \def\mathaccent##1##2{                                              %
    \rel@kern{0.8}                                                    %
    \overline{\rel@kern{-0.8}\macc@nucleus\rel@kern{0.2}}             %
    \rel@kern{-0.2}                                                   %
  }                                                                   %
  \macc@depth\@ne                                                     %
  \let\math@bgroup\@empty \let\math@egroup\macc@set@skewchar          %
  \mathsurround\z@ \frozen@everymath{\mathgroup\macc@group\relax}     %
  \macc@set@skewchar\relax                                            %
  \let\mathaccentV\macc@nested@a                                      %
  \macc@nested@a\relax111{#1}                                         %
  \endgroup                                                           %
}                                                                     %
\makeatother                                                          %

\begin{document}

\preprint{\leftline{KCL-PH-TH/2020-{\bf 71}}}

\title{Stringy-Running-Vacuum-Model Inflation: \\from primordial Gravitational Waves and stiff Axion Matter to Dynamical Dark Energy}

\author{{\bf Nick~E.~Mavromatos$^{a}$} and \vspace{0.5cm} {\bf Joan~Sol\`a~Peracaula$^b$}}

\affiliation{$^a$Theoretical Particle Physics and Cosmology Group, Physics Department, King's College London, Strand, London WC2R 2LS.
\vspace{0.5cm}\\
 $^{b}$Departament de F\'\i sica Qu\`antica i Astrof\'\i sica, \\ and \\ Institute of Cosmos Sciences (ICCUB), Universitat de Barcelona, Av. Diagonal 647 E-08028 Barcelona, Catalonia, Spain.\vspace{0.5cm}
}


\begin{abstract}
\vspace{0.05cm}

In previous works we have derived a Running Vacuum Model (RVM) for a string Universe, which provides an effective description
of the evolution of  4-dimensional string-inspired cosmologies from inflation till the present epoch.
In the context of  this ``stringy RVM'' version, it is assumed that the early Universe is characterised by purely gravitational degrees of freedom, from the massless gravitational string multiplet, including the antisymmetric tensor field. The latter  plays an important role, since its dual gives rise to a `stiff' gravitational-axion ``matter'', which in turn couples to the gravitational anomaly terms, assumed to be non-trivial at early epochs. In the presence of primordial gravitational wave (GW) perturbations, such anomalous couplings lead to an RVM-like dynamical inflation, without external inflatons.
We review here this framework and discuss potential scenarios  for the generation of such primordial GW, among which the formation of unstable domain walls, which eventually collapse in a non-spherical-symmetric manner, giving rise to GW. We
also remark that the same type of ``stiff'' axionic matter could provide, upon the generation of appropriate potentials 
during the post-inflationary eras,  (part of) the Dark Matter (DM) in the Universe, which could well be ultralight, depending on the parameters of the string-inspired model.
All in all, the new (stringy) mechanism for RVM-inflation preserves the basic structure of the original (and more phenomenological)  RVM, as well as its main advantages: namely, a mechanism for graceful exit and for generating a huge amount of entropy  capable of explaining the horizon problem.  It also predicts axionic DM and the existence of mild dynamical Dark Energy  (DE) of quintessence type in the present universe, both being  ``living fossils'' of the inflationary stages  of the cosmic evolution. Altogether the modern RVM  appears to be  a theoretically sound (string-based) approach to cosmology with a variety of  phenomenologically testable consequences.

\end{abstract}

\maketitle


\tableofcontents


\section{Introduction \label{sec:intro}}

In the string-inspired effective gravitational field theory for the very early Universe, proposed in~\cite{bms1,bms2,bms3},  and  further discussed from the point of view of the swampland criteria and the weak gravity conjecture in \cite{msb},  it was assumed that the only degrees of freedom present are those from the
massless bosonic  gravitational multiplet of the (super)string, consisting of dilaton, gravitons and antisymmetric tensor (Kalb-Ramond (KR)) fields. The latter can be dualised by means of a massless gravitational KR axion field, which is characterised by a ``stiff'' equation of state. Upon assuming constant dilatons, which are consistent string backgrounds, it was shown that condensation of primordial gravitational waves (GW) leads to  a ``running-vacuum-model''(RVM)-type cosmology~\cite{ShapSol1,Fossil2008,ShapSol2}, with a dynamically-induced (approximately) de Sitter era, without the need for external inflatons. Crucial to this, was the fact that gravitational anomalies are present in the early phases of this string Universe, which couple to the KR (and other stringy) axions via CP-Violating anomalous gravitational Chern-Simons couplings. The condensation of GW perturbations imply, in turn, condensation of such anomalous terms, and an approximate-de-Sitter era, in which the vacuum energy density resembles that of the RVM.  The GW condensates are triggered by the  `cosmological birefringence' of the GW during inflation and, as shown in ~\cite{bms1,bms2}, are responsible for the generation of terms in the vacuum energy density proportional to the fourth power of the Hubble rate  $H^4$, which induce inflation without the need for external inflaton fields. We term this phase ``GW-induced stringy RVM inflation''.

During this inflationary era, KR-axion backgrounds of a specific type (varying linearly with cosmic time) remain undiluted, leading to eventual matter-antimatter asymmetries (baryogenesis through leptogenesis) in the post-inflationary radiation era. During the radiation and matter eras, the gravitational anomalies cancel, due to the generation of chiral matter at the exit phase from the GW-induced stringy RVM inflation. However, chiral anomalies remain, which lead, through non-perturbative effects (e.g. instantons in the Gluon sector of Quantum Chromodynamics part of the matter action) to potentials, and thus masses, for these axions, which in general can mix with other stringy axions, leading to significant components of axionic Dark Matter  (DM) in the late eras of this string Universe. As discussed in \cite{bms1}, the latter can also be ultralight, depending on the parameters of the model.

 In this study,  we revisit these ideas and  elaborate further  on the properties of such KR axions, and other stringy-type axions that may characterise  the very early stages of the string Universe. We propose the possibility that there is a pre-inflationary era of (cold) {\it stiff-axionic-matter} dominance, which then, upon condensation of primordial gravitational waves (GW), leads - through the gravitational anomalous couplings of the axions- to  ``stringy  RVM-inflation''. {Such form of inflation preserves all of the virtues of the original, more phenomenological,  RVM proposal  (see \cite{JSPRev2013,JSPRev2015}  for a review)  but it has a more solid theoretical formulation.

Moreover, we discuss potential origins of the primordial GW, by presenting several pre-inflationary scenarios for their production,
One of them, involves dynamical broken supergravity, which are models embeddable in string theory. The breaking occurs through
condensation of gravitino fields, the partners of gravitons,  whose double-well potential may eventually be deformed
by `bias' induced due to, say, percolation effects of vacuum bubbles in the effective theory. This leads to the formation of unstable domain walls, whose collapse in a non-spherically-symmetric manner leads to primordial GW. In this minimal scenario, only gravitational degrees of freedom are encountered in the early Universe, in agreement with the assumption of \cite{bms1,bms2}, given that the gravitino is the
supersymmetry partner of the graviton.

To be complete, and give as much information as possible to the reader who might not be familiar with our previous works~\cite{bms1,bms2,bms3,msb}, we also review here the conventional RVM~\cite{ShapSol1,Fossil2008,ShapSol2} and compare it with its modern `stringy' version, stressing the essential similarities but also the important differences, which might have important, and observable in principle, phenomenological consequences.

The structure of the article is as follows: in the next section \ref{sec:rvm}, we describe the essential features of the conventional RVM model, discuss its dynamical inflation, without external inflatons, stressing important differences from other dynamical-inflation scenarios like the Starobinsky model, and review the thermodynamical aspects of the framework. In regards to the latter topic, we
review in detail the mechanism~\cite{rvmhistory}  underlying the generation of an enormous amount of entropy during the exit phase from the RVM inflation, which provides also an explanation of the horizon problem in cosmology\,\cite{JSPRev2015,GRF2015} and shows that the RVM  satisfies the Generalized Second Law of Thermodynamics\,\cite{solaentropy}.
In section \ref{sec:string}, we describe the stringy version of the RVM.  We first discuss in detail the kind of stringy axions that arise in the model, which apart from the KR axion, contain other axions arising from compactification schemes in string theory. All these axions have non trivial anomalous couplings to the gravitational Chern-Simons terms. They constitute a form of `stiff' ``matter'', the evolution of which is discussed within the RVM framework.  We then study a phase of dynamical inflation induced by gravitational anomaly condensates induced by primordial GW perturbations. We explain how the latter lead to an RVM energy density, but we stress the essential differences of the string version from the conventional RVM approach, some of which might have observable consequences. In section \ref{sec:origin}, we discuss potential scenarios for the origin of such primordial GW in pre-RVM-inflationary phases of the string Universe. Among the discussed scenarios, there is a minimal one, in which there is a dynamically broken supergravity phase during this pre-RVM inflationary phase, which occurs at the end of a first (unobservable) hill-top inflation, induced by the gravitino-condensate field. We discuss physical mechanisms, due to percolation effects among vacuum bubbles during this early phase of the string Universe, which lead to unstable domain walls, whose collapse produces the primordial GW responsible for the second (and observable) ``GW-induced-RVM inflation''. In section \ref{sec:swamp} we review the swampland criteria for embedding the RVM model in ultraviolet complete theories of quantum gravity, and discuss~\cite{msb} how these criteria are evaded in the case of
the GW-condensate-induced composite inflation that characterises the stringy RVM. This is consistent with the phenomenological
agreement of the RVM inflation with slow-roll data. In this section we also mention briefly a mechanism for entropy generation at the last stages of the stringy RVM inflation, which is due to string states that in general  fail to decouple from the low-energy effective field theory. This provides the stringy RVM mechanism for entropy production, which calls for comparison with the field-theoretic RVM version reviewed in section \ref{sec:rvm}. Requiring the satisfaction of the swampland criteria in models with fundamental inflatons is consistent with the thermodynamics of such states being treated within a local effective field theory approach.
However, the composite nature of the condensate field that leads to the GW-induced stringy inflation enables the evasion of the swampland restrictions in this case. Finally, conclusions and outlook are discussed in section \ref{sec:concl}.

\section{Essentials of the Running Vacuum Model (RVM)} \label{sec:rvm}

The original RVM\cite{ShapSol1,Fossil2008,ShapSol2} --  see \cite{JSPRev2013,JSPRev2015} for  detailed reviews --  constitutes  a theoretically (renormalization-group inspired) and phenomenologically  compelling alternative to the standard concordance $\Lambda$CDM model,   providing  an effective description of the cosmological evolution of the Universe from inflation till the present era.  It also helps alleviate the current tensions with the data, as shown in a variety of fitting analyses~\cite{rvmpheno1,rvmpheno2,rvmpheno3}. These tensions must be overcome as they  are indeed  a potentially important headache for  the phenomenological  consistency of the standard model of the cosmic evolution  (the $\CC$CDM model)\,\cite{tensions}. Their stubborn persistence may point to the existence of physics  beyond $\CC$CDM. Whatever the nature of the new physics might be, we expect that it should not imply a drastic modification of the $\CC$CDM since the latter provides already a fairly good description of the overall cosmological data.  At the same time we expect that the needed corrections to the concordance model should be sensitive to the new  features of both, the late and the early RVM universe~\cite{rvmhistory,GRF2015,solaentropy}.

\subsection{General structure of the RVM and its connection to the Renormalization Group in Curved Spacetime}

The RVM  evolution of a cosmological Universe is usually formulated on a Friedmann-Lema\^\i tre-Robertson-Walker (FLRW) background space-time metric, with scale factor $a(t)$,  in the context of General Relativity (GR). Let us, however, note that one may also obtain an effective RVM evolution in a Brans-Dicke (BD) context with a cosmological constant,   hence within a gravity paradigm different from  GR. It turns out that in this latter form  (in which there is an evolution of the effective gravitational coupling as well) the RVM is  particularly efficient in solving the main tensions, above all the one associated with the local value of the Hubble parameter, $H_0$\,\cite{BDpapers}.    In either framework (GR or BD), the dynamical vacuum energy density  associated with the RVM is based on the following general renormalisation-group (RG)-like  form of the vacuum energy density  in terms of powers of the Hubble parameter $H(t)={\dot a}/{a}$, which is a function of the cosmic time $t$,
and its  cosmic-time derivative $\dot H$~\cite{JSPRev2013,JSPRev2015}:
\begin{equation}\label{runningmu}
\frac{d\, \rho_\Lambda (\mu)}{d\, {\rm ln}\mu^2 } = \frac{1}{(4\pi)^2}
\sum_i \Big[A_i M_i^2 \mu^2 + B_i \mu^4 +C_i \mu^6+... \Big]\,,
\end{equation}
where the coefficients $A_i, B_i$ are dimensionless, whereas the $C_i$ ones  and higher are dimensionfull.  They receive contributions from loop corrections of boson  and fermion  matter fields with different masses $M_i$.  The above RG equation  provides the rate of change of the quantum effects on the CC as a function of some characteristic cosmological scale  $\mu$.  The leading effects are controlled by the ``soft-decoupling'' terms of the form $\sim M_i^2\,\mu^2$. Notice that the $M_i^4$ terms are absent, as they would trigger a too fast a running of $\rho_\Lambda (\mu)$ as a function of the scale $\mu$. In fact, these effects are ruled out by the RG
formulation itself, since only the fields satisfying $\mu>M_i$ are to be included as active degrees of freedom contributing to the running.

The  association of $\mu$ with some representative cosmological scale can be a matter of debate, but the ansatz $\mu={\cal O}(H)$  (i.e. $\mu$ being of order of the Hubble scale at each epoch) has been fostered since long ago \cite{ShapSol1,Fossil2008}. If we adhere to it,  it is obvious that  the condition $\mu>M_i$ (i.e. $H>M_i$)  cannot be currently satisfied by the SM particles. Therefore, the leading effects on the running of $\rL$ are, according to Eq.\,(\ref{runningmu}), of order $M_i^2\mu^2\sim M_i^2\,H^2$, and hence  dominated by the heaviest fields at disposal.  In the context of a typical GUT near the Planck scale,  $\mpl\sim 10^{19}$ GeV,  the main contribution comes from  fields with masses $M_i\sim M_X\lesssim \mpl$.

While we agreed that  $\mu\sim H$ can be a natural association of the RG-scale in cosmology,  a more general option is to associate $\mu^2$
to a linear combination of $H^2$ and $\dot{H}$ (both terms being dimensionally homogeneous). Adopting this setting and integrating (\ref{runningmu}) up to the terms of ${\cal O}(\mu^4)$  it is
easy to see that we can express the result as follows ~\cite{JSPRev2013,JSPRev2015}:
\be
\rvm(H,\dot{H},\ddot{H},\dots )
=a_0+a_1\,\dot{H}+a_2\,H^2+a_3\,\dot{H}^2+a_4\,H^4+a_5\,\dot{H}\,H^2+ a_6H\ddot{H} \dots,
\label{GRVE}
\ee
where  the coefficients $a_i$ are real,  having different dimensions in natural units, we work with here.  From the foregoing discussion we can see that the  form \eqref{GRVE} for the vacuum energy density in cosmology has been derived from general RG qualitative arguments. It is worth noticing, though,  that it can actually be supported  by explicit calculations as well. This is in fact the result presented very recently in \cite{Cristian2020}, based on computing the  quantum corrections  to a classical action with  a scalar field non-minimally coupled to gravity (see next section \ref{Sec:RVMQFT}, for a summarized discussion).  In \cite{Cristian2020} it is shown from the calculation of quantum effects in QFT in curved spacetime (specifically on a FLRW background)  that the $\sim M_i^4$ terms are in fact absent from the correctly renormalized vacuum energy density. Furthermore, such calculation confirms  that it is indeed the cosmological scale $H$  that controls the size of the quantum corrections and, in addition,  that the ``soft-decoupling'' terms  $\sim M_i^2\,\mu^2$  are the leading quantum corrections to the vacuum energy density. At the end of the day, it is reassuring to see that the  banishing of the quartic mass terms from the RG equation \eqref{runningmu}  is not an \textit{ad hoc} procedure,  and also that  $\mu\sim H$ is in fact a sensible ansatz for gauging the size of the quantum effects in cosmology.  The physical results, therefore,  are no longer  based on the useful (but  qualitative) arguments originally proposed in \cite{ShapSol1}.

Because of the general covariance of the effective action of quantum field theories,  which must characterise all gravitational field theories, all the terms in the RVM form \eqref{GRVE}  must appear as being of even  adiabatic order, and therefore only  an even number  of derivatives of the scale factor is possible.  For example, apart from the $H^2$ and $\dot{H}$ terms which constitute the leading terms at low energies, the next-to-leading ones would be  of the three forms $\dot{H}^2,\, H^2 \dot{H}$  and  $ H\ddot{H}$,  all of them of adiabatic order $4$. Despite most of these structures can actually be derived from the aforesaid  QFT calculation in FLRW spacetime~\cite{Cristian2020}, no specific structure of the form $\sim H^4$ appears in it, despite being suggested as one of the expected terms in the general solution \eqref{GRVE} of the RG equation\,\eqref{runningmu}.  As a result,  all of the terms of order $4$ vanish for $H=$constant. Thus, if these were the only ones  available at this adiabatic order, inflation could not  have been triggered from a transitory period where $H=$constant.   This is of course no drama, since inflation can alternatively be triggered from a short period where $\dot{H}=$constant. This is exactly the situation, for example, with Starobinsky inflation\,\cite{staro,staro2}, where the variation of the $R^2$ term in the Starobinsky action produces precisely the aforementioned structures, as will be reviewed in more detail in Sec.\,\ref{sec:Staro-inflation}.

\subsection{The RVM and its connection to Quantum Field Theory in Curved Spacetime}\label{Sec:RVMQFT}}

As noted,  we would like to substantiate the RVM form \eqref{GRVE} of the vacuum energy density on more explicit calculational grounds and, in addition, we  would  like to use it to produce inflation with an alternative mechanism,  e.g. one in which  $H=$constant for an initial period.  For this to occur,  a new  term of order $4$ should enter the adiabatic expansion \eqref{GRVE}, namely the term $\sim H^4$ with no derivatives of the Hubble rate. Unfortunately,  as previously indicated,  such a term does not appear if  one considers just the quantum effects of QFT in curved spacetime within a  scalar field theory coupled to curvature ~\cite{Cristian2020}.  Indeed, let us  consider  the action of a free neutral scalar field non-minimally coupled to gravity\footnote{We use here the following geometric  conventions:  metric signature  $g_{\mu\nu}$, $(+, -,-,- )$; Riemann tensor,
$R^\lambda_{\,\,\,\,\mu \nu \sigma} = \partial_\nu \, \Gamma^\lambda_{\,\,\mu\sigma} + \Gamma^\rho_{\,\, \mu\sigma} \, \Gamma^\lambda_{\,\, \rho\nu} - (\nu \leftrightarrow \sigma)$; Ricci tensor, $R_{\mu\nu} = R^\lambda_{\,\,\,\,\mu \lambda \nu}$; and Ricci scalar,  $R = g^{\mu\nu} R_{\mu\nu}$.  Overall, these correspond to the $(-, +, +,+)$ conventions in the classification by Misner-Thorn-Wheeler\,\cite{MTW}. }:
\begin{equation}\label{eq:Sphi}
  S[\phi]=\int d^4x \sqrt{-g}\left(\frac{1}{2}g^{\mu \nu}\partial_{\nu} \phi \partial_{\mu} \phi-\frac{1}{2}(m^2-\xi R)\phi^2 \right)\,,
\end{equation}
where $\xi$ is the non-minimal coupling of the quantum matter field $\phi$ to curvature.  It is well-known that for  $\xi=1/6$, the massless ($m=0$)  action is conformally invariant.  Since $\phi$ is a quantum matter field, we can expand it around its background value $\phi(\eta,x)=\phi_b(\eta)+\delta \phi (\eta,x)$, where $\delta \phi (\eta,x)$  denote the quantum fluctuations and $\eta$ is the conformal time.  Because of these fluctuations  one has to add the higher derivative (HD) terms of the vacuum action, since they are generated at the quantum level  and are therefore needed for renormalizability\,\cite{BirrellDavies82}. The HD vacuum action  is composed of  the ${\cal O}(R^2)$ terms, i.e. the squares of the curvature and Ricci tensors:  $R^2$ and $R_{\mu\nu}R^{\mu\nu}$. No addittional HD terms are needed in $4$ dimensions since the square of the Riemann tensor, $R_{\mu\nu\rho\sigma}R^{\mu\nu\rho\sigma}$,  is not independent owing to the topological (total derivative) nature of the Euler's density (or Gauss-Bonnet (GB) term, as is otherwise called): GB=$R_{\mu\nu\rho\sigma} \, R^{\mu\nu\rho\sigma}  - 4 R_{\mu\nu}\, R^{\mu\nu} + R^2 = \mathcal J^\mu_{\,\,\,\,; \mu}$, $\mu,\nu,\rho,\sigma=0, \dots 3,$ where the semicolon in the last expression denotes as usual the gravitational covariant derivative\,\footnote{Here, for concreteness and brevity, we do not discuss situations, like the one encountered in 4-dimensional effective low-energy field theories coming from string theory, where the scalar (dilaton) field couples to the Euler invariant. In such cases, the dilaton and graviton fields are part of the string gravitational vacuum, and the inclusion of dilaton-Riemann-curvature-square terms play an important r\^ole on the underlying physics, for instance, they may lead to black holes with (secondary) scalar dilaton hair~\cite{kanti}.}

Thus, the full action consists of the Einstein-Hilbert (EH)  action with cosmological constant, the HD action and finally  the matter part, which in this case boils down to Eq.\,(\ref{eq:Sphi}): $S=S_{\rm EH}+S_{\rm HD}+S[\phi]$.  Since  the HD terms are included,  the variation of $S$  leads to  the  modified Einstein's equations:
\begin{equation}
\frac{1}{8\pi G_N}G_{\mu \nu}+\rho_\Lambda g_{\mu \nu}+a_1 H_{\mu \nu}^{(1)}=\langle T_{\mu \nu}^{\delta \phi} \rangle +T_{\mu \nu}^{\phi_b}\,,
\end{equation}
where  $T_{\mu \nu}^{\phi_{b}} $ is the  contribution to the energy-momentum tensor (EMT) from the classical or background part,  whereas  $\langle T_{\mu \nu}^{\delta_\phi}\rangle$  is  the contribution from the vacuum fluctuations of  $\phi$. The $00$-component of the latter is connected with the zero-point energy (ZPE) density of the scalar field in the FLRW background. This is of course the genuine effect of the vacuum energy-density we are after. Finally,
\begin{equation}\label{eq:H1munu}
  H_{\mu\nu}^{(1)}=\frac{1}{\sqrt{-g}}  \frac{\delta}{\delta g^{\mu\nu}}  \int d^4x  \sqrt{-g} R^2=-2\nabla_\mu\nabla_\nu R - 2 g_{\mu\nu} \Box R - \frac12 g_{\mu\nu} R^2 +2 R R_{\mu\nu}\,,
\end{equation}
with $\Box $ the gravitationally-covarfiant D' Alembertian, comes from the functional differentiation of the $R^2$ term in the HD vacuum action\footnote{Recall that the corresponding term associated with the functional differentiation of the square of the Ricci tensor,  $H_{\mu \nu}^{(2)}$, is not necessary since it is not independent of $H_{\mu \nu}^{(1)}$  for  FLRW spacetimes\,\cite{BirrellDavies82}.}.
 Thus, the total vacuum contribution reads
\begin{equation}\label{EMTvacuum}
\langle T_{\mu \nu}^{\rm vac} \rangle= T_{\mu \nu}^\Lambda+\langle T_{\mu \nu}^{\delta \phi}\rangle=-\rho_\Lambda g_{\mu \nu}+\langle T_{\mu \nu}^{\delta \phi}\rangle\,.
\end{equation}
The above equation  states that the total vacuum EMT  is made out of the contributions from the cosmological term and  the quantum fluctuations of the field. 

The computation of all these quantities has been performed in \,\cite{Cristian2020}, where an adiabatic regularization and renormalization procedure has been used in order to produce finite quantities.
If we define the fundamental parameters at the characteristic scale of a generic  Grand Unified Theory (GUT),  typically at $M_X\sim 10^{16}$  GeV, the renormalized vacuum energy density at low energy emerging from explicit QFT calculation reads as follows\,\cite{Cristian2020}
\begin{equation}\label{eq:RVM2}
\rho_{vac}(H)\simeq \rvo+\frac{3\nu}{8\pi}\,(H^2-H_0^2)\,\mpl^2\,,
\end{equation}
where $\nu$ is a dimensionless parameter given by\,\cite{Cristian2020}
\begin{equation}\label{eq:nueff}
\nu=\frac{1}{2\pi}\,\left(\frac{1}{6}-\xi\right)\,\frac{M_X^2}{\mpl^2}\left(1+\frac{m^2}{M_X^2}\ln \frac{H_0^{2}}{M_X^2}\right)\,.
\end{equation}
It is obvious that $|\nu|\ll1$ since  $M_X^2/\mpl^2\ll1$, with $\mpl=1/\sqrt{G}\simeq 1.22\times 10^{19} GeV $  the Planck mass defined in terms of Newton's constant, $G$,  and $H_0$ stands for the current value of the Hubble parameter\footnote{Let us stress that the result \eqref{eq:RVM2} was foreseen long ago on the basis of general renormalization group arguments~\cite{ShapSol1}, which however were only merely indicative, since the renormalization procedure in curved spacetime is not so straightforward, especially if using off-shell renormalization schemes (such as  Minimal Subtraction (MS)). The latter  do not lead to the correct answer in the infrared and  generate large (but spurious!) $\sim m^4$ terms. These terms  have been a well-known problem for a longtime, as they lead to huge fine tuning in the value of $\rvo$.  The RVM  form \eqref{eq:RVM2}, which contains no such unwanted contributions, had also been predicted in ~\cite{Fossil2008} in the context of anomaly-induced inflation and it was further discussed in the general context of quantum fields in curved spacetime in \cite{JSPRev2013,JSPRev2015}.  However, as indicated,  it  was only recently  that the RVM has been accounted for in full detail from explicit QFT calculation in curved spacetime based on the adiabatic regularization and renormalization of the EMT corresponding to the action \eqref{eq:Sphi}~\cite{Cristian2020}. In such a context all the above mentioned spurious contributions can be disposed of.}. We will also use at convenience the reduced Planck mass $\MPl=\mpl/\sqrt{8\pi}\simeq 2.435 \times 10^{18}~{\rm GeV}$ in other parts of the paper. The reader should also notice from \eqref{eq:nueff} that,
in the conformal case, $\xi=1/6$ the coefficient $\nu=0$, as expected.   However, as noted in ~\cite{Cristian2020}, this  does not mean that the effective value of this coefficient should be zero,  even  for conformal fields, as $\nu$ really receives contributions not only from fundamental scalar particles ($s$) but also from fundamental fermions ($f$) and vector bosons ($v$).  In other words,  the final value for such coefficient is $\nu_{\rm eff}=\nu_s+\nu_f+\nu_v$. The calculation performed in ~\cite{Cristian2020} accounts  for the contributions from a single scalar field particle non-minimally coupled to gravity,  i.e. it partially accounts for the value of  $\nu_s$. This is nevertheless sufficient to demonstrate that the structure \eqref{eq:RVM2} can be derived from QFT calculations on a FLRW background.  As argued in that reference, we expect that all fields would contribute formally the same with only differences in the values of $\nu_i$ for each spin.

As  already mentioned, the above result  applies for the present universe since it involves the constant term $\rvo$ (the current value of the vacuum energy density (VED)) and  the corrections of order  ${\cal O}(H^2)$.  The obtained result conforms with the expansion \eqref{GRVE} up to this order, since we have neglected at this point all the terms of order  ${\cal O}(H^4)$, which, however, will play an important role in the early universe.  If such terms are included, their contribution reads\,\cite{Cristian2020}
\begin{equation}\label{RenormalizedVE}
\rho_{vac}(H)
=\frac{9}{16\pi^2} \left( \xi-\frac{1}{6}\right)^2 \left( 2 H\ddot{H} + 6 H^2 \dot{H}-\dot{H}^2 \right)\ln \frac{H^2}{M_X^2}\,.
\end{equation}
and again most of  the new terms conform with the general expansion \eqref{GRVE} up to fourth adiabatic order. No  $\sim H^4$-term, though, appears  in \eqref{RenormalizedVE}. The reader should again notice the vanishing of the vacuum energy for the conformal case $\xi=1/6$, which constitutes a nice consistency check of the approach.

In contradistinction to the higher order corrections found in the  QFT case\,\cite{Cristian2020} ,  the  $\sim H^4$-term appears in the context of a stringy-dominated era of the Universe at scales above the  effective RVM inflationary scale~ \cite{bms1,bms2,bms3,msb}.  All other degrees of freedom such as e.g. the gauge ones, appear as virtual quantum fluctuations, or in hidden sectors of the string-inspired model.  In such a stringy RVM framework, scenarios can be conceived leading to the formation of primordial gravitational waves (GW) and other metric (tensor) fluctuations.  The  supermassive transplanckian string modes decouple from the effective field theory during the expansion  and an effective action is left  involving only the massless degrees of freedom of the bosonic gravitational multiplet of the string in a broken supergravity phase, see \,\cite{bms1}-\cite{msb} and  discussion in subsequent sections of this paper.

What is important for this consideration is that the string induced primordial GW generate the necessary $\sim H^4$ terms, which are the trademark of the new (stringy) mechanism of RVM inflation that cannot be produced from QFT effects. We stress that the gravitational-anomaly terms in the string-inspired RVM couple to fundamental massless gravitational axion fields, which, together with the dilaton and graviton fields, constitute the massless ground-state gravitational multiplet of the string. Such CP-violating couplings of the gravitational axion fields with anomalies are crucial for the aforementioned GW-induced condensates. In the absence of axion fields, the gravitational anomaly terms are irrelevant, being total derivatives (topological).
The characteristic of this new formulation of the RVM,  therefore, is that $\sim H^4$ -inflation can be accounted for on string-based theoretical grounds.   The  $\sim H^4$ -inflation mechanism is obviously different from Starobinsky inflation\,\cite{JSPRev2015}.  Its compatibility with observations\,\cite{bms1,bms2}, make it worth of detailed studies. In subsequent sections of this paper we will  devise new detailed scenarios aimed at explaining the transition from the string era into the QFT one  within this modern, stringy RVM, formulation.

\vspace{0.5cm}

\subsection{Basics of RVM-inflation}\label{sec:RVM-inflation}

In simple cosmological models, which  suffice to describe phenomenologically-realistic
cosmic evolution, from inflation to the present epoch, the various epochs are described with
approximately constant deceleration parameter $q$ per era, in which case one can write
\be\label{Hdot}
\dot H \simeq - (q + 1) \, H^2.
\ee
The above relation is exact if $q(t)$ is the instantaneous value at cosmic time $t$, but is approximate for each epoch if $q$ is taken to be as constant. Recall that $q=(1,1/2,-1)$ for radiation, matter and vacuum energy, respectively.
From \eqref{GRVE}, then,
one can then use for all practical purposes~\cite{rvmhistory,GRF2015,solaentropy}:
\begin{equation}\label{rLRVM}
\rho^{\Lambda}_{\rm RVM}(H) = \frac{\Lambda(H)}{\kappa^2}=
\frac{3}{\kappa^2}\left(c_0 + \nu H^{2} + \alpha
\frac{H^{4}}{H_{I}^{2}} + \dots \right) \;,
\end{equation}
where $\kappa^2 = 8\pi G=\frac{1}{M^2_{\rm Pl}}$ is the four-dimensional gravitational constant, with $M_{\rm Pl}$ the aforementioned reduced Planck mass. As pointed out before, this structure can also be motivated from  RG argments on assuming that  $\mu^2$ can be associated with a linear combination of the homogeneous quantities  $H^2$ and $\dot{H}$.
The  $H^4$ terms and higher  do not play any significant r\^ole in fitting the current data.  We shall only keep the  $H^4$ ones in order to investigate the physics of the early universe, where they could provide e.g. a new inflationary mechanism as an alternative to the standard inflaton models and the like.  The dots in \eqref{rLRVM} denote terms of  order $H^6$ and higher,  which we expect  to remain further suppressed. Recall that the expansion \eqref{GRVE}  is supposed to emerge from solving the aforementioned  RG  equation \eqref{runningmu}.  Since the terms  $\sim H^6$ and higher in that RG equation are  accompanied by coefficients with negative dimension of mass,  such terms should follow the fate dictated by the decoupling theorem  in QFT\,\cite{AppelquistCarazzone75}, and  hence they will be ignored from now on. The notation $\Lambda (H)$ is used in \eqref{rLRVM}, in order to stress the connection of the RVM with a ``running cosmological  vacuum energy term'' with an equation of state (EoS) identical to that of a cosmological constant:\nit{\footnote{We note for completeness, that a similar situation, with an EoS of the form \eqref{wl} but a vacuum energy $\Lambda (t)$ depending on cosmic time, also appears to characterise the vacuum of some non-critical-string-theory cosmologies, with space-time brane defects~\cite{emntime}, where the induced running of the cosmological constant with the cosmic time is a direct consequence of the interpretation of the target time as a local renormalization scale on the world-sheet of the non-critical string. The considerations of \cite{bms1,bms2,bms3,msb} provide another non-trivial connection of RVM with rather generic critical, this time, string cosmological models with gravitational anomalies.}}
\be\label{wl}
w_{\rm RVM} = -1.
\ee

Within the standard RVM framework a smooth evolution of the Universe is assumed, where the numerical coefficients of the various terms in
\eqref{rLRVM} are assumed the same at various epochs~\cite{rvmhistory,GRF2015,solaentropy}. In this respect, the late Universe evolution is dominated by the true cosmological constant $c_0$ and $H^2$ terms, which imply a slight, but phenomenologically important and observable deviation from the $\Lambda$CDM evolution, which mimics quintessence behavior and helps to smooth out the aforementioned tensions existing in the $\CC$CDM~\cite{rvmpheno1,rvmpheno2,rvmpheno3}.  On the other hand, the early-Universe eras are dominated by the $\sim H^4$-terms, which can lead to dynamical inflation~\cite{rvmhistory,GRF2015,solaentropy}, without the need for external inflaton fields~\cite{inflaton}.

To understand this, let us one consider  a generic RVM model, which includes matter/radiation excitations of the running vacuum, with energy density $\rho_m$ and pressure
\be\label{eosmatter}
p_m = w_m \, \rho_m~,
\ee
where  $w_m$ denotes the relevant EOS,  with $w_{m}=1/3$ for  radiation (relativistic matter in general), and $w_{m}=0$ for non-relativistic matter.
The pertinent
cosmological (Friedmann) equations in the presence of a running $\Lambda (t)$ are given by:
\be
 \kappa^{2}\rho_{\rm tot}=\kappa^2 \rho_{m}+\Lambda (t)= 3H^2 \;,
\label{friedr} \ee
\be
\kappa^{2}p_{\rm tot}=\kappa^2 p_{m}-\Lambda (t) =-2{\dot H}-3H^2,
\label{friedr2}
\ee
where the overdot denotes derivative with respect to cosmic time
$t$, and
\be\label{tot}
\rho_{\rm
tot}=\rho_{m}+\rho_{\Lambda}, \quad
p_{\rm
tot}=p_{m}+p_{\Lambda}=w_m \, \rho_m -\rho_{\Lambda},
\ee
are the total energy density and pressure density
of both the vacuum (suffix $\Lambda$) and matter/radiation (suffix $m$) terms, and in the second equation we used  \eqref{wl}.

It is important to stress that, unlike in the standard  $\Lambda$CDM model of cosmology,  where $\Lambda=$constant,
in the RVM, for which $\Lambda(t)$ depends on cosmic time ({\it cf.} \eqref{GRVE})
 there are nontrivial interactions between radiation/matter and the
vacuum, which are manifested in the modified conservation equation for the matter/radiation energy density $\rho_m$, obtained from the corresponding Bianchi identities of the RVM Universe:
\begin{equation}\label{frie33}
\dot{\rho}_{m}+3(1+\omega_m )H\rho_{m}=-{\dot{\rho}}^{\Lambda}_{\rm RVM}\,.
\end{equation}

The alert reader should notice that, in view of \eqref{Hdot}, the right-hand side of \eqref{frie33} consists of terms of order $H^3$ and higher, and also suppressed by factors $q + 1$ which, during the inflationary phase are almost zero (in fact, during the inflationary phase, for which the $H^4$ term in \eqref{rLRVM} dominates, one has
$\dot \rho^\Lambda_{\rm RVM} =  \mathcal O \Big((q+1) H^5\Big)$).

Taking into account the RVM expression \eqref{rLRVM} and  using  equations \eqref{friedr} and \eqref{friedr2}, one arrives at:
\be\label{Hdot2}
\dot H + \frac{3}{2} \, (1 + \omega_m) \, H^2 \, \Big( 1 - \nu - \frac{c_0}{H^2} - \alpha \, \frac{H^2}{H_I^2} \Big) =0~,
\ee
which, on using
\eqref{frie33}, leads to a solution for $H(a)$ as a function of the scale factor and the equations of state of `matter' in RVM~\cite{rvmhistory}:
\begin{equation}\label{HS1}
 H(a)=\left(\frac{1-\nu}{\alpha}\right)^{1/2}\,\frac{H_{I}}{\sqrt{D\,a^{3(1-\nu)(1+\omega_m)}+1}}\,,
\end{equation}
where $D>0$ is an integration constant.  Notice that in arriving at \eqref{HS1}, we ignore the $c_0$-dependent term in front of 1, but we keep the order $H^4$ terms in the expansion
\eqref{rLRVM}, as they play a crucial r\^ole in the early Universe.  On assuming $|\nu|\ll1$, which is consistent with the standard RVM phenomenology~~\cite{rvmpheno1,rvmpheno2,rvmpheno3} (which implies $\nu \sim 10^{-3}$, consistent with previously-existing theoretical estimates~\cite{Fossil2008}), one observes
that for early epochs of the Universe, where the scale factor $a \ll 1$, one has
$D\,a^{3(1-\nu)(1+\omega_m)} \ll 1$, and thus
an (unstable) dynamical de Sitter phase~\cite{rvmhistory}, characterised by
an approximately constant
\be\label{desitter}
H_{\rm de~Sitter}  \simeq \left(\frac{1-\nu}{\alpha}\right)^{1/2}\,\, H_{I}
\ee
emerges.

\begin{figure}
\begin{center}
\includegraphics[scale=0.55]{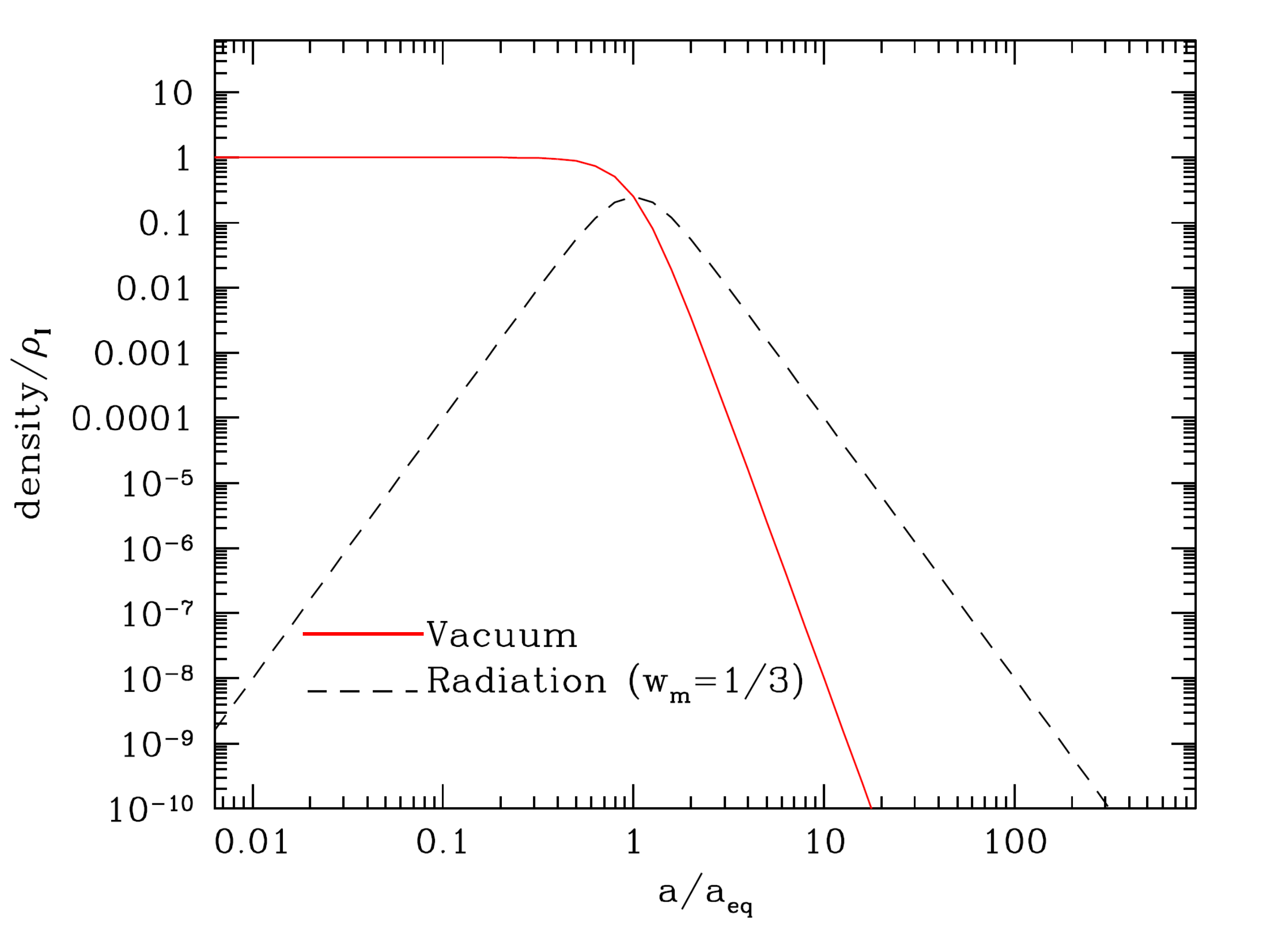}
\end{center}
  \caption{\it Evolution of matter and vacuum  energy densities \eqref{eq:rhorfinal}-\eqref{eq:rhoLfinal} as a function of the scale factor during the primeval inflationary epoch  in RVM inflation, and their transition into the FLRW radiation era.  Both densities are normalized with respect to the primeval vacuum energy density $\rho_I$, and the scale factor with respect to the equality point, $a_{eq}$,
for which $\rho_{\Lambda} = \rho_r$ (see the text).  The relativistic matter component is  shown as a black dashed line,
whereas the  vacuum energy density as a red solid line. \label{Fig:RVM_Inflation}}

\end{figure}

On the other hand, in radiation-dominated epochs of generic RVM models, with an EOS  $w_m =1/3$,  one obtains from \eqref{HS1}:
\begin{equation}\label{HS1rad}
 H(a)=\left(\frac{1}{\alpha}\right)^{1/2}\,\frac{H_{I}}{\sqrt{D\,a^{4}+1}}\,,
\end{equation}
for $|\nu | \ll 1$,
which connect smoothly with the early (unstable) de Sitter phase at $Da^{4} \ll 1$, during which $H$ remains approximately constant \eqref{desitter}.

 The above expressions for the RVM Hubble function describe the early universe,  when the scale factor is  $a\simeq 0$. To make the connection with the current universe ($a\simeq 1$, in units of the present-day scale factor) more apparent,  it is convenient to rescale some quantities. In particular, it is convenient to rewrite  $D$ in terms of a more physical parameter of the early stages  of the cosmic evolution: the point where inflation stops. To determine this point we have to compute the energy densities of matter and vacuum and find the equality point  $a_{\rm eq}$  between them.  Such a point is defined by the condition $\rho_r(a_{\rm eq})=\rho_\Lambda(a_{\rm eq})$ and can then be used to trade $D$ for  $a_{\rm eq}$.  In fact, it is even more convenient  to use a rescaled form of the latter:
 \begin{equation}\label{eq:ahat}
  \ha\equiv \frac{a}{\astar}\,,
\end{equation}
where $\astar$ is related to $a_{\rm eq}$ through\,\cite{solaentropy}
\begin{eqnarray}\label{aeq0}
D=\frac{1}{1-2\nu}\,a_{\rm eq}^{-4(1-\nu)}\equiv\astar^{-4(1-\nu)}\,.
\end{eqnarray}
It is clear that  $\astar$ is essentially the same as  $a_{\rm eq}$ since $|\nu|\ll1$ but the former  is  a more convenient notation simplifying  the writing of the formulae.
Thus the Hubble function and the associated energy densities of matter and vacuum energy read, respectively\,\cite{rvmhistory,GRF2015,solaentropy}:
\begin{eqnarray}\label{hubbleeq0}
H(\ha)=\frac{\tHI}{\sqrt{1 +\ha^{4(1-\nu)}}}\,,
\end{eqnarray}
\begin{eqnarray}\label{eq:rhorfinal}
\rho_r(\ha)&=&\trI (1-\nu)\,\frac{\ha^{4(1-\nu)}}{\left[1+  \ha^{4(1-\nu)}\right]^2}
\end{eqnarray}
and
\begin{eqnarray}\label{eq:rhoLfinal}
\rho_\Lambda(\ha)&=&\trI\, \frac{1+\nu \ha^{4(1-\nu)}}{\left[1+  \ha^{4(1-\nu)}\right]^2}\,.
\end{eqnarray}
In the above equations we have also rescaled  $H_I$ and $\rho_I={3 H_I^2}/{\kappa^2}$  as follows:
\begin{equation}\label{eq:tHI}
\tHI=\sqrt\frac{1-\nu}{\alpha}\,H_I\,,\ \ \ \ \ \ \trI=\frac{3}{\kappa^2}\,\tHI^2\,.
\end{equation}
As it is evident from  \eqref{eq:rhorfinal} and (\ref{eq:rhoLfinal}), the values of $\tHI$ and  $\trI$ are precisely the Hubble rate and  vacuum energy density at $a=0$:  $H(0)=\tHI$,  $\rL(0)=\trI$, i.e. at the start of the inflationary epoch.  We had initially called the former quantity the de Sitter scale value of the Hubble rate, see  \eqref{desitter}. 

The  previous formulae  clearly demonstrate the existence of a transfer of energy from  vacuum into matter during the cosmic evolution, as clearly illustrated in Fig.\,\ref{Fig:RVM_Inflation} (see \cite{rvmhistory} and \cite{JSPRev2015,GRF2015,solaentropy} for more details).  Such process is especially pronounced in the early stages of the cosmic evolution.   At $\ha=0$, we confirm that the vacuum energy is maximal, whilst the matter density is zero. From this point onwards the vacuum decay continues until reaching an equality point at $a_{\rm eq}$, where $\rho_r(a_{\rm eq})=\rho_\Lambda(a_{\rm eq})$.   An order-of-magnitude estimate of the point  $a_{\rm eq}$ $\simeq \astar$ can be easily obtained by  taking into account that,  in the asymptotic limit ($\ha\gg1$, i.e.  $a\gg a_{\rm eq}$) deep into the radiation epoch, the radiation density (\ref{eq:rhorfinal}) behaves as:
\begin{eqnarray}\label{eq:rhorfinal2}
\rho_r(a)&=&\trI (1-\nu) \ha^{-4(1-\nu)}=\trI (1-\nu) \astar^{4(1-\nu)}\,a^{-4(1-\nu)}\,.
\end{eqnarray}
This form should be familiar, given that we are able recover the standard behavior of the radiation density,  $\rho_r(a)\sim\rho_{r0} a^{-4(1-\nu)}$,  from  the exact expression \eqref{eq:rhorfinal},  up to a tiny correction which depends on $\nu$ (recall that $|\nu|\ll1$).
Imposing that \eqref{eq:rhorfinal2}  must reproduce  the radiation density at present: $\rho(a=1)=\rho_{r0}$ gives a handle to estimate  $\astar$ and hence $a_{\rm eq}$.  Indeed, using also that the energy density at the inflationary period must be of order of the GUT one, $\trI\sim M_X^4$,  with  $M_X\sim 10^{16}$ GeV and that the current radiation energy density in units of the critical density  is of order  $\Oro=\rro/\rco\sim 10^{-4}$, we can easily derive
\begin{equation}\label{eq:astar}
  a_{\rm eq}\simeq  \astar\simeq \left(\Oro\,\frac{\rco}{\rho_I}\right)^{1/4} \sim \left(\Oro\right)^{1/4}\frac{\left(H_0\,\MPl\right)^{1/2}}{M_X} \sim  10^{-29}\,.
 \end{equation}
This numerical value places the balance  point between radiation and vacuum energy densities in the very early universe, virtually at the end of inflation or, equivalently, the incipient radiation-dominated epoch\,\footnote{It may be illustrative to compare it  in order of magnitude   with  the (much more recent) equality point between radiation and nonrelativistic matter:  $a_{\rm EQ}\sim 10^{-4}$\cite{KolbTurner}.}.

However, as we discussed in \cite{bms1,bms2}, and shall  further address below, in the context of a specific string-inspired RVM model the `matter content' is different from that of relativistic matter. Moreover, there is no perfectly smooth evolution from the de Sitter inflationary eras to the current era, as there are phase transitions at the exit from inflation, which result in new degrees of freedom entering the effective field theory, although qualitatively the main features of RVM are largely preserved.  In fact, the above discussion shows that a smooth evolution can lead to a  reasonable picture, in which the standard radiation dominated epoch ($\rr\sim a^{-4}$) follows continuously from the inflationary one. A more realistic scenario, however, requires an intermediate step (phase transition)  in which the  Kalb-Ramond (KR)  axion from the effective low-energy string theory will play a significant role, see Sec.\,\ref{sec:string}. Needless to say, this is an important point of the stringy version of the RVM, which was absent in its original form, and is under discussion  in this article.

At this point, some important remarks are in order, which help elucidate the connection between the RVM physics of the early universe and the one expected at the present era.
From the generic RVM expression for the vacuum energy density \eqref{rLRVM}, one might expect that the connection with the current universe is obtained in the limit  $\alpha\to 0$. However, such a limit is undefined for both the Hubble rate \eqref{HS1rad} and the energy densities \eqref{eq:rhorfinal}-\eqref{eq:rhoLfinal} and hence it cannot be implemented ~\cite{solaentropy}.  Indeed, as can be inferred from  \eqref{HS1}, a crucial virtue of the RVM approach is that the initial value of the Hubble rate for a scale factor $a \to 0$ ({\it cf.} \eqref{desitter}), $H(0)=\tHI\simeq {H_I}/\sqrt{\alpha}$,  is {\it finite}
and hence there is  {\it no initial singularity} for the RVM Universe. To ensure this feature, it is indispensable that $\alpha>0$ (strictly) in \eqref{rLRVM}.

Par contrast, when $\alpha \to 0$,  so that the $H^4$ term is subdominant compared to $H^2$ in \eqref{rLRVM},
the entire RVM physics of the early universe disappears since no non-singular solution  can exist at $a=0$,  except the trivial one corresponding to a static Universe ($H=0$). Indeed, in such a case,
for matter/radiation dominance, obtained from \eqref{Hdot2} by setting $\alpha=0$ {\it and} assuming $c_0 \ll H^2$, which justifies ignoring the $c_0$ term in \eqref{Hdot2}, the solution for $H(a)$ exhibits an initial singularity, as $a \to 0$, in the form:
\be\label{lcdmH}
H(a)_{\rm H^4-ignored}^{\rm matter/radiation~dominance} \sim a^{-\frac{3}{2}(1 + w_m)(1-\nu)},
\ee
with the the standard $\Lambda$CDM case corresponding to $\nu=0$. We shall come back to this point  in section \ref{sec:string}, when we discuss early Universe phases in the context of string-inspired RVM.

In other words, it is only when the term $H^4$  is present, and carries a positive coefficient, that nonsingular solutions to the cosmological equations can exist.  A nonvanishing value for $\alpha$ is mandatory and hence the way to connect the early universe and the current universe in the context of the RVM model \eqref{rLRVM} is \textit{not} by performing a zero limit of the parameters $\nu, \alpha$  but by just letting the evolution of $H$ to interpolate between the different epochs.  The  two coefficients must be present and nonvanishing in the entire cosmic history.  The connection between epochs  is implemented dynamically through  the relative strength of  $H^4$ vs $H^2$  that changes when moving from early epochs to the current ones, in which  the former term is completely negligible compared to the latter.  The function \eqref{rLRVM} is indeed a continuous function of $H$ and  moves from $H^4$ dominance into $H^2$ dominance, and finally we are left with a mixture of a constant (and dominant CC) term plus a tiny correction $\sim\nu H^2$. This means that, according to the RVM,  the dark energy in the current universe is evolving, as  there is still a mild dynamical vacuum energy $\sim H^2$  on top of the dominant term (the cosmological constant).  Although it may create the illusion of quintessence,  it is just a residual dynamical vacuum energy that helps to improve the fit to the data\,\cite{rvmpheno1,rvmpheno2,rvmpheno3}.

While this is the basics of the standard picture  within the RVM\,\cite{JSPRev2013,JSPRev2015},  in a stringy RVM formulation the contributions to the current-era cosmological constant may come from condensation of much-weaker GW, and the evolution cannot be described by a smooth solution \eqref{HS1}, connecting the initial inflation to the current epoch\,\cite{bms1,bms2}. More details will be given here.  Basically, the GW condensation leading to the initial and current-era (approximately) de Sitter space times are viewed as dynamical phase transitions, whose presence affect the smoothness of the evolution of the stringy Universe. In this respect, the RVM can be seen as providing an effective description within each epoch, with non-trivial coefficients of the various $H$-powers in the string-inspired RVM analogue of \eqref{rLRVM}, which are computed microscopically in the various eras, as we discussed in \cite{bms1} and revisit below.

Having said that, though, we also stress that in the stringy RVM the $H^4$ term in the vacuum energy density arises from condensation of GW, and is linked to the gravitational anomalies, which are non zero only in the presence of GW~\cite{bms1}. When the latter are absent, as, for instance, may happen deep in a pre-RVM-inflationary phase of the Universe, the $H^4$ terms vanish, consistent with the fact that in the stringy RVM, the GW-induced inflationary phase is associated with a phase transition, that of the formation of the anomaly condensate.  Nonetheless, even in such a case, the initial singularities at the Big-Bang point may be absent due to the higher-order-curvature corrections of the string-inspired  gravitational action~\cite{art}, which must be taken into account at such an early epoch. So in this respect, the spirit of the RVM, as implying the absence of initial singularities in the Universe, is maintained by its stringy version. We shall discuss such issues in sections \ref{sec:string} and \ref{sec:origin}.

In the following two subsections, we shall discuss some additional features of the standard RVM, before proceeding to a discussion of the stringy RVM version in section \ref{sec:string}. Such features are shared by both the conventional and stringy RVM frameworks, and their description will help the reader appreciate the connection between the two formalisms at an effective field theory level. In the next subsection \ref{sec:Staro-inflation} we shall review a comparison/contrast of the inflation within the RVM framework with that of the Starobinsky model. Although both models do not require external inflation fields, nonetheless there are important differences, which we will  discuss in some detail.  In subsection \ref{sec:ThermoAspects} we shall discuss the thermodynamical aspects of the RVM framework. We stress that both of these features are important for the compatibility of the (stringy) RVM approach with the swampland criteria~\cite{msb} of embedding it properly in an UltraViolet (UV) complete theory of quantum gravity, such as strings, discussed briefly in  section \ref{sec:swamp}.

\subsection{Short comparison  of  RVM inflation with Starobinsky inflation}\label{sec:Staro-inflation}

In this subsection we compare the RVM-inflation  with Starobinsky's inflation\,\cite{staro,staro2}, which  is based on adding a classical term $R^2$ to the usual EH action:
\begin{equation}\label{eq:StarobinskyAction}
S=\int d^4x\sqrt{-g}\left(-\frac{R}{16\pi G}+ \tilde{b}R^2\right)+S_{\rm
matter}\,.
\end{equation}
We will follow closely the discussion of \cite{JSPRev2015}, except that we denote the (dimensionless) coefficient of the $R^2$ term by $\tilde{b}$ in order not to be confused with the KR field $b(x)$ which will appear later on.  The $R^2$ term is  present  explicitly  in the Starobinsky  classical action.  It is usually written as $\tilde{b}=\mpl^2/(6M_{\rm sc}^2)$,
where $M_{\rm sc}$  is a parameter of mass dimension [+1] -- playing the role of the scalaron mass in the original model\,\cite{staro}.  In the case of the scalar-field model non-minimally coupled to gravity, Eq.\,\eqref{eq:Sphi}, the $R^2$ term is part of the vacuum action,  in order to absorb the divergences of the renormalization procedure, but the value of its coefficient is not needed, only the renormalization shift (i.e. the counterterm)  associated with its variation ($a_1\rightarrow a_1+\delta a_1$)  should be specified~\cite{Cristian2020}. However, in both cases the variation term, or, more properly, the functional differentiation of $R^2$ with respect to the metric, is involved in the effective action, which gives the result \eqref{eq:H1munu}.

\begin{figure}
\begin{center}
\includegraphics[scale=0.35]{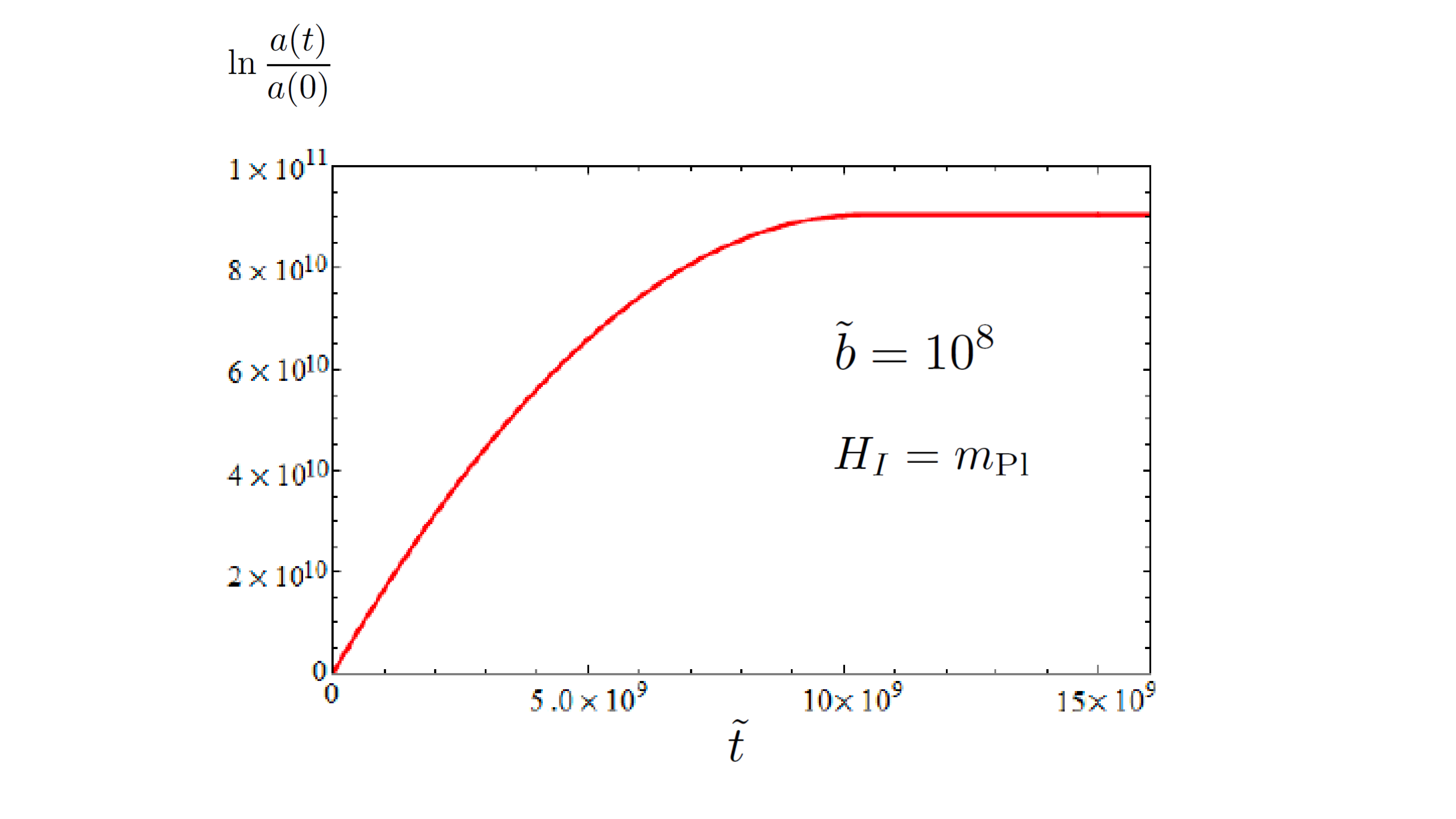}\\
\includegraphics[scale=0.35]{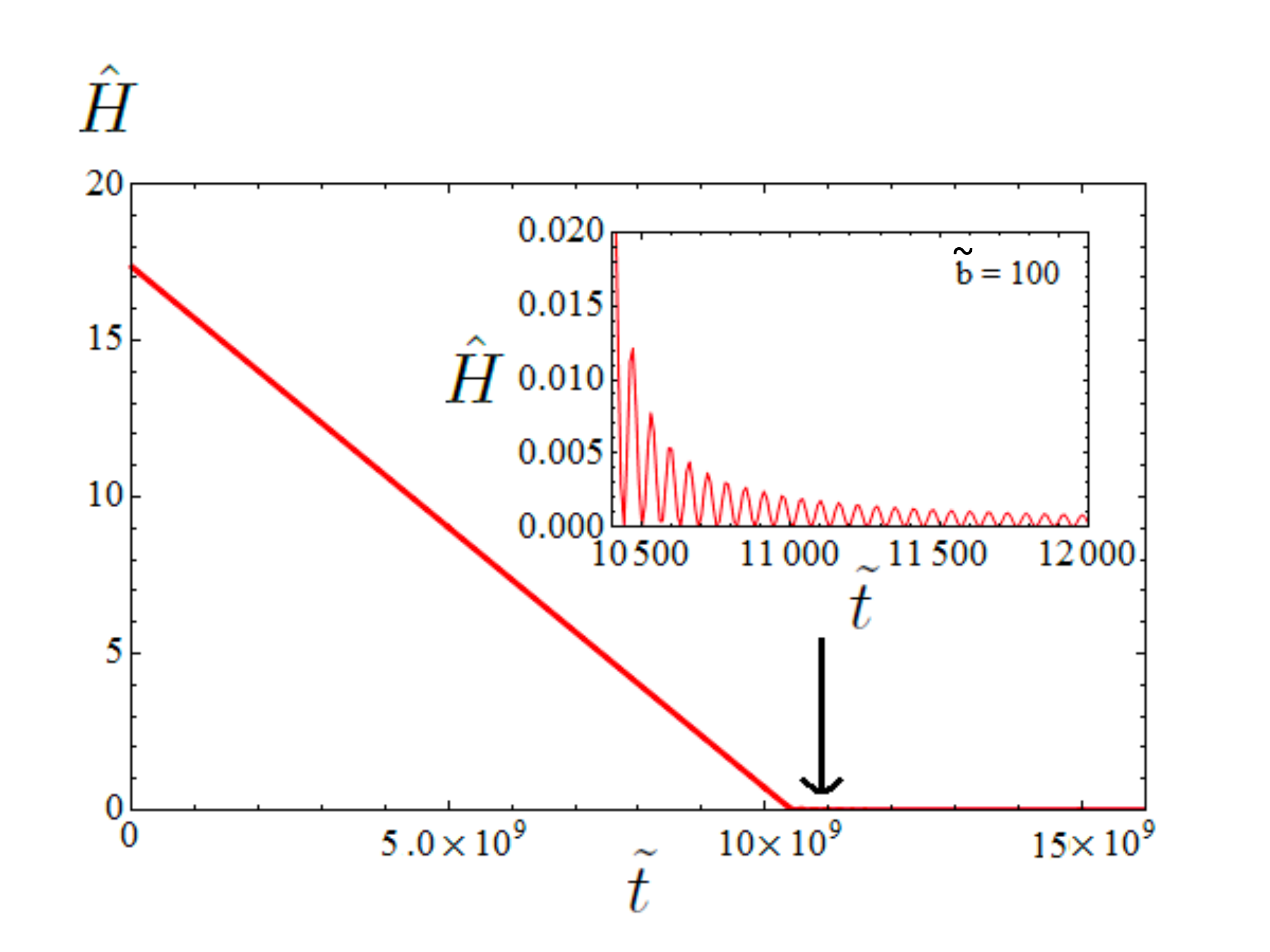}
\end{center}
\caption{\it The exact (numerical) inflationary solution of Eq.\,\eqref{eq:HubbleEq} corresponding to the Starobinsky model  \eqref{eq:StarobinskyAction}.  On the upper panel it is shown the initial exponential growth $a\sim e^{H_I\,t}$ of the scale factor and its stabilization into the radiation regime $a\sim t^{1/2}$. We have taken $H_I=\mpl$  and $\tilde{b}=10^8$\,\cite{Copeland2015}. On the lower panel we display the corresponding behavior of the Hubble function.  The straight line is described by the approximate inflationary solution \eqref{eq:ScaleFactor1P}.  In  the inner window we show the characteristic oscillations when the Universe leaves the inflationary phase and enters the radiation epoch in the form $a\sim t^{1/2}+{\it oscillations}$. In that window we have now  set $\tilde{b}=100$ to make the oscillations more apparent. Time has been rescaled as $\hat{t}=(M_P/\sqrt{96\pi})\,t$, and $\hat{H}=\hat{\dot{a}}/a$ is the (dimensionless) Hubble rate in the rescaled time.}
\label{Fig:Starobinsky}
\end{figure}

The field equations associated with the variation of the action \ref{eq:StarobinskyAction} are easily  obtained. Assuming a single matter  component behaving as an ideal fluid of density  $\rho$ and pressure $p$ ,  they read
\begin{equation}\label{eq:FieldEquationsStaro}
G_{\mu\nu}-32\pi G\tilde{b}(\nabla_\mu\nabla_\nu R + g_{\mu\nu}\Box R+RR_{\mu\nu}-\frac{g_{\mu\nu}}{4}R^2)
=8\pi G\, T_{\mu\nu}\,,
\end{equation}
with
\begin{equation}\label{eq:Tmatter}
T_{\mu\nu}=-p g_{\mu\nu}+(\rho+p)U_\mu U_\nu
\end{equation}
the EMT of the matter fluid and $U_\mu$ its four-velocity field.  In the early universe we may assume that the latter is a relativistic fluid, so we have   $p_r=\rho_r/3$ for the matter EoS.
Next, writing  down the $(\mu,\nu)=(0,0)$ and $(\mu,\nu)=(i,j)$ components of the field equations \eqref{eq:FieldEquationsStaro}  in the  spatially flat FLRW metric,  as usually done in the GR case, we may combine them to find:
\begin{equation}\label{eq:HubbleEq}
2H^2+\dot{H}+48\pi
G\tilde{b}(2\dddot{H}+14\ddot{H}H+24H^2\dot{H}+8\dot{H}^2)=0\,.
\end{equation}
One can easily see that for $\tilde{b}=0$ we recover the expected GR equation $2H^2+\dot{H}=0$ characteristic  of the pure radiation era ($a\sim t^{1/2}$). However, when $\tilde{b}\neq0$,  solving the nonlinear Eq.\,\eqref{eq:HubbleEq} can be  a challenge.   Even before making any attempt  in this direction, it is pretty obvious that no $H=$constant solution is possible.  So a steady  Hubble rate  is not the trademark of Starobinsky inflation. Nonetheless, inflation can still be triggered by an initial phase characterized by   $\dot{H}=$constant, instead of  $H=$ constant.  This is confirmed  by the exact numerical solution given in Fig.\,\ref{Fig:Starobinsky}\, (see \cite{JSPRev2015} for details). Since $\dot{H}$ remains essentially constant until we are very near the end of the inflationary phase (as it is obvious from the straight line in the plot on the lower panel in Fig.\,\ref{Fig:Starobinsky}), we can solve (\ref{eq:HubbleEq}) by
neglecting $\dot{H}/H^2\ll1$ and all higher derivative terms.  This yields  $576\pi G_N\,b\,\dot{H}=-1$,  which is solved by $H(t)=H_I-{\mpl^2t}/{576\pi b}$ (the equation of the aforesaid straight line). Integrating once more we get the approximate solution for the scale factor:
\begin{equation}\label{eq:ScaleFactor1P}
a(t)\sim e^{H(t) t}\sim\,e^{H_It}e^{-\frac{M_{\rm sc}^2t^2}{192\pi}}\,.
\end{equation}
Obviously, we must have $\tilde{b}>0$ (hence a well-defined scalaron mass, $M_{\rm sc} >0$) in order to have a stable inflationary solution until the inflationary phase is extinguished at around $t_{f}\simeq192\,\pi\,H_I/M_{\rm sc}^2$. The larger the $\tilde{b}$ ({\it i.e.} the lighter the scalaron) the longer the inflationary time.

When  the $\dot{H}\simeq$constant period is over,  a final phase,  characterised by rapid oscillations of the gravitational field, produces a reheating period (see Fig.\,\ref{Fig:Starobinsky}).  This period is usually associated to the slow roll of the scalaron prior to its decay into relativistic particles, which reheat the universe (note that the intermediate state dominated by scalarons is not hot, but cold, since the scalarons are heavy particles).  This is how the standard reheating  picture  in Starobinsky's inflation proceeds and leads the universe into the radiation-dominated epoch. During that epoch, the scalar curvature vanishes identically  $R=0$, and the higher order terms cease to be relevant from that point onwards in the cosmic evolution.  This also happens in the RVM case discussed in the previous section, except that for Starobinsky inflation nothing is left of the inflationary phase at the present time, whereas after  RVM-inflation, e.g. during the matter-dominated period and the DE epoch, the RVM still provides terms proportional to  $H^2$  in \eqref{rLRVM}, which remain currently active and make DE a dynamical quantity, see particularly the form \eqref{eq:RVM2}.

Unfortunately, the missing  $H^4$ terms in Starobinsky inflation, are also missing in the explicit QFT calculation of \cite{Cristian2020} aiming at reproducing the entire RVM structure \eqref{rLRVM}. In that calculation only the $\sim H^2$ terms are found plus the higher order contributions  given in Eq.\,\eqref{RenormalizedVE}, which depends on the same structures $\dot{H}^2,\, H^2 \dot{H}$  and  $ H\ddot{H}$  appearing in Starobinsky inflation,  with no pure $H$-term of the form $\sim H^4$.    So the non-minimally coupled scalar field  action\eqref{eq:Sphi} with quantum corrections and the Starobinsky action \eqref{eq:StarobinskyAction} both lead to common terms of adiabatic order $4$ which vanish  for $H=$constant, as we have just seen from the preceding discussion.
The missing term $\sim H^4$,  which  is  the hallmark of RVM inflation as compared to Starobinsky inflation\,\cite{staro}  and is crucial for generating the characteristic form of RVM-inflation based on a period of  $H=$constant  rather than a period of $\dot{H}=$ constant,  though, will be finally generated in the stringy RVM scenario~\cite{bms1}, to be discussed in Sec.\,\ref{sec:string}. 

Obviously, once the $H^4$ term is secured,  the associated thermodynamical features to each type of inflation are very different in the two inflationary mechanisms (Starobinsky and RVM).  In the original version of RVM inflation there is no ``reheating'';  the vacuum decays into particles through a continuous ``heating up'' period  rather than through an intervening state of material particles (inflatons or scalarons)\,\footnote{That the RVM inflation, when formulated as a scalar quantum field theory, cannot be described as a typical scalaron-induced inflation has been recently discussed in \cite{vacuumon}.}
  This is obvious from the cosmological solution of the RVM equations presented in the previous section and can be appraised graphically in Fig. \,\ref{Fig:RVM_Inflation}.
 In the next subsection we shall close the discussion on the conventional RVM by reviewing  some of its specific thermodynamical implications. These will be retained in full  in the stringy version of the RVM, of course,  since the string inspired formulation actually provides a raison d'\^etre for the preeminent $H^4$ power in the RVM structure \eqref{rLRVM}.

\subsection{Some thermodynamical aspects of RVM-inflation and the GSL \label{sec:ThermoAspects}}

Because $H=$const characterizes  the initial period of RVM-inflation, it is obvious that the RVM can provide an explanation for  the horizon problem since the particle horizon (essentially $H^{-1}$) remains much larger than the size of the universe when we recede to  very early times where  $a\to 0$.  This was already amply clarified in \cite{JSPRev2015}.  Here, however,  we would like to emphasize the solution of the horizon problem from the point of view of the large production of entropy and the fulfilment of the Generalized Second Law (GSL) by the RVM universe\,\cite{solaentropy}.  The GSL ideas for the universe are inspired from the situation with black holes (BH).  The GSL for BH's asserts that in all physical processes in which BH's are involved, the sum of the BH entropy, $S_{\rm BH}$, and the ordinary entropy of matter and radiation fields  in the BH exterior volume, collectively denoted as $\SV$, cannot decrease:
\begin{equation}\label{eq:GSLBekenstein}
\SVp+S^\prime_{\rm BH}\geq 0\,,
\end{equation}
where the prime indicates differentiation with respect to a convenient variable defining the evolution of the process.
The idea stems from Bekenstein~\cite{Bekenstein} who conjectured a proportionality between the BH entropy and the horizon area, which is based on Hawking's area theorem stating that the BH surface cannot decrease~\cite{Hawking71}.
The proposed BH entropy formula is the famous Bekenstein-Hawking formula:
\begin{eqnarray}\label{BHentropy}
S_{\rm BH}=\frac{k_B\,A}{4\ell_{\rm Pl}^2}=\frac{k_B c^3\,A}{4\, \hbar\, G}\ \ \ \longrightarrow\ \ \ S_{\rm BH}=\frac{A}{4\, G}\ \ \ \ ({\rm natural\ units)}\,.
\end{eqnarray}
For a  Schwarzschild's BH of mass $M$ the surface area is  ${A}=4\pi r_S^2$,  where  $r_S=2GM/c^2$ is the Schwarzschild radius.

These notions were later extended to cosmology  for the entire universe\,\cite{GH},  see e.g. \cite{Bousso2002,Faraoni2015} for a review. In this case the Schwarzschild radius is replaced by the apparent horizon (AH), let us call it $\ell_h$. So, formally the same Eq.\,\eqref{BHentropy} applies, but now the area $A$ is replaced by that of the AH: $\A=4\pi\ell_h^2$.  Since for spatially  flat FLRW geometries (the only ones we shall use here)  the AH coincides with the inverse of the Hubble rate\footnote{Although we can spare the reader a  formal definition of  AH here (see e.g. the above mentioned reviews), physically speaking, we can say that beyond the cosmological AH all null geodesics recede from the observer and no information can reach us. The closest intuitive notion to it may be the Hubble sphere, but the latter is only a particular case for spatially flat spacetime.  The edge of that sphere  is the ultimate lightcone for spatially flat universes since in it the galaxies recede at the velocity of light.  The  AH  is generally different from the event horizon. The latter is a null surface, whereas the former generally is not. When the event horizon exists, the AH is usually contained in it, or coincides with it. In cosmology, the AH is dynamical, and in the important case of the $\CC$CDM, and also for the RVM,  the presence of the cosmological constant term $\CC$, is such that the AH becomes eventually
an event horizon with $\ell_h=1/H_\CC$, where $H_\CC$ is the limiting value of Eq.\,\eqref{eq:lateH}, below, for $a\to\infty$. In the $\CC$CDM case,  one sets $\nu=0$.  The AH is  generally considered more suitable
for thermodynamics discussions than the event or particle horizons\,\cite{BakRey2000,CaiKim2005,CaiCao2007}. }
\begin{eqnarray}\label{eq;SAentropy}
\SA=\frac{\pi}{G\,H^2(t)}=\frac{\pi \mpl^2}{H^2(t)}\ \ \ \ ({\rm natural\ units)}\,.
\end{eqnarray}

Of the two entropy contributions in Eq.\,\eqref{eq:GSLBekenstein}, the last  ($S^\prime_{\rm BH}$) is the biggest in the RVM universe. Using \eqref{hubbleeq0}, we find
\begin{eqnarray}\label{Hradentropy}
\SA^{\rm early}(\ha)=\frac{\pi \,\mpl^2  \left[ 1+\hat{a}^{4(1-\nu)} \right]}{\tHI^2}\sim \frac{\mpl^4}{M_ X^4}\   \ha^{4(1-\nu)}\ \  \ (\ha\gg1)\,,
\end{eqnarray}
where the fast growing $\SA\sim a^{4(1-\nu)}$   holds deep in the radiation epoch. Insofar as  the radiation entropy from relativistic particles inside the horizon is concerned, it involves the radiation temperature,  $T_r$, related to the radiation density through $\rho_r=\frac{\pi^2}{30}g_\ast T_r^4$.  The calculation is similar to that of the comoving entropy\,\cite{KolbTurner},  except that we  have to replace the comoving volume by the horizon volume ($a^3\rightarrow V_h= \ell_h^3$), whence
\begin{eqnarray}\label{Rentropy}
\SV^{\rm rad}(\hat {a})&=&\frac43\frac{\rho_r}{T_r}\,V_h=\frac43\left(\frac{\pi^2 g_\ast}{30}\right)^{1/4} \frac{4\pi}{3}\frac{\rho_r^{3/4}}{H^3(a)} \sim \frac{\mpl^3}{M_ X^3} \  a^{3(1-\nu)}\,\ \  \ (\ha\gg1)\,.
\end{eqnarray}
In the last step we used (\ref{hubbleeq0}) and (\ref{eq:rhorfinal}).

Both $\SA^{\rm early}$ and $\SV^{\rm rad}$  increase very rapidly with the scale factor at this epoch, but the former is clearly dominant.  Particularly noteworthy is the following observation: even though  $S_{\cal A, V}'>0$ in both cases, the convex behavior  $S_{\cal A, V}''>0$  shows that the Universe does \textit{not} tend to equilibrium at this early stage.  For this we would need overall concave behavior :  $S_{\cal A}''+S_{\cal V}''<0$.  So the question arises as to whether the RVM  universe eventually reaches thermodynamical equilibrium.  One can answer this question in the affirmative, and the existence of a positive cosmological constant plays a crucial r\^ole in this.
To see that, let us note that the late time behaviour is different from that displayed in the above equations.  Differentiating  the expression \eqref{eq:RVM2} describing the running of the VED at low energies, we find $d\rv=(3\nu/8\pi) \mpl^2 dH^2= \nu\,(d\rho_m+d\rv)$, where in the second step  Friedmann's equation has been used. Thus, we  we arrive at the following relation between the infinitesimal variations of the  energy densities of vacuum and matter in the RVM framework:
\begin{equation}\label{eq:ratiodiff}
  d\rv=\frac{\nu}{1-\nu}\,d\rho_m\,.
\end{equation}
Substituting on the \textit{r.h.s.} of Eq.\,\eqref{frie33} and trading the cosmic time differentiation for the differentiation with respect to the scale factor ($df/dt=aH df/da\equiv aHf'(a)$ for any function $f$)   we arrive at the following differential equation for the matter density with respect to the scale factor:
\begin{equation}\label{eq:diffconservationMatter}
\rho_m'+\frac{3}{a}(1+w)(1-\nu)\,\rho_m=0\,.
\end{equation}
Its integration provides
\begin{equation}\label{eq:rhoMatter}
\rho_m(a)=\rho_{m 0}\, a^{-3(1+w)(1-\nu)}\,.
\end{equation}
This result now can be used back into \eqref{frie33} to derive the evolution of the vacuum energy density explicitly in terms of the scale factor:
\begin{eqnarray}
\rv(a) &=& \rvo + \frac{\nu\,\rho_{m 0}}{1-\nu}\left(a^{-3(1-\nu)}-1\right)\,. \label{eq:rhoVRVM}
\end{eqnarray}

From these densities  we can compute  the (square) of the  late-eras Hubble rate at low energies (in the matter and DE epochs, for which $w=0$ for the dust component):
\begin{eqnarray}\label{eq:lateH}
H_{\rm late}^2=\frac{H_0^2}{1-\nu}\,\left[\Omo\,a^{-3(1-\nu)}+\OLo-\nu\right]\,,
\end{eqnarray}
where  $\Omo+\OLo=1$,  with $\Omega_{i\,0}$, $i=m,\Lambda$, denoting present-day energy densities for matter ($i=m$) and vacuum ($\CC$) in units of the critical density of the Universe.  Notice that  $H_{\rm late}\to$ constant as the cosmic time evolves.  As expected, it boils down to  the RVM tending to  $\CC$CDM for $\nu\to 0$.
Upon inserting Eq.~\eqref{eq:lateH} in Eq.~\eqref{eq;SAentropy},  we can calculate the entropy of the AH near our time and into the future:
\begin{eqnarray}\label{eq:SAreason}
\SA^{\rm late}(a)=\pi\,\frac{\mpl^2}{H_{\rm late}^2(a)}=\frac{\pi \mpl^2\,(1-\nu)}
{H_0^2\,\left[\Omo\,a^{-3(1-\nu)}+\OLo-\nu\right]}\,,
\end{eqnarray}
This contribution is overwhelmingly large  as compared to that of the radiation energy density in the same late epochs of the cosmological evolution.  Recall that in these epochs the radiation energy density behaves as in  Eq.\,\eqref{eq:rhorfinal2}.  Furthermore, substituting $H\rightarrow H_{\rm late}$  in \eqref{Rentropy} and taking into account $H_{\rm late}\to$ constant,  we find a fast drop of the radiation entropy  inside the AH when the universe evolves into the future:
\begin{eqnarray}\label{RentropyLate}
\SV^{\rm rad late}(a)\sim \frac{\rho_r^{3/4}}{H_{\rm late}^3(a)} \sim  a^{-3(1-\nu)}\,.
\end{eqnarray}

\begin{figure*}
\begin{center}
\includegraphics[width=11cm,height=5.5cm]{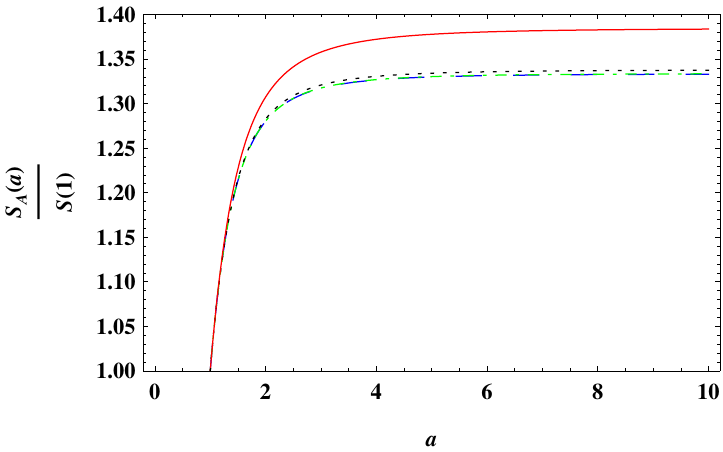}
\includegraphics[width=11cm,height=5.5cm]{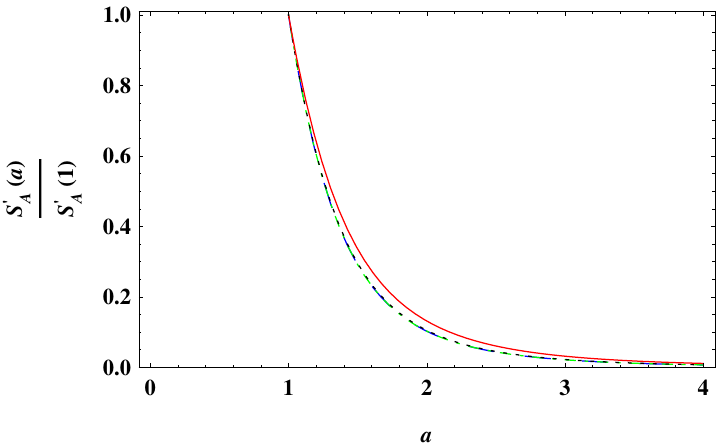}
\end{center}
\caption{\it Late time evolution of the the horizon entropy $\SA(a)$, as gven in Eq.\,\eqref{eq:SAreason} and its first  derivative, from the current universe into the future. The upper panel shows the ratio $\SA(a)/\SA(a=1)$, normalized to its current value. The lower panel shows the first  derivative with respect to the scale factor, also normalized in the same way.  We can see quite evidently  from the plots (and confirmed by the calculations) that  $S_{\cal A}' >0$ and $S_{\cal A}''<0$. Therefore, we can assert that  the GSL is ultimately preserved by the RVM evolution  since the numerical significance of $\SV$ is comparatively negligible, as explained in the text.  Plots are shown for four different values of the parameter $\nu$: $\nu=0.1$ (red solid), $\nu=0.01$ (black dotted), $\nu=0.001$ (green dash-dotted), and $\nu=0.0001$ (blue dashed).}\label{fig:EntropyHorizon}
\end{figure*}

At the same time we can neglect the  entropy from the material (nonrelativistic) particles, which is given by the product of the particle number density ($n$) times the specific entropy per material particle (typically taken to be one Boltzmann unit $\kappa_B$, hence $1$ in natural units) times the volume of the AH ($V_h=(4\pi/3)\ell_h^3$). In the asymptotic regime, we find
\begin{equation}\label{Mentropy}
  \SV^{\rm mat}(a)=n(a)\kappa_B V_h=\frac{4\pi}{3}\frac{n(a)}{H_{\rm late}^3(a)}\sim  a^{-3(1-\nu)}\,,
\end{equation}
where we used that $H_{\rm late}\to$ constant  for $a\gg1$  (late de Sitter era) and we accounted for the particle dilution law $n(a)\propto  a^{-3(1-\nu)}$ (involving a small $\nu$-correction), in accordance with Eq.\,\eqref{eq:rhoMatter}.
We find that both the radiaton entropy and that from the material nonrelativistic particles can be utterly neglected.
They definitely play no r\^ole  on judging  the fulfilment of the GSL in the late universe, as the GSL is in fact completely controlled by the holographic contribution from  the AH\,\eqref{eq:SAreason}.

Thus, it is enough to focus on the late time behavior of \eqref{eq:SAreason}, which is much more tamed than the rampant behavior of the early times, Eq.\,\eqref{Hradentropy} -- which was essential for the huge initial production of entropy.  The two behaviours are of course connected by the continuous vacuum energy density function \eqref{rLRVM}.  Explicit calculation shows that the dominant holographic contribution from the AH fulfils  $\SA'(a)>0$ \underline{\textit{and}} the concave condition $\SA''(a)<0$. These results can be appraised graphically in the plots of Fig.\,\ref{fig:EntropyHorizon}, see \cite{solaentropy} for an extended discussion.
Thus, the entropy rise eventually enters the  correct  behavior required by the Generalized Second Law applied to the universe with an apparent horizon.  The entropy finally reaches a maximum\,\cite{solaentropy}
\begin{equation}\label{eq:SAmax}
\SA^{\rm max}(a\to\infty)=\frac{\pi \mpl^2\,(1-\nu)} {H_0^2 (\OLo-\nu)}\sim 10^{122}\,.
\end{equation}
The RVM universe is thus  granted to eventually attain a state of  thermodynamical equilibrium  carrying an enormous amount of entropy, which is far bigger than in the standard $\CC$CDM model\,\cite{KolbTurner}. This solves comfortably the entropy problem in the $\CC$CDM and a fortiori the horizon problem\,\cite{solaentropy}.   In the absence of a cosmological constant term, we would have $\OLo-\nu=c_0/H_0^2=0$ in \eqref{eq:SAreason} and  the horizon  entropy  would still grow as $\sim\,a^{3(1-\nu)}$,  hence preventing the Universe from ever attaining  thermodynamical equilibrium within the  GSL.

\section{String-Inspired RVM:  primordial GW's  and stiff-axion ``matter''  \label{sec:string}}

In the previous sections we have summarized the RVM as a unified model for the cosmic evolution and we have described a variety of implications for the early universe and for the phenomenology of the current universe, including some thermodynamical considerations, which show the  consistency of the model with the Generalized Second Law. In the string-inspired RVM, all these appealing features remain essentially intact, but as discussed in \cite{bms1,bms2,bms3,msb} the various coefficients  in \eqref{rLRVM} {\it depend} on the era.  For instance, during the early inflationary epochs of the string-inspired model, when only degrees of freedom from the massless gravitational string multiplet are present, the coefficient $\nu_{\rm infl} < 0$~\cite{bms1}, while at later (radiation, matter and current) epochs, for which matter and cosmic gauge fields are present, the coefficient becomes positive, $\nu_{\rm rad/matt} > 0$, as the result of these additional contributions. Nonetheless, during each era, from the inflationary one to the present, the basic dominant features of the RVM are preserved.

However, there is an important feature for the every-early-Universe string-inspired RVM, which is not predicted in a generic RVM framework, but seems to be a specific feature of the string inspired model. This is the existence of {\it stiff} ``matter '' comprising of the KR and other stringy axions that may exist in string theories, as a result of compactification to four space-time dimensions. The presence of such stringy axions may lead to a stiff-matter era dominating the pre-inflationary era, in analogy with suggestions made in~\cite{stiff,stiff2}, but with a very different microscopic origin and properties of the stiff-matter, which comprises of stringy axions in our model, unlike the cold-fermion (baryon) gas of \cite{stiff}. Our axions are electrically neutral, but they do couple to gravitational-anomaly terms via CP-violating interactions. The latter play an important role in inducing dynamically an inflationary era, as we discussed in detail in \cite{bms1}.

Below we shall first review the basic features of this string-inspired cosmological model (``stringy RVM''), which is essential for the reader to understand the emergence of axionic stiff matter. Then we shall go one important step ahead of the discussion in our previous papers~\cite{bms1}-\cite{msb}, to study the emergence of a pre-inflationary stiff-matter-dominated era, and its consequences, under certain  circumstances, for the absence of an initial cosmological singularity.

\subsection{Types of Stringy Axions \label{sec:axions}}

In \cite{bms1}-\cite{msb} we have considered a four-dimensional string-inspired cosmological model, based on critical-string low-energy effective actions of the graviton,  $g_{\mu\nu}=g_{\nu\mu}$, and antisymmetric tensor (spin-one) Kalb-Ramond (KR) fields , $B_{\mu\nu}=-B_{\nu\mu}$, of the massless (bosonic) string gravitational multiplet~\cite{gsw,string,kaloper}, after compactification to (3+1)-dimensions. In our studies, and here, we ignore non constant dilaton fields, assuming that a constant dilaton $\Phi=\Phi_0$ is a consistent solution to our equations of motion, which has been checked explicitly.~\footnote{Specifically, a constant dilaton is assumed to be the result of, say, quantum-string physics, possibly non perturbative, which results in a potential for the dilaton. The constant dilaton may then be seen as a configuration that minimizes this potential. In the context of our string effective actions, this imposes constraints in the pertinent equations of motion, which however have been implemented consistently (see discussion in \cite{bms1,bms2} and references therein.}
 A crucial ingredient for the embedding of RVM formalism into our string effective theory is  the
presence of the KR axion field.  In four space-time dimensions, the latter is equivalent to a pseudoscalar massless excitation, the KR axion field $b(x)$. Such a field couples to gravitational anomalies, through the effective low-energy string-inspired gravitational action, which in \cite{bms1,bms2,bms3} has been assumed to describe fully the early Universe dynamics:
\begin{align}\label{sea4}
S^{\rm eff}_B =\; \int d^{4}x\sqrt{-g}\Big[ -\dfrac{1}{2\kappa^{2}}\, R + \frac{1}{2}\, \partial_\mu b \, \partial^\mu b
+   \sqrt{\frac{2}{3}}\,
\frac{\alpha^\prime}{96 \, \kappa} \, b(x) \, R_{\mu\nu\rho\sigma}\, \widetilde R^{\mu\nu\rho\sigma} + \dots \Big]~,
\end{align}
where $\alpha^\prime = 1/M_s^2$ is the Regge slope, with $M_s$ the string mass scale, which is in general different from the four-dimensional Planck mass~\cite{gsw}. Greek indices refer to the four-dimensional space-time,  and the last term in the right-hand side of this equation is the gravitational Chern-Simons term, associated with
a CP-violating gravitational anomaly~\cite{eguchi}.  The tilde above the Riemann tensor denotes its dual, defined as:
\begin{align}\label{duals}
\widetilde R_{\mu\nu\rho\sigma} = \frac{1}{2} \varepsilon_{\mu\nu\lambda\pi} R_{\,\,\,\,\,\,\,\rho\sigma}^{\lambda\pi}\, ,
\end{align}
with $\varepsilon^{\mu\nu\rho\sigma}$ being the four-dimensional
covariant Levi-Civita tensor density in curved space time,  totally antisymmetric in its indices:
\begin{equation}\label{leviC}
\varepsilon_{\mu\nu\rho\sigma} = \sqrt{-g}\,  \epsilon_{\mu\nu\rho\sigma}, \quad \varepsilon^{\mu\nu\rho\sigma} = - \frac{1}{\sqrt{-g}}\,  \epsilon^{\mu\nu\rho\sigma},
\end{equation}
with $\epsilon^{0123} = +1$, {\emph etc.}, the totally antisymmetric Levi-Civita symbol in Minkowski space time.

We now remark that in string theory~\cite{gsw}, the KR axion is associated with a {\it dualisation procedure} of the field strength $\mathcal H_{\mu\nu\rho}$ of the spin-one field $B_{\mu\nu}=-B_{\nu\mu}$~\cite{kaloper,witten},  and  it
is only one type of the several kinds of axions allowed in the landscape of string theory (`{\it Axiverse}'~\cite{axiverse}). The KR axion constitutes the so-called {\it string-model-independent} axion, present in all string theories. There is a plethora of other axions, however, associated with Kaluza-Klein zero modes of appropriate $p$-forms that appear in the spectrum of strings compactified to four space-time dimensions, that is string theories formulated on target-space-time manifolds of the form  $M_{1,3} \times \mathcal X_6$, with $M_{1,3}$ the uncompactified (3+1)-dimensional space time, and $\mathcal X_6$ the extra-dimensional space, assumed to be a smooth compact manifold ({\it e.g}. Calabi-Yau~\cite{gsw}). Such axions are therefore dependent on the microscopic string-theory model considered and are called {\it model-dependent} axions. In the heterotic string theory, for intance~\cite{gsw}, one has the (Neveu-Schwarz(NS)-type)  two-form  $\mathbf{\mathcal B}$ of the Kalb-Ramond field in ten dimensions, which, upon compactification on a Calabi-Yau
 six-dimensional compact space, $\mathcal X_6$, can be written as:
\be\label{KRBfield}
\mathbf{\mathcal B}  = B_{\mu\nu} (x) \, dx^\mu \, dx^\nu + \frac{1}{2\pi} \, b^I(x) \, \omega^I_{ij} (z,\bar z) \, dz^i \, d\bar z^j ~, \quad \mu, \nu=0, \dots 3, \quad i, j =1,2,3
 \ee
where  $z^i$, $i=1,2,3$ are complex coordinates parametrising the compact manifold.
The $B_{\mu\nu}(x)$ field yields, upon the  aforementioned dualisation procedure,  the KR axion $b(x)$, whilst
the quantities $\omega^I_{ij}(z, \bar z)$, $I=1, \dots h^{1,1}$ (in standard notation for the Hodge numbers $h^{1,1}$), represent harmonic (1,1) forms that depend only on the coordinates of the complex manifold, and are linked to the aforementioned KK zero modes. One uses the normalisation~\cite{witten}
\be\label{normalisation}
\int_{\mathcal C^J} \omega^I = \delta^{IJ}
\ee
where $\mathcal C^I$ is a 2-cycle in the compact manifold. In other words, the harmonic forms $\omega^I$ span the integer (1,1) cohomology group of the target space~\cite{eguchi}. The quantities $b^I(x)$, $I=1, \dots h^{1,1}$, represent dimensionless pseudoscalar fields on the uncompactified space-time, and the factor $\frac{1}{2\pi}$ has been inserted so that the fields $b^I(x)$ have a period $2\pi$, as is conventional for axions~\cite{kim}.
The kinetic terms of the two-form \eqref{KRBfield}, in the ten-dimensional-targert-space-time action,
yield, upon compactification, the four-space-time-dimensional kinetic terms of the $b^I(x)$ fields.

In generic string or D-brane models, {\it model-dependent} axion fields $a^I(x)$ are also obtained as KK zero models of other appropriate $p$-form fields, $C_p$, {\it e.g.} the Ramond-Ramond(RR)-type $p=0,2,4$-forms of type IIB string theory, or the $p=1,3$-forms of type IA~\cite{witten}:
 \be\label{pformaxion}
 a^I (x) = \frac{1}{2\pi} \int_{\mathcal C^{(p)}_I} \, C_p~,  \quad  I=1, \dots M,
 \ee
where $\mathcal C^{(p)}_I \ \subset \mathcal X_6$  are appropriate homologically-non-equivalent $p$-cycles in the compact manifold, and we normalised again the axion so as to have period $2\pi$.

 The pertinent kinetic terms for the axions $b^I(x)$ in (3+1)-dimensions
stem from terms in the ten-dimensional Lagrangian of the form
\be\label{tendim}
\mathcal L^{\rm 10D} \ni \mathcal H_{MNP} \, \mathcal H^{MNP}~,
\ee
where ${\mathcal H}_{MNP} = \partial_{[M} \, B_{NP]} + {\rm Chern-Simons~(gauge~and~gravitational)~terms}$, $M,N,P=0, \dots 9$ (with $[ \dots ]$ denoting complete antisymmetrisation of the respective indices).
 Upon compactification down to (3+1)-dimensions, with $B_{MN}$ being given in differential-form language by \eqref{KRBfield}, the  structures \eqref{tendim} yield, apart from model-independent-KR-axion-$b(x)$ kinetic  terms
 ({\it cf.} \eqref{sea4}), also kinetic terms for the model-dependent axions, of the form
{\small \begin{align}\label{kin4dim}
S_{\rm 10D} &\ni \int \sqrt{-g} \, d^4 x \int_{\mathcal X_6} \partial_\mu B_{ij} \, \partial^\mu B^{ij} = \int \sqrt{-g} \, d^4 x  \, \partial_\mu b^I (x) \partial^\mu b^J (x) \,\int_{\mathcal X_6}  \omega_{ij}^I (z,\bar z) \, \omega^{J\, ij} (z,\bar z)~,
\nonumber \\ & \equiv \int \sqrt{-g} \, d^4 x  \, \partial_\mu b^I (x) \partial^\mu b^J (x) \, \gamma^{IJ}
\quad \mu=0, \dots 3, \quad I, J=1, \dots h^{1,1},
\end{align}}where, for brevity,  we only indicated the structures, omitting numerical coefficients. The reader should observe the non-trivial kinetic mixing $\gamma^{IJ} \,  \ne \,  \delta^{IJ}$ of the model-dependent stringy axions $b(x)^I$, which depends on details of the compact manifold.

The axion coupling constants $f_{b^I}$, $I=1, \dots M$, where $M$ are the species of such axions in a given string theory model,  which couple the model-dependent-axions to anomalies, are determined~\cite{witten} by the (one-loop) counterterms required for Green-Schwarz (GS) anomaly-cancellation mechanism in string theory~\cite{gsw}.
To see this, let one consider, as an example,  the $E_8 \times E_8$ heterotic string, formulated on $M_{1,3} \times \mathcal X_6$,  with the Standard Model gauge group $SU(3)_c \times SU(2) \times U_Y(1)$
embedded, say, in the first $E_8$ group factor. It was shown in
\cite{witten}, that in such a case, the GS counterterms in the string effective action, yield four-space-time dumensional anomaly terms for the axion-$b^I(x)$ fields:
{\small \begin{align} \label{axionanom}
&S_{\rm anom~string~axion} = \nonumber \\
& \Big( \frac{1}{16\, \pi^2} \,\int_{\mathcal X_6} \omega^I (z,\bar z) \, \wedge \, \Big[{\rm Tr_1} \mathbf F \wedge \mathbf F - \frac{1}{2} \, \mathbf R \wedge \mathbf R \Big] \Big) \, \int d^4 x \, b^I(x) \, \Big(\nit{- \, } \frac{1}{16\pi^2} \, {\rm Tr_1} \mathbf F \wedge \mathbf F \, + \, \dots  \Big)
\end{align}}
where for brevity we did not write down explicitly the gravitational anomaly terms, denoted above by $\dots$,
 which have the same structure as in \eqref{sea4}. The first term inside the parentheses on the right-hand side of \eqref{axionanom} expresses mixed anomalies in the compact manifold $\mathcal X_6$, with $\mathbf F$ the appropriate gauge-field-strength two-form over the compact space $\mathcal X_6$ and $\mathbf R$ the corresponding compact-space-$\mathcal X_6$ curvature two form. The symbol $\wedge$ denotes  the appropriate exterior product among differential forms~\cite{eguchi}, and the trace ${\rm Tr_1}$ pertains to the first $E_8$ gauge group.

The form \eqref{axionanom}, defines the axion coupling to the anomaly terms, and thus the corresponding model-dependent axion coupling constant, which thus depends on the details of compactification. Moreover, by diagonalising the kinetic-mixing terms, upon appropriate redefinition of the axion field, one arrives at the generic conclusion that in string theory, in the presence of both model-independent and model-dependent axions, one is having several anomalous couplings of the various axions to the anomaly terms, which correspond to a variety of axion coupling constants, $f_{b^I}$. That is, there are the following anomalous couplings of the model-dependent axions in the effective action, which should be considered in addition to the KR-model-independent axion terms
\be\label{modep}
\mathcal S^{\rm model-depnd}_{\rm4-dim} \ni  \sum_{I=1}^M \, \int d^4x \, \sqrt{-g} \, \frac{1}{f_{b^I}} b^{\prime \, I} (x) \, \Big(c_1  R_{\mu\nu\rho\sigma} \, {\widetilde R}^{\mu\nu\rho\sigma}   - \dots \Big)
\ee
where $c_1$ are numerical constants, which can be absorbed in the definition of $f_{b^I}$, and the $\dots$ denote gauge terms. The fields
$b^{\prime \, I}$ here denote appropriately redefined dimensionful (of mass dimension +1) model-dependent stringy axions with canonical (diagonalised) kinetic terms $ \sum_{I=1}^M \,  \mathcal \int d^4 x \, \sqrt{-g}\, \frac{1}{2} \,  \partial_\mu b^{\prime\, I} (x) \, \partial^\mu b^{\prime \, I}(x)$.

Including the KR axion in the set, we may write the relevant gravitational (3+1)-dimensional, string-inspired, low-energy effective action in the form:
{\small \be\label{seamulti}
S^{\rm eff \, {\rm multi-axion}}_B =\; \int d^{4}x\sqrt{-g}\Big[ -\dfrac{1}{2\kappa^{2}}\, R + \sum_{I=1}^{M+1} \, \Big( \frac{1}{2} \,  \partial_\mu b^{\prime\, I} (x) \, \partial^\mu b^{\prime \, I}(x) +  \frac{1}{f_{b^I}} b^{\prime \, I} (x) \, R_{\mu\nu\rho\sigma} \, {\widetilde R}^{\mu\nu\rho\sigma}   \Big)  + \dots \Big]
\ee}
where
\be\label{krcoupl}
f_{b^{M+1}} \equiv f_b =  \Big(\sqrt{\frac{2}{3}}\,
\frac{\alpha^\prime}{96 \, \kappa}\Big)^{-1}
\ee
is the axion coupling of the KR axion field $b^{\prime \, M+1}(x) \equiv b(x)$ ({\it cf.}\eqref{sea4}).

In our analysis in \cite{bms1}, we only considered the model-independent action \eqref{sea4} as describing the dynamics of the early Universe. However, one could extend rather straightforwardly such an analysis to include the model dependent axions. The reader can easily see that this will not affect the results of \cite{bms1} qualitatively, given that, in most formulae  discussed in \cite{bms1}, one simply replaces
\be\label{multiaxion}
b(x) \, \Rightarrow \, \sum_{I=1}^{M+1} b^{\prime\, I}~.
\ee
Hence from now on we only restrict ourselves to the single (model-independent) KR axion only \eqref{sea4}, although whenever appropriate, we shall also make explicit reference to the multi-axion case. We do note though that, despite the formal simplicity of \eqref{multiaxion}, which allows for the passage from the single- to the multi- axion cases, the presence of more than one species of stringy axions have much richer phenomenological implications for Dark Matter in the string-inspired Universe~\cite{bms1,axiverse}. Nonetheless, it should be stressed that only one of these axions, the KR-axion, is the dual of the antisymmetric-tensor field strength $\mathcal H_{\mu\nu\rho}$, which plays the r\^ole of torsion in the string effective actions~\cite{kaloper,gsw}, and thus the corresponding Dark matter, upon the development of a non-perturbative potential for it by instanton effects during the matter era~\cite{bms1},  admits a geometrical `torsion' interpretation.

\subsection{Stiff stringy axion matter and Gravitational-Wave Contributions to Anomaly condensates \label{sec:stiff}}

The interactions of the $b$ field (or, in that matter, also of all the model-dependent axion fields $b^{\prime I}$ \eqref{modep}) with the gravitational anomaly terms in the early Universe, where gauge fields are assumed absent in the model of \cite{bms1}, vanish for a Friedmann-Lemaitre-Robertson-Walker background. This is because the gravitational Chern-Simons term $ R_{\mu\nu\rho\sigma} \, {\widetilde R}^{\mu\nu\rho\sigma} $ identically vanishes for FLRW spacetime.
In that case, from the effective action \eqref{sea4}, we observe that the massless KR and model-dependent axions, without any potential terms, have a stress tensor
\begin{align}\label{stressb}
T_{\mu\nu}^b = \frac{2}{\sqrt{-g}} \, \frac{\delta S^{b}(b,\, g_{\alpha\beta})}{\delta g^{\mu\nu}}
  =   \partial_\mu b \, \partial_\nu b - \frac{1}{2}g_{\mu\nu} (\partial_\alpha b\, \partial^\alpha b).
\end{align}
and thus constitute~\cite{bms1} a type of  `{\it stiff matter}'  with stiff EOS~\cite{bms1}\footnote{Stiff matter was originally introduced by Zeldovich in the context of a phenomenological cold gas of baryons\,\cite{stiff} -- see also \cite{stiff2} for considerations along similar lines. Our context is completely different to that;  it is connected with properties of the KR axion in the early universe. Moreover, in  RVM- inflation there is no singularity at $a=0$ since all energy densities are finite at that point (cf Sec.\,\ref{sec:RVM-inflation}).  This is actually the reason why the huge amount of entropy generated in the RVM universe is calculable and can explain the entropy and horizon problems, see Sec.\,\ref{sec:ThermoAspects}. We  will come back to the role played by stiff matter in our context in Sec.\,\ref{sec:stiffevol}.}
\be\label{stiff}
w^{\rm stiff}_{\rm m-string-RVM}=\frac{p}{\rho}=+1\,.
\ee
This kind of EoS characterizes a fluid in which the velocity of sound in it equals the velocity of light:  $c_s=\delta p/\delta\rho=1$.

We stress again that the stiff character of  the EoS for the KR axion  is valid only within the model of \cite{bms1}, in which gauge field terms, that could generate an axion potential through instantons, are assumed absent in early epochs of the Universe. Such terms, of course, are generated at the post inflationary period, as discussed in detail in \cite{bms1,bms2}, where we refer the interested reader for details.

If one includes formally, {\it i.e.} before specifying the space-time background,
the interactions of the b-field and the gravitational anomalies, \eqref{sea4}, then this
yields a modified, {\it conserved} stress tensor, as a result of the non-trivial variation of the gravitational Chern-Simons anomalous terms with respect to the variation of the metric tensor:
\begin{align}\label{cons}
\kappa^2 \, {\widetilde T}_{b + {\rm gCS}}^{\mu\nu} \equiv \sqrt{\frac{2}{3}}\,\frac{\alpha^\prime\, \kappa}{12} \mathcal C^{\mu\nu} + \kappa^2 T_b^{\mu\nu}    \quad \Rightarrow \quad  {\widetilde T}_{b + \Lambda + {\rm gCS} \,; \mu}^{\mu\nu} =0~,\end{align}
the extra terms, proportional to the Cotton tensor $C^{\mu\nu}$, describing energy exchange between the axion and gravitational field. The Cotton tensor is defined as~\cite{jackiw}
\begin{align}\label{cotton}
&{\mathcal C}^{\mu\nu} \equiv  -\frac{1}{2}\, \Big[v_\sigma \, \Big( \varepsilon^{\sigma\mu\alpha\beta} R^\nu_{\, \, \beta;\alpha} +
\varepsilon^{\sigma\nu\alpha\beta} R^\mu_{\, \, \beta;\alpha}\Big) + v_{\sigma\tau} \, \Big(\widetilde R^{\tau\mu\sigma\nu} +
\widetilde R^{\tau\nu\sigma\mu} \Big)\Big]\, , \nonumber \\
&= - \frac{1}{2} \Big[\Big(v_\sigma \, \widetilde R^{\lambda\mu\sigma\nu}\Big)_{;\lambda}  + \, (\mu \leftrightarrow \nu)\Big]\, ,
\nonumber \\
&v_{\sigma} \equiv \partial_\sigma b = b_{;\sigma}, \,\,v_{\sigma\tau} \equiv  v_{\tau; \sigma} = b_{;\tau;\sigma}.
\end{align}
and, due to properties of the Riemann tensor, it is gravitationally traceless
\begin{align}\label{tracecot}
g_{\mu\nu}\, \mathcal C^{\mu\nu}= 0~,
\end{align}
and obeys
\begin{equation}\label{csder}
{\mathcal C}^{\mu\nu}_{\,\,\,\,\,\,\,;\mu} = -\frac{1}{8} v^\nu \, R^{\alpha\beta\gamma\delta} \, \widetilde R_{\alpha\beta\gamma\delta}.
\end{equation}
Eq. \eqref{csder} implies that the $b$-matter stress tensor \eqref{stressb} is {\it not} conserved, which is to be expected due to the non-trivial exchange of energy between the axion and the gravitational field due to the anomaly terms in \eqref{sea4}. Nonetheless, there is no issue with general covariance, in view of the existence of the improved stress tensor \eqref{cons}, where such interactions are taken correctly into account.

In the multi-axion case, similar results hold, upon the replacement of $C^{\mu\nu}$ by $C^{\mu\nu\, I}$, for each axion $b^{\prime\, I}$, $I=1, \dots N$, which now replaces $b$ in the above expressions. The multi-axion improved and conserved stress tensor, generalising \eqref{cons}, is evaluated from \eqref{seamulti}, and yields
\be\label{multicons}
\kappa^2 \, {\widetilde T}_{b^\prime + {\rm gCS}}^{\mu\nu} =  \sum_{I=1}^{M+1} \, \Big(8\, \frac{1}{f_{b^I}} \, \mathcal C^{\mu\nu\, I} + \kappa^2 \, \partial^\mu b^{\prime \, I} \, \partial^\nu b^{\prime\, I} - \frac{1}{2}g^{\mu\nu} (\partial_\alpha b^{\prime \, I} \, \partial^\alpha b^{\prime \, I}) \Big)
\ee

For flat or FLRW space-time backgrounds, the Cotton tensor {\it vanishes}, as already mentioned, and in such a case
the stress tensor \eqref{cons} reduces to the stress tensor of the KR axion field \eqref{stressb}.
However, in the presence of CP-violating primordial gravitational waves (GW) in the early Universe, which perturb the
FLRW metric background, the CP-violating gravitational anomaly term is non trivial, as a result of GW condensation~\cite{stephon,bms1}, which yields:
 \begin{align}\label{rrt}
  \langle R_{\mu\nu\rho\sigma}\, \widetilde R^{\mu\nu\rho\sigma} \rangle  = \frac{16}{a^4} \, \kappa^2\int \frac{d^3k}{(2\pi)^3} \, \frac{H^2}{2\, k^3} \, k^4 \, \Theta + {\rm O}(\Theta^3) ,
 \end{align}
 where $\langle \dots \rangle$ indicates the condensation, in which graviton fluctuations of momentum $k$ are integrated out. The Fourier integral over $k$ is cutoff at an Ultraviolet (UV) momentum scale $\mu \lesssim \kappa^{-1}$. The result \eqref{rrt} holds to leading order in $k \, \eta \gg 1$, where $k$ is the standard Fourier scale variable, and $\eta$ is the conformal time, defined as $d\eta =  \frac{dt}{a(t)} \,  \Rightarrow \, \eta = \frac{1}{H}\, \exp (-Ht)$.  The quantity $\Theta$ is given by
   \begin{align}\label{theta}
 \Theta = \sqrt{\frac{2}{3}}\, \frac{\alpha^\prime \, \kappa}{12} \, H \,  {\dot {b}} \, = 8\, \sqrt{\frac{2}{3}}\,
 \frac{1}{f_{b^{M+1}}} \, \kappa^2 \, H \, {\dot{b}}~,
 \end{align}
  using \eqref{krcoupl},
and is assumed small,  $|\Theta |  \ll 1$, which is phenomenologically consistent~\cite{bms1}, allows for a perturbative treatment of the induced anomalies.

We should remark at this stage that the physical mechanism behind the GW condensate  \eqref{rrt}  as computed in ~\cite{stephon,bms1}  is the  `cosmological birefringence' of the GW's during inflation.  The different behavior of the  left  ($h_L$) and right handed ($h_R$)  chiral components  of the GW's leads to attenuation of the former  and amplification of the latter  in the early universe. This means that one can distinguish the effects from chiral gravitational components having different dispersion relations,  which explains the name. Such effect is of quantum origin since the Chern-Simons condensate \eqref{rrt} is a quantum vacuum expectation value of the quantity  $R_{\mu\nu\rho\sigma}\, \widetilde R^{\mu\nu\rho\sigma}$, which is computed from the two-point Green's function  $\langle h_L(x)\,h_R(x')\rangle$  associated to the left and right handed chiral components  of the GW's. In the absence of such cosmological birefringence during inflation, the aforementioned VEV would vanish and no gravitational condensate would occur.

We remark at this point, that in  the stringy multi-axion case, the anomaly condensate assumes a similar form as
in \eqref{rrt}, but wth the parameter $\Theta$ now replaced by ({\it cf.} \eqref{multiaxion}):
\be\label{multitheta}
\Theta \, \Rightarrow \, \Theta_{\rm multi} = 8\, \sqrt{\frac{2}{3}}\, \kappa^2 \, H\, \sum_{I=1}^{M+1} \,
 \frac{1}{f_{b^{I}}} \, {\dot{b}^{\prime \, I}}
\ee
which can also be assumed small $|\Theta_{\rm multi} | \ll 1$.

As discussed in detail in \cite{bms1,bms2}, the GW-induced anomaly condensate \eqref{rrt}, \eqref{theta} (or \eqref{multitheta} in the multi-axion case), leads to the possibility of an approximately constant anomaly,
upon appropriate restrictions of the string scale. For a dominant KR axion, such restrictions read
\be\label{stringrange}
M_{\rm Pl} \, \gtrsim \, M_s \,  \gtrsim \,  10^{-3} \, M_{\rm Pl},
\ee
which guarantees a Lorentz-violating solution of the equations of motion for the KR axion, for a cosmological background $b(t)$ in a FLRW space-time,  with metric $g_{\mu\nu} = g_{\mu\nu} (t)$ of the form~\cite{bms1}:
\begin{align}\label{LV}
\frac{d}{dt} \Big(\sqrt{-g} \, \Big[ \dot b - \frac{1}{f_b} \, \mathcal K^0 \Big] \Big) =0 \, \quad  & \Rightarrow \, \quad {\dot b} =  f_b^{-1} \, \mathcal K^0 \simeq {\rm constant}~, \nonumber  \\ &\Rightarrow \quad  b(t) = b(0) + {\rm (constant)} \, t ~,
\end{align}
where $b(0)$ is an initial value of the KR axion field at the beginning of inflation.
In \eqref{LV}, $\mathcal K^0$ denotes the temporal component of the (GW-induced condensate of the) total derivative
$\mathcal K^\mu$, in terms of which the gravitational anomaly can be expressed~\cite{eguchi}. In our case we have approximately, for weak GW perturbations~\cite{bms1}:
\be\label{anomaly}
\langle \sqrt{-g}\, R_{\mu\nu\rho\sigma}\, \widetilde R^{\mu\nu\rho\sigma} \rangle \simeq \sqrt{-g} \langle R_{\mu\nu\rho\sigma}\, \widetilde R^{\mu\nu\rho\sigma} \rangle \simeq
\frac{d}{dt} \Big( \sqrt{-g}\, K^0 \Big)
\ee
where the anomaly condensate on the left-hand side is given by \eqref{rrt}. Under the formation of condensates \eqref{anomaly}, the (approximately) constant $\mathcal K^0$ arises as a consistent solution of \eqref{LV}~\cite{bms1}:
\begin{align}\label{k02}
{\mathcal K}^0 (t)  \sim {\mathcal K}^0_{\rm egin} (t(\eta=H^{-1})) \, \exp\Big[  - 3H\, t(\eta) \, \Big( 1  -  \frac{1}{3\,\pi^2 \times 18 \times  96}\,  \,  \Big(\frac{H}{M_{\rm Pl}}\Big)^2 \, \Big(\frac{\mu}{M_s}\Big)^4 \Big)\Big]~,
\end{align}
where $t$ ($\eta$) denotes the cosmic (conformal)  time, and ${\mathcal K}^0_{\rm egin} (t(\eta=H^{-1}))$ is the value of the anomaly
condensate at the onset of the RVM inflation. From \eqref{k02} one observes that one derives an approximately constant $\mathcal K^0$ during inflation, for an UV cut-off of the modes of the GW perturbations of order
\be\label{cutoff1}
\mu\simeq \left(3\,\pi^2 \times 18 \times  96\right)^{1/4} \Big(\frac{H}{M_{\rm Pl}}\Big)^{-1/2} \, M_{s}\simeq 15\, \Big(\frac{H}{M_{\rm Pl}}\Big)^{-1/2} \, M_{s}~,
\ee
We now remark that, for the multi-axion case, which involve several axion couplings, which depend on details of the compactitication (see, {\it e.g.}, \eqref{axionanom}), the corresponding restrictions \eqref{stringrange} are compactication-model dependent, but the of the conclusions of \cite{bms1}, based on the single-axion case, are largely  maintained.
The Lorentz-violating solution for the axions now read
\be\label{LVmulti}
\frac{d}{dt}  \Big(\sqrt{-g} \, \Big[ {\dot b}^{\prime \, I}  -  \, \frac{1}{f_{b^{I}}} \, \mathcal K^0 \Big] \Big) =0 \, \quad  \Rightarrow \, \quad {\dot b}^{\prime \, I} =  \frac{1}{f_b^{\prime\, I}} \, \mathcal K^0 \simeq {\rm contant}~,
\quad I = 1, \dots M+1~,
\ee
including the KR axions $b \equiv b^{\prime \, M+1}$.

For a period where a scalar field drives the cosmological evolution, the connection between the rate of change of the Hubble function and that of the scalar field is given in very good approximation  by $\dot{\varphi}^2 \simeq -2\MPl^2\dot{H}$. This follows immediately from differentiating Friedmann's equation and using the Klein-Gordon equation satisfied by the (homogeneous) scalar field.   Thus,  the standard slow-roll parameter $\epsilon$, which characterises the inflation period, can be written as
\begin{equation}\label{eq:SlowRoll}
\epsilon \equiv -\frac{\dot{H}}{H^2}=\frac12\,\frac{\dot{\varphi}^2}{(\MPl H)^2}\ll1\,.
\end{equation}
Such parameter must be small during inflation, as  in that period the rate of change of the Hubble function is small.   The previous equation, which we may apply to the undiluted background of the KR axion feld  $b$ at the end of inflation\,\cite{bms1}),  tells us that
\be\label{dotearly}
\dot{b}=\sqrt{2\epsilon} \, M_{\rm Pl} H~,
\ee
where  $H \simeq  H_I$ is the approximately constant value of the Hubble parameter in the de Sitter phase and hence the parameter setting the
inflationary scale.  Comparing the obtained relation  with  \eqref{LV},  which is  the inflationary solution for the KR background,  we see that  one may take the constant coefficient in the solution for  $b$   to be of order
\be\label{valconst}
{\rm constant} = \sqrt{2\epsilon} \, M_{\rm Pl} \, H_I\,.
\ee
Fitting  the data~\cite{Planck} requires $\epsilon \sim 10^{-2}$ and
\be\label{HIscale}
\frac{H_I}{M_{\rm Pl}} \in [10^{-5}\, ,\, 10^{-4} )~.
\ee
In view of \eqref{cutoff1}, one obtains for the mode UV cutoff $\mu$: $ \mu \sim 10^3 \, M_s $. If one imposes that, for a consistent low-energy theory of quantum gravity, transplanckian modes must decouple (although strictly speaking this restriction might be avoided, if one attributes transplanckian values of the UV cutoff to modes in the deep (stringy) quantum-gravity regime~\cite{bms1}), then $\mu \lesssim m_{\rm Pl}= \sqrt{8\pi} \, M_{\rm Pl}$, which would restrict the allowed string-mass-scale range \eqref{stringrange} to
\be\label{eq:MsRange}
 10^{-3} \lesssim \frac{M_s}{M_{\rm Pl,}  }\lesssim 10^{-2}\,.
\ee
Taking into account the value of the reduced Planck mass,  $\MPl\simeq 2.4\times 10^{18}$ GeV, it follows that the working range for our string scale estimate would be close to the typycal GUT scale $M_X\sim 10^{16}$ GeV -- up to the above qualification on avoiding  decoupling of the  transplankian modes in the stringy regime, which could soften the limits \eqref{eq:MsRange} and push them a bit higher.
The phenomenology of the multi-axion case \eqref{LVmulti}, required to
match the cosmological data~\cite{Planck}, is similar to the single-KR axion case \eqref{LV}, and will not be discussed further here.

\subsection{GW condensates and RVM-like dynamical Inflation \label{rvminflation}}

As discussed in \cite{bms1}, the GW condensation leads, apart from the anomaly condensate,
implying the existence of the spontaneously-Lorentz-symmetry-breaking solutions \eqref{LV} (or \eqref{LVmulti}),
also to a cosmological-constant-type term in the effective action,\footnote{In the multi-axion case, this term would read:
\begin{align}\label{lambdamulti}
\mathcal S_\Lambda  &=
\sqrt{\frac{2}{3}}\,\sum_{I=1}^{M+1} \, \frac{1}{f_{b^I}} \, \int d^4 x \sqrt{-g} \, \langle b^{\prime \, I}  \, R_{\mu\mu\rho\sigma}\, \widetilde R^{\mu\nu\rho\sigma} \rangle~.
\end{align}
This also leads (upon taking into account \eqref{multitheta}) to a cosmological-constant contribution in the effective action of qualitatively similar form as \eqref{lambda}.}
\begin{align}\label{lambda}
\mathcal S_\Lambda  &=
\sqrt{\frac{2}{3}}\,
\frac{\alpha^\prime}{96 \, \kappa} \, \int d^4 x \sqrt{-g} \, \langle \overline b \, R_{\mu\mu\rho\sigma}\, \widetilde R^{\mu\nu\rho\sigma} \rangle  \equiv  -  \int d^4x \, \sqrt{-g} \, \frac{\Lambda (H)}{\kappa^2} \nonumber \\ & \simeq   \int d^4 x \, \sqrt{-g}\, \Big(5.86 \times 10^7 \, \, \sqrt{2\, \epsilon} \,
\Big[\frac{\overline b(0)}{\MPl} + \sqrt{2\, \epsilon} \,  \mathcal N \Big] \, H^4 \Big) \,
~.
\end{align}
Above, the symbol $\simeq $ indicates and order of magnitude estimate, and we took into account that $H\, t $ is bounded from above by $(H\, t)_{\rm max}$ evaluated at the end of the inflationary period, for which $(H\, t)_{\rm max} = H\, t_{\rm end} \sim {\mathcal N} = 60-70 $, with ${\mathcal N}$ the number of e-foldings. We also set $\epsilon \sim 10^{-2}$, as required by inflationary phenomenology. The notation $\Lambda (H)$ implies that the term is (approximately) constant during the de Sitter phase, in which the Hubble parameter is approximately constant, $H \simeq H_I$.

We next notice that~\cite{bms1,bms2}, on assuming
\be\label{b0val}
|b(0)| \gtrsim \sqrt{2\, \epsilon}\, {\mathcal N} \, \MPl \sim 10 \, M_{\rm Pl}~,
\ee
the quantity $\Lambda > 0$ in \eqref{lambda} does {\it not} change order of magnitude during the entire inflationary period, for which $H \simeq $ constant, and thus it can be approximated by a constant. In that case, the term \eqref{lambda} behaves as a {\it positive-cosmological-constant} (de Sitter) type term, which is responsible for inducing inflation. Quantum fluctuations of the condensate are then responsible for deviations from scale invariance, providing a novel mechanism for cosmological perturbations.

The corresponding modified stress tensor \eqref{cons} now acquires a $\Lambda$-vacuum contribution, but its conservation \eqref{cons} is not of course affected by the presence of a constant $\Lambda$:
\begin{align}\label{cons2}
\kappa^2 \, {\widetilde T}_{b + \Lambda + {\rm gCS} + \Lambda}^{\mu\nu} \equiv \sqrt{\frac{2}{3}}\,\frac{\alpha^\prime\, \kappa}{12} \mathcal C^{\mu\nu} + \kappa^2 T_b^{\mu\nu}  + \Lambda g^{\mu\nu} ~,\end{align}

As demonstrated in \cite{bms1}, the presence of a de-Sitter-like term \eqref{lambda} is crucial for ensuring
the positive-definiteness of the total vacuum energy density of the string Universe obtained from (the temporal component of) \eqref{cons2}, including contributions from axions and their interactions with the GW-induced gravitational-anomaly condensates,
which has the form:
\begin{align}\label{toten}
\rho_{\rm total} &  \simeq
3\kappa^{-4} \, \Big[ -1.65 \times 10^{-3} \Big(\kappa\, H \Big)^2
+ \frac{\sqrt{2}}{3} \, |\overline b(0)| \, \kappa \, \times {5.86\, \times} \, 10^6 \, \left(\kappa\, H \right)^4 \Big] > 0~.
\end{align}
Remarkably, this has a structure similar to the vacuum energy density in a RVM~\eqref{rLRVM}, but here, in contrast to the conventional RVM~\cite{JSPRev2013}, the $\nu$ coefficient of the $H^2$ term is {\it negative}, due to negative contributions from the gravitational Chern-Simons anomaly term, which overcome the positive $\nu$ contributions from the `stiff' KR axion~\cite{bms1,bms2}.  In the early Universe, however, this has no dramatic consequences since the dominant term is the $H^4$ term, with a positive coefficient, arising from the GW condensate, and thus the energy density $\rho_{\rm total}$ is {\it positive} and drives an almost de Sitter (inflationary) phase in that period~\cite{bms1,rvmhistory}. From \eqref{b0val}, one can easily check that the corresponding coefficient $\alpha$ in \eqref{rLRVM} is of order $0.1$, in agreement with a RVM associated  with a typical GUT scale $M_X\sim 10^{16}$ GeV.

At the inflationary exit period, massless chiral fermionic matter, as well as gauge degrees of freedom, are assumed to be created~\cite{bms1,bms2}, which enter the effective action via the appropriate fermion kinetic terms and interaction with the gravitational and gauge anomalies.
The primordial gravitational anomaly terms {\it  are cancelled} by the chiral matter contributions~\cite{bms1,bms2}, but the triangular (chiral)  anomalies (electromagnetic and of QCD type) in general remain.  In fact, they must remain,  in orderto explain important features of  particle phenomenology ({\it e.g.} $\pi^0\to \gamma\gamma$ through the anomalous chiral term in the  Axial-Vector-Vector (AVV) amplitude).

In the post inflationary phase
the KR axion acquires, through instanton effects, a non perturbative mass, and may play the role of Dark Matter~\cite{bms2}. It can be shown~\cite{bms1} that, due to the presence of cosmic gauge fields and other effects, the late-era vacuum energy density acquires a standard RVM form \eqref{rLRVM}, with positive coefficient $\nu_{\rm late} \sim \mathcal O(10^{-3})$,  consistent with phenomenology. At late eras, higher than $H^2$ terms in the energy density are not phenomenologically relevant and thus can be safely ignored. The $\nu_{\rm late} H^2$ corrections to the standard current-era cosmological constant term $c_0$ lead to distinctive signatures of a ``running'' dark energy, which helps to alleviate the aforementioned tensions in the cosmological data with the predictions of the
standard $\Lambda$CDM\,\cite{rvmpheno1,rvmpheno2,rvmpheno3}.

\subsection{String-inspired-RVM Evolution in the presence of stiff matter \label{sec:stiffevol}}

There are some interesting features of the RVM with stiff matter, that we feel we must stress at this point.
In our stringy RVM model~\cite{bms1,bms2}, after condensation of GW, the anomalous interactions of axions with gravity, through the CP-violating terms in the string-inspired gravitational effective action \eqref{sea4} (or the generalized one \eqref{seamulti}) obscures the r\^ole of such  gravitational-in-origin axions  as pure matter.

Indeed, in the absence of GW condensates, our effective gravitational theory \eqref{sea4} would have been dominated by the KR stiff axion matter, with with EOS \eqref{stiff}. The solution of the classical equations \eqref{LV} (or \eqref{LVmulti}) would be of the form
\be\label{stiffsol}
{\dot b} = \frac{c_0}{a^3(t)}, \quad c_0 = {\rm constant} \ne 0.
\ee
The stress tensor \eqref{stressb} of such axions would thus scale with $a^{-6}$ which is the scaling
power of stiff matter, as follows from \eqref{frie33}, upon setting $w_m=1$, and  ${\dot{\rho}}^{\Lambda}_{\rm RVM} = 0$, since the anomaly term  in \eqref{sea4} vanishes if  GW metric fluctuations would not occur, and hence the vacuum energy density is either zero, or, at most, has the form of some cosmological-constant vacuum contribution, due to some yet unknown quantum string/brane physics.
The solution of \eqref{frie33}  in such a case,  with null right hand side and for $\rho_m\to \rho_{\rm stiff}$, would yield the scaling
\be\label{stiffenden}
\rho_{\rm stiff} \sim a^{-6},
\ee
for the corresponding stiff-matter energy density,  in agreement with \eqref{stiffsol}  (or, equivalently, \eqref{lcdmH} with $\nu=0$, following from the corresponding Friedman equation \eqref{friedr} with $\Lambda (t)=0$).
In this case, the stringy axions, although gravitational in origin, nonetheless would behave as true stiff-matter excitations~\cite{stiff,stiff2}.\footnote{We remark in passing that, naively, one would think that, if such phase of the string Universe with a scaling \eqref{stiffenden} exists, which would precede the RVM-GW-condensate inflationary period, it would imply that there should be an initial (Big-Bang) singularity, as $a \to 0$, if the effective action \eqref{sea4} was valid up to such early pre-inflationary eras. This, however, is not so, given that at very early epochs of the Universe, higher-curvature and higher-than-two-derivative terms dominate the effective theory. In such a case, one
may encounter situations in which there is {\it no initial singularity}. For example, higher-curvature Gauss Bonnet terms, in the presence of non trivial (time-dependent) dilatons, are known to produce initial-singularity-free cosmologies~\cite{art}, and the incorporation of Kalb-Ramond gravitational axions $b(x)$ in such theories is expected to be characterised by terms in the corresponding effective actions that contain higher-than-quadratic derivatives of the field $b(x)$. Thus, it is  not unlikely that, in the presence of such higher-order-derivatives axion terms, combined with the dilaton-Gauss-Bonnet ones, the singularity-free situation of \cite{art} is maintained.  The nonexistence of an initial singularity, which was already the hallmark of the original RVM-inflation -- cf. Sec.\,\ref{sec:RVM-inflation} --  could then  be realized as well  in the stringy version of it.}.  We note that such a stiff-axion dominated phase will be described by free KR-axion stress energy tensor \eqref{stressb}, which, in view of \eqref{stiffsol} and \eqref{stiffenden},  leads to an energy density of a running-vacuum form \eqref{rLRVM}, with
{\it positive} coefficient of the $H^2$ term:
\be\label{stiffnu}
\nu_{\rm stiff~axion} \, > \, 0~.
\ee
Specifically, since such contribution emerges  from the $b$-axion field stress tensor $T^{\mu\nu}_b$  upon ignoring the Chern-Simons terms, we simply equate $\rho_b = T_{b}^{00}$ and use \eqref{dotearly}  and we find
\begin{equation}\label{nub}
\nu_{\rm stiff~axion} =\frac{\kappa^2\,\dot{b}^2}{6H^2}= \frac{\epsilon}{3}\simeq 3 \times 10^{-3} > 0\,.
\end{equation}
We note that the validity of this equation hinges on that of  \eqref{dotearly} and in general we cannot associate that estimate to the value of $\nu$ at post-inflationary epochs. However,  a new, more conventional, actor comes on stage when we enter the regular FLRW regime, i.e.  when we cross  the point where the stringy inflationary phase associated to the $H^4$  power no longer feeds the cosmological evolution.  From that point onwards, it will be the turn of the QFT effects, discussed in Sec. \ref{Sec:RVMQFT}, to generate the coefficient $\nu$ of the $\sim H^2$ term.  The reader should recall that the ${\cal O}(H^4)$ effects predicted by the QFT corrections vanish for $H=$constant, since they all depend on time derivatives of the Hubble rate, and hence all these terms are subdominant during the inflationary phase, which is primarily driven by the the $H^4$ terms. However, once the inflationary period $H=$const ends, all the ${\cal O}(H^4)$ effects fade away and we enter the radiation epoch. At this point the stringy features of the early universe,  associated with the gravitational anomaly terms, have no longer influence on the dynamics of the vacuum and the main effects are determined by the more pedestrian context of  QFT in curved spacetime.  In fact, as discussed in \cite{bms1}, the gravitational anomalies cancel out during the post inflationary period by the chiral matter generated at the end of the stringy-RVM inflation.
 A typical QFT contribution to $\nu$ from a scalar field non-minimally coupled to curvature  is given by Eq.\,\eqref{eq:nueff}. The effective final value of $\nu$, though, as we have pointed out  in Sec. \ref{Sec:RVMQFT}, actually receives contributions from all possible (fundamental) scalars, fermions and vector bosons: $\nu_{\rm eff}=\nu_s+\nu_f+\nu_v$. So ultimately  such effective value is to be fitted to experiment. Interestingly, this task has been performed in the literature and a wealth of remarkable results have been obtained. The fitting analyses to the modern cosmological data indeed suggest that the RVM can be competitive, if not superior,  to the $\CC$CDM in describing the overall cosmological observations  (and in alleviating the tensions afflicting the latter) provided the effective value of $\nu_{\rm eff}$ is positive and of order $\nu_{\rm eff} ={\cal O}(10^{-3})>0$~\cite{rvmpheno1,rvmpheno2,rvmpheno3,BDpapers}.   Amazingly, this value is also of the order of the predicted one in  \eqref{nub}. 

Let us now come back to the regime where the KR axion has emerged from the inflationary phase and let us compare the behavior of its energy density with that of the traditional stiff matter\,\cite{stiff}.  As we have seen above, in the presence of GW-induced anomaly condensates, the classical solution to the axion equation of motion \eqref{LV} (and \eqref{LVmulti} in the multi-axion case) corresponds to  a (approximately)  constant $\dot b$. This leads to constant contributions to the `running vacuum' energy density \eqref{toten}, which are viewed as vacuum contributions, leading to dynamical inflation.
In such a case, during the early de Sitter era, in view of the aforementioned fact that $|\nu|={\cal O}(10^{-3}) \ll 1$ in \eqref{toten}, one would have from \eqref{HS1}
\be\label{stringHdS}
H(a) _{\rm early~string~RVM} \simeq
\left(\frac{1}{\alpha}\right)^{1/2}\,\frac{H_{I}}{\sqrt{D_{\rm string} \,a^{6}+1}}\,,
\ee
to be contrasted with the result \eqref{HS1rad} in the standard-RVM with relativistic matter, as far as the scaling with the scale factor $a(t)$ of the FLRW Universe is concerned.

The corresponding energy densities of (stiff) axionic ``matter'' and ``vacuum'' can  be readily found, within the same approximation, from \eqref{toten} (\eqref{rLRVM}) and \eqref{frie33}, with $w_m=+1$.
We obtain  the following scaling for the energy density of the ``stiff-axionic'' matter:
\be\label{stringrohrdS}
\rho_{\rm stiff}(a) \simeq
\frac{3H_I^2}{\kappa^2\alpha}\frac{D_{\rm string} a^6}{(D_{\rm string} \,a^{6}+1)^2}
\ee
while for that of the ``running vacuum'' we have:
\be\label{stringrohLdS}
\rho_{\CC}(a) \simeq
\frac{3H_I^2}{\kappa^2\alpha}\frac{1}{(D_{\rm string} \,a^{6}+1)^2}
\ee
These equations characterize what we may call  the axionic pre-heating phase, and can be compared with the corresponding  ones in the original RVM, see Eqs.~\eqref{hubbleeq0}-\eqref{eq:rhoLfinal}.  In the present context, such pre-heating phase precedes the ordinary RVM-inflationary stage and hence acts as a pre-inflationary stage.  The steeper behavior of the above formulae is of course caused by the stiff matter EoS \eqref{stiff} governing such pre-heating phase. It should be noted that the energy densities \eqref{stringrohrdS} and  \eqref{stringrohLdS}
are equally well-behaved at $a=0$ as their counterparts  in the conventional RVM.
In stark contrast with the ordinary situation of self-conserved stiff-matter\, \cite{stiff,stiff2}  or  the particular situation with gravitational-anomaly-and-potential-free stiff axions \eqref{stiffenden} in our stringy context, the scaling behavior \eqref{stringrohrdS} is \textit{not} of the ordinary form \eqref{stiffenden}.  As it turns out, when  matter axions are in interaction with a  $H^4$-driven  vacuum phase,   the smoother behavior \eqref{stringrohLdS} with no singularity at $a=0$ is warranted.  The steeper scaling $\rho_{\rm stiff}(a)\sim a^{-6}$ appears only later on, for sufficiently large values of the scale factor.  In ordinary stiff-matter frameworks, such more drastic behavior is there right from the start and cannot avoid the initial singularity at $a=0$, unless one assumes that the energy density of the stiff fluid is negative\,\cite{stiff2}.  We do not assume exotic situations with negative energy densities anywhere.

In the stringy RVM scenario under discussion all physical quantities are smooth,  the stiff-matter-energy density is perfectly positive and vanishes at the Big Bang ($a \to 0$), while the Hubble function and the  vacuum energy density are both finite at that point and remain approximately constant in the early inflationary stages.  We believe that this is much more natural, and suggests that the de Sitter phase would start immediately after the Big Bang. The initial singularity is fully averted, at least as far as the above physical quantities are concerned.

Nonetheless, as we shall discuss later on, in section \ref{sec:origin}, one needs to devise appropriate mechanisms for the generation of the primordial GW that are responsible for inducing the RVM inlfation. In some of them, there is a
pre-inflationary phase, before the formation of GW, in which the anomaly terms, and hence the $H^4$ terms in the energy density, are absent. Nevertheless, even in such cases there is no initial singularity, because, as we have already mentioned, higher-curvature corrections in the string-inspired effective gravitational action become important, which are capable of removing the initial singularity~\cite{art}. As we shall discuss in section \ref{sec:origin}, there are also models characterised by a first hill-top inflationary phase, that precedes the GW-induced inflation, and occurs shortly after the Big-Bang. Such a phase might also be described by an RVM-type energy density~\cite{sugraRVM}, which is finite at the Big-Bang point, hence no initial singularity in such models either.

The coefficient $D_{\rm string}$, depends on the underlying microscopic string model.  In a similar way as we did in Eq.\,\ref{aeq0},  it  can be related to the equality point between the density of stiff matter and vacuum, i.e. $\rho_{\rm stiff}(\tilde{a}_{\rm eq}) =\rho_{\CC}(\tilde{a}_{\rm eq})$, where we use $\tilde{a}_{\rm eq}$ to distinguish it from other equality points previously defined . If we neglect once more the $\nu$ effects at these early times,  we find from \eqref{stringrohrdS} and\eqref {stringrohLdS}:
\be\label{dstring}
D_{\rm string}=\left(\tilde{a}_{\rm eq}\right)^{-6}~.
\ee
In our string-inspired RVM the very early (unstable) de Sitter phase is therefore characterised by
$\left(a/\tilde{a}_{\rm eq}\right)^6 \ll 1$.  Naively, the $a^{-6}$ scaling of the energy density \eqref{stiffenden} would occur for $D_{\rm string}\, a^6 \gg 1$. However, we note that the above equations cannot actually be applied for $a\gg \tilde{a}_{\rm eq}$ because the vacuum {\it cannot fully decay} into massless axions. The ordinary radiation-dominated phase must be generated during the late inflationary period, as proposed in \cite{bms1}, when the scale factor lies in the range
\be\label{radiation}
\tilde{a}_{\rm eq}\, < \, a \, < \, a_{\rm eq}~,
\ee
where $a_{\rm eq}$ is the ordinary equality point between vacuum and radiation in the ordinary version of the RVM~\cite{rvmhistory,GRF2015}, see Sec.\,\ref{sec:RVM-inflation}.
Thus, in this scenario, the ordinary radiation-dominated phase of the Universe follows the first period of vacuum decay into massless axions.

Let us now compare the string-inspired RVM with the ideas of \cite{stiff}, where it was postulated that  a pre-inflationary stiff-matter-dominated phase occurs immediately after the Big Bang. In \cite{stiff}, however, the stiff matter was associated with baryons (charged fermions). Par contrast, in our case, although at early eras after the Big bang stiff matter also exists, nonetheless it has a gravitational origin, as it consists of axions that exist in the massless gravitational multiplet of strings (plus other, compoactification-related axion-like particles, as we discussed in this work).  In \cite{stiff} a matter-antimatter asymmetry was axiomatised for such stiff baryons. In our case, one does not need to do this. The gravitational action \eqref{sea4} (or \eqref{seamulti} for the multiaxion case)  is CPT invariant, and the KR axions are their own antiparticles. In the model of \cite{bms1} the early Universe is characterised by gravitational degrees of freedom only, appearing in the massless gravitational multiplet of the string spectrum. (Relativistic) Matter is generated only at the
final stages of inflation, and then, as a consequence of the background \eqref{LV} (or \eqref{LVmulti}), which remains undiluted at the exit from inflation~\cite{bms1}, one encounters a {\it matter-antimatter-asymmetric} Universe in the radiation era, due, {\it e.g.} of leptogenesis induced by the decay of heavy right-handed neutrinos into standard model leptons and antileptons in the backgrounds \eqref{LV}~\cite{decesare}. Such lepton asymmetries can then be communicated to the baryon sector (baryogenesis) through Baryon-minus-Lepton(B-L)-symmetry-preserving sphaleron processes~\cite{rubakov} in the standard model sector of the effective field theory.

Moreover,  as previously emphasized, our stringy stiff-axion `matter' has always a positive energy density, and  a simple equation of state \eqref{stiff}. We do not consider here polytrope EOS, which would even involve non linear dependence of the pressure density on higher (e.g. quadratic) powers of the energy density. Such EOS appear in some cosmological models with stiff-matter discussed in \cite{stiff2},\footnote{It is also remarked that polytrope EOS might characterise stiff-matter which forms a Bose-Einstein Condensate in the early Universe. Our early-Universe axions, which are massless, and not characterised by any potential, cannot form such condensates. This might be the case of the axions in the post-inflationary matter and radiation eras, where as  we have explained in \cite{bms1,bms2}, can develop masses and potentials due to instanton effects, and under certain conditions play the r\^ole of dark matter. The condition for the formation of cosmic Bose-Einstein condensates for such dark matter axions, including ultralight stringy ones, through their gravitational self interactions, have been discussed in \cite{sikivie}.}
 where cases with even negative energy densities for stiff matter have been considered, leading to initial-singularity-free Universe. In our case \eqref{stringrohrdS}, \eqref{stringrohLdS} the initial singularity is avoided if one assumes an RVM right up to the Big Bang (i.e. when the scale factor of the Universe vanishes). In the context of our string-inspired model, where the RVM energy density \eqref{toten} arises due to GW condensates, this means that we assume the existence of such primordial GW right up to the Big-Bang point.

\section{Potential Origins of GW and a pre-inflationary era for the string-inspired RVM Universe? \label{sec:origin}}

The origin of the primordial GW, which lead to the anomaly condensate, is not precisely known. If such GW appear
shortly after the Big-Bang, then an inflationary era follows the initial singularity.
However, there could well be, a very short pre-inflationary epoch, covering the intermediate period between the Big-Bang and the inflationary era. It is in that period that GW can be generated in a variety of ways, which we now proceed to discuss in the context of our stringy RVM. There are many ways in the literature to generate DW, but in our models we shall try to maintain their basic feature that only gravitational in origin degrees of freedom are dominant~\cite{bms1,bms2}.

\subsection{GW from primordial Black Hole mergers \label{sec:pBH}}

As a first relevant scenario for the formation of GW, we consider merging of primordial black holes~\cite{pBH}, which, in the context of our string-inspired RVM, could be formed by the collapse of massive space-time defects that could populate the very early Universe. Indeed, in string/brane-Universe models, one has  space-time brane defects, for instance compactified 3-branes which are wrapped around 3-cycles or other appropriate Calabi-Yau compact spaces. From the point of view of a three-large-dimension brane world, such defects might look like effective point-like ``D-particle defects''~\cite{dparticle,shiu}. Such defects are massive, with masses of order $M_s/g_s$, where $g_s$ is the string coupling, assumed weak ($g_s < 1$). The non-spherically-symmetric gravitational collapse of populations of such massive stringy objects, say in brane Universes, may lead to the formation of primordial black holes in the very early (pre-inflationary) brane Universe, on which our gravitational effective field theory \eqref{sea4} lives.

These primordial  black holes may coalesce, and thus produce GW, which in turn will condense, provided the appropriate conditions for such a condensation exist in this very early Universe. Such a GW condensation will lead to inflation, as we discussed in \cite{bms1} and reviewed above. Inflation will dilute beyond trace these primordial black holes.\footnote{We cannot exclude however the possibility that for some fine-tuned situations, a percentage of such primordial black holes remains during the post inflationary period, which then could play the role of some component of dark matter~\cite{pBHDM}. This is not a possibility we pursue, however, in the context of our string-inspired models, where we believe that the axions could play such a role, as discussed in \cite{bms1,bms2}.}

\subsection{GW from Unstable Domain Walls  (DW) in pre-inflationary Universe \label{sec:unstDW}}

Another scenario that could be in operation in our models is that of the formation of Domain Walls (DW) in the pre-inflationary Universe, which are {\it unstable}, and either annihilate each other, or collapse {\it non-spherically}, leading to the production  of GW~\cite{gwwalls}. DW are known to  appear in theories with broken discrete symmetry~\cite{zeld}. One mechanism for the production of {\it unstable} DW
is proposed in \cite{zeld},\footnote{Although not relevant to our purposes here, given that any DW, that could be produced in our models, would be produced during the pre-inflationary era, and hence would be diluted by inflation, nonetheless we mention, for completion, that the presence of stable DW would be incompatible with Big-Bang cosmology, leading to a power-law expansion of the Universe. Therefore, if DW were created during the early Universe eras, they have to disappear somehow. The same mechanisms that make DW unstable, are also responsible for the production of GW in our case, hence our interest in such unstable DW.} and also in~\cite{kibble} and \cite{vil},  and requires the existence of {\it only an approximate } discrete symmetry, which, {\it e.g}. could be due to a {\it bias} between the two minima of the double-well potential of the (pseudo)scalar field that gives rise to DW. The difference in the energy density between the two vacua in such asymmetric situations generates a pressure force, which is responsible for the DW annihilation. Cosmological studies
of such biased-discrete-symmetry-induced unstable DW, discussing detailed ways of their annihilation, can be found in
\cite{gel}.

Axion models, of interest to us in this work, are also known to produce unstable domain walls, via appropriate
discrete-symmetry-breaking terms~\cite{siki}. However this mechanism requires the presence of matter fermions (quark) interacting (via Yukawa couplings) with massive axions, with potentials generated by gauge instantons, responsible for the spontaneous breaking of the U(1)-Peccei-Quinn symmetry down to a discrete subgroup $Z_N$, with $N$ the number of fermion (quark) flavours. This will imply the presence of $N$-degenerate vacua, leading to stable DW. Hence, such a scenario, would not be suitable for the case of our effective field theories at (pre)inflationary eras, which comprise only gravitational degrees of freedom, and in which the (gravitational-in-origin stringy) axions in the very early Universe have no potential~\cite{bms1,bms2}. Nonetheless, in our models, gravitational axions acquire instanton-induced potentials {\it only} after inflation~\cite{bms2}, since matter and radiation (gauge fields) are generated at the end of inflation~\cite{bms1}. In this sense, the generation of stable DW would be incompatible with the standard Cosmology. Fortunately, as shown in \cite{siki}, the DW generated in the above way become unstable in the presence of four-fermion (quark) matter interactions, which, if sufficiently strong at an appropriate energy scale, close, e.g. to QCD scale, can lead to quark vacuum condensates,  and thus terms $<\overline q \, q >^2 \ne 0$ in the energy density. The latter are responsible for lifting the degeneracy of the $Z_N$ vacua, by energy shifts proportional to the (square of the) quark condensates, which leads to {\it instabilities and eventual collapse} of the DW, as per the arguments of \cite{zeld}. This would solve the DW problem, but in addition, if the collapse is non-spherically symmetric, will produce GW in such models, which could be present in the radiation era as well, of relevance to phenomenogy.

\subsection{GW from gaugino-condensate-induced DW in Supergravity Model with Hidden gauge sectors \label{sec:gaugino}}

However, in string theories, of the type considered in our approach, one may have supergravity in the early phases of the Universe, which may be broken spontaneously  by means of vacuum {\it condensates} of  the supersymmetric partners of gauge fields ({\it gauginos}) pertaining to the gauge sector of the pertinent supergravity models~\cite{sugra,sugra2}. Such {\it gaugino condensates} provide the scale of supersymmetry (supergravity) breaking, and give rise to massive gravitinos, with masses $m_{3/2}$ proportional to the gaugino mass.  Although in our string-inspired models we assume only gravitational degrees of freedom, nonetheless one could allow for such gaugino vacuum condensates in early, pre-inflationary phases of the Universe.
 Below we briefly discuss the scenario of \cite{gluino1} involving an SU($\mathcal N$) super Yang-Mills gauge (hidden) sector, responsible for supersymmetry breaking, which may be assumed not to have sizeable couplings with the other sectors of the supergravity theory, interacting with them primarily gravitationally~\cite{gluino1}. This sector is
 characterised by an R symmetry, which is spontaneously broken down to a discrete subgroup $Z_{2\mathcal N}$  by non-perturbative instanton effects~\cite{wittenYM,vainst}, and then down to $Z_2$ due to condensation of gaugino $\lambda^a$ fields, with $a=1, \dots \mathcal N^2 -1,$ an SU($\mathcal N$) index:
 \be\label{cond}
 < \lambda^a \, \lambda^a >_k = - 32\pi^2 \, \Lambda_{gc}^3\, e^{i 2\pi\, k/\mathcal N} \quad, \quad k=1, \dots \mathcal N~,
 \ee
where  $\Lambda_{gc}$~\cite{gluino1} is the energy scale at which the gauge interactions become strong, leading to the {\it gaugino condensation} (gc). There are $
\mathcal N$ degenerate vacua, as a result of \eqref{cond}.
The corresponding effective superpotential, below the scale $\Lambda_{gc}$, is given by~\cite{gluino1}
\be\label{superpot}
\mathcal W_{gc} = \mathcal N\,  \Lambda_{gc}^3\, e^{i 2\pi\, k/\mathcal N} ~.
\ee
However, in order to cancel the cosmological constant that ``afflicts'' the vacuum of the corresponding
broken supergravity theory, one should add a constant $w_0$ to the above superpotential \eqref{superpot}:
$W_{gc}  + w_0$, which implies a scalar potential~\cite{kovner}
\be\label{scalar}
V =- 3\frac{\mathcal N}{M^2_{\rm Pl}}\,  \Lambda_{gc}^3\, w_0^\star \, e^{i 2\pi\, k/\mathcal N} + {\rm h.c.}
\ee
where $M_{\rm Pl}$  is again the reduced Planck mass, h.c. denotes hermitian conjugate and $\star$ denotes complex conjugate, and the cancellation of the cosmological constant in the scalar potential of supergravity requires $|w_0| =  m_{3/2} \, M_{\rm Pl}^2$, where $m_{3/2}$ is the gravitino mass, which is connected to the gaugino mass $m_\lambda $ via~\cite{gluino1}:
$m_\lambda = 3 \mathcal N \, g^2/(16\, \pi^2) \, m_{3/2} $.

The important thing to notice is that the presence of the factor $w_0^\star$ on the right-hand-side of \eqref{scalar} implies that the degeneracy of the vacua is now {\it lifted}, because the scalar potential now takes on different values
at the various vacua.
Thus, the discrete symmetry $Z_{2\mathcal N}$ is broken down to $Z_2$ by the formation of {\it gaugino condensates} (gc), due to gauge interactions that become strong at a scale. This leads to the formation of unstable DW, since the pertinent energy shifts would contribute to pressure force and eventual collapse of DW. In this respect, the r\^ole of the gaugino condensates is in some way analogous to the quark condensate formation in QCD-like axion models~\cite{siki}, which also destabilises the pertinent DW,  as discussed above.

\subsection{DW in axion-dilaton models \label{sec:axdil}}

DW formation have also been discussed in generic
axion-dilaton models~\cite{townsend}, which can characterise early Universe phase of string-inspired models, like ours. Indeed, it is possible that during a pre-inflationary phase of our Universe, interpolating between the Big-Bang and the GW-induced de Sitter (inflationary) phase, one has non-trivial dilatons, with say exponential potential, arising, e.g. in the so-called Liouville cosmologies, as a result of the non-criticality of the string~\cite{aben}, or in other contexts, such as supergravty models, including string-inspired ones~\cite{ortin} (in units where the (3+1)-gravitational constant is : $2\kappa^2=1$, for notational brevity):
\be\label{axiondil}
\mathcal S_{a-\Phi} = \int d^4 x \Big (-R + \frac{1}{2} \, \partial_\mu \Phi \, \partial^\mu \Phi  + \frac{1}{2} \, e^{\mu \, \Phi} \, \partial_\mu a \, \partial^\mu a  - \Lambda_0 \, e^{-\lambda \, \Phi} \Big)
\ee
where $\Phi$ is the (canonically) normalised dilaton, and $a$ a pseudoscalar axion field, which in our stringy-RVM-context would be a stringy axion. For our string-inspired cases, we may take the scale $\Lambda_0 > 0$, and the constant parameters $\lambda > 0, \, \mu \ge 0$. Domain wall solutions for the case of the action \eqref{axiondil} have been explicitly constructed in \cite{townsend}, including the case of constant axions. But such structures are stable, and hence of no direct relevance to us, as we are interested in mechanisms for generating GW from unstable domain walls.

However, one may embed such models in supergravities with gauge sectors, which are characterised by gaugino condensation, for instance, as we discussed above. To this end one may consider more general dilaton potentials
$V(\Phi)$  in \eqref{axiondil}, and identify the latter with the scalar potential of the appropriate supergravity model,
which involves both axions and dilatons, as the real and imaginary components of complex scalar fields
\be\label{taufields}
\tau = a + i \Sigma(\Phi),
\ee
respectively, taking values in a K\"ahler target space, with metric  $G = \partial_\tau \, \partial_{\overline \tau} \, \mathcal K$, with $\mathcal K$ the K\"ahler potential~\cite{townsend}.  The function $\Sigma(\Phi)$ is to be determined by
demanding that the Lagrangian density of \eqref{axiondil}, with a given dilaton potential $V(\Phi)$, can be derived from the appropriate supergravity action~\cite{townsend}
\be\label{sugralag}
\mathcal L  = \sqrt{-g} \Big( -R + 2 \, G\, \partial_\mu \, \tau \,  \partial^\mu \,  \overline \tau  + V (\tau, \overline \tau)\Big)
\ee
where the scalar potential $V (\tau, \overline \tau)$ is expressed in terms of  $\mathcal K$ and a holomorphic superpotential as usual~\cite{sugra}. For our purposes here it is not necessary to give its explicit expression.
We only mention that one can then construct the appropriate superportential $W$ by identifying $V$ with the dilaton potential in \eqref{axiondil}.  In \cite{townsend}, an explicit construction of the Lagrangian \eqref{axiondil}, with an exponential dilaton potential,  from such supergravities, has been given.

We next remark that, on considering hidden-sector supersymmetry breaking, via
gaugino condensation, then, as we discussed above, one may induce instabilities in the formed DW, whose collapse would lead to GW. In such a case, the dilaton field could then be relaxed to a constant value, $\Phi \to \Phi_0$,  through its potential minimum, and the Universe \eqref{axiondil} could lead to \eqref{sea4}, in the presence of a ``bare'' cosmological constant term $\Lambda \, e^{-\lambda \, \Phi_0} > 0.$ Such a term will not affect the analysis of \cite{bms1,bms2}, provided $\Lambda_0$ is smaller than the dynamically induced cosmological constant $\Lambda$ due to the GW condensates, which would then drive inflation \`a la RVM. The bare $\Lambda_0$ could contribute to the cosmological constant today, given that the RVM evolution is impervious to it.

\subsection{Gravitino condensation and DW formation in pre-inflationary RVM-like Universe? \label{sec:gravitino}}

Finally, let us close this section with the remark that in \cite{houston} we have discussed  dynamical breaking of N=1 four-dimensional supergravity by means of a {\it gravitino} condensate in the early Universe, which lead to a double-well
potential for the gravitino scalar condensate $\sigma(x)$.
In that work we have discussed special conditions to allow for Starobinsky inflation, which are not necessary in view of our GW-condensate-induced RVM scenario \cite{bms1,bms2}. However, the dynamical scenario for supergravity breaking without the need for gaugino condensation we presented in \cite{houston}  might be used in our context as providing a pre-inflationary era, which is also of RVM type as far as the cosmic evolution of the energy density is concerned,  as shown explicitly in \cite{sugraRVM}. The massive gravitinos, which can have a mass of order up to the Planck scale, as discussed in \cite{houston,sugraRVM}, can then be integrated out of the spectrum of the pre-inflationary supergravity action, which includes only light gravitational degrees of freedom (axions, dilatons and gravitons), plus the scalar condensate of gravitinos. The latter, with its double-well potential, might provide the seeds for the formation of DW,  interpolating  between vacuum bubbles,  in the interior of which the condensate field acquires either of its expectation value $\pm v$.
The system has a $Z_2$ symmetry. The bare cosmological constant that is added as a regulator for the analysis~\cite{houston}, does not play an important role in our arguments, for the same reason that the scale $\Lambda_0$ in section \ref{sec:axdil} plays no role.

At present, though, we are not aware of a microscopic mechanism for inducing a {\it bias} in these two degenerate vacua, that would make the $Z_2$ symmetry not exact, and thus cause the DW to collapse, due to pressure exerted by the bias energy shift~\cite{zeld,kibble,vil} (unless of course we consider extending the supergravity model to incorporate (hidden) gauge sectors, which, via gaugino condensations, as mentioned previously, can lead to unstable DW). If such a mechanism existed, then this scenario would be the simplest extension of our string-inspired RVM in which GW could be generated in a pre-inflationary era. As discussed in \cite{sugra} one has an RVM type of vacuum evolution for the energy density in such models, which then, upon the appearance of GW, and their subsequent condensation, would be connected smoothly to the RVM inflationary phase.
In addition, one could also couple the N=1 four-dimensional supergravity to a chiral superfield, so as to incorporate axions and dilatons ({\it cf.} \eqref{sugralag}), which in turn can also form their own DW according to the arguments of \cite{townsend,ortin}. But again to induce instabilities to such walls, so as to produce GW by their decays, one needs to introduce gaugino condensates.

\subsubsection{Gravitino Condensation as an out-of-Equilibrium Phase transition and statistical origin of
bias leading to GW \label{sec:biasgvino}}

Before closing the section we would like to discuss one more scenario to introduce bias between the two vacua of the double-well potential of the gravitino condensation, which may be of {\it statistical origin}, due to {\it biased non-equilibrium phase transitions} in the early Universe~\cite{ross}. In our case that would be the pre-GW-induced-inflation era. If one considers the one-loop double-well-shaped gravitino potential in such a case, given in \cite{houston}, whose real part we call $\widetilde V$, then, in a FLRW background space-time, with Hubble parameter $H$, the equation of motion for the (homogeneous and isotropic) gravitino condensate field $\sigma$  is
\be\label{potgrad}
\ddot \sigma + 3 \, H \, \dot \sigma = \frac{\partial \widetilde V}{\partial \, \sigma}.
\ee
The potential can generate a hill-top ``first'' inflation near the origin $\sigma=0$~\cite{ellis}, which, for reasons to become clear below, we may assume to take place
in this early epoch, that precedes the GW-condensate-induced RVM inflation of \cite{bms1}.\footnote{We note for completeness that the potential's imaginary parts~\cite{houston}, expressing the decay of the condensate field after the first hill-top inflationary epoch, will not play a r\^ole in our qualitative arguments here, and hence they may be ignored. One may, e.g., assume that the scale of such imaginary parts is much smaller than the scale of the real parts during the hill-top `first' inflation.}
By assuming such an early first inflation, one ensures that at its end, any spatial inhomogeneities of the gravitino-condensate scalar field $\sigma$ have been washed out, and hence \eqref{potgrad} is valid to an excellent approximation. This will be important in what follows.

It goes without saying that we assume here that the gravitino condensation phase takes place after the string-dominated phase of the Universe, near the Big-Bang, where as mentioned above, higher curvature and purely string effects are in operation and might be responsible for the absence of any initial-singularity in the Big-Bang Universe.
The gravitino condenation phase  is an intermediate phase between the Big-Bang and the
GW-induced-RVM inflationary phase. The duration of such epochs depend on the details of the microcsopic string theory, and are of no direct relevance to our arguments. In general, we may assume this phase to be short compared with the subsequent phases of the RVM universe. 

It is important to note that the gravitino condensation process is viewed as a {\it non-thermal equilibrium} phase transition in this early supersymmetric/superstring-inspired Universe, since only gravitational interactions are involved in the formation of the condensate (the relevant four-gravitino interactions, which give rise to the condensate~\cite{houston}, characterise any supergravity theory, as a result of the inherent fermionic `torsion' terms of supergravity models~\cite{sugra}). The formation of condensates results in massive gravitinos, while gravitons remain massless. This breaks local supersymmetry at a given high scale~\cite{houston,ellis}, which may be taken to be much higher than the inflationary scale of the second RVM GW-induced inflation of \cite{bms1}, that occurs at a subsequent phase of this Universe evolution, after GW are formed due to the collapse of unstable DW, as we shall discuss below. This implies that the scale of the first
inflation is also much higher, close to Planck scale.

We may also assume that, after the hill-top first inflation, during the decay of the gravitino, stringy KR axions are produced (assuming that the model is  viewed as an effective field theory embedded appropriately in string models). The latter have a stiff equation of state \eqref{stiff} and a scaling of the Hubble parameter of the form (following from \eqref{stiffenden} via the Friedmann equation in a stiff-axion dominated phase):
\be\label{Hstiff}
H_{\rm stiff~axion} = \frac{H_i}{a(t)^3} ~,
\ee
where $H_i$ denotes the constant Hubble parameter during the first hill top inflation induced by the gravitino-condensation potential, and we normalise \eqref{Hstiff} such that $a(t_i)=1$, where $t_i$ is the cosmic time corresponding to the exit from the first hill-top inflation.
We note at this stage that, since this `first' inflation does not lead to observable effects in the CMB~\cite{Planck}, given the existence of the second RVM-like GW-induced inflationary phase, we need not worry about fine tuning the parameters to ensure the right phenomenology, and hence, as already mentioned, its scale, $H_i$, could be assumed as lying higher than that of the second inflation, close to Planck. Moreover in our approach here, we also {\it do not} consider the possibility of a Starobinsky inflation around the non trivial minima of the one-loop gravitino potental, as suggested  in \cite{houston2}. We stress that in our case, any observable effects on the CMB should come from the RVM-like GW-condensate-induced inflation that succeeds the stiff-axion phase.  We have already discussed in Sec.\,\ref{sec:Staro-inflation} that Starobinsky inflation is intrinsically very different from the RVM inflation, both being compatible with the CMB data but subject to very different inflationary mechanisms,  which means that these two inflationary models should be eventually distinguishable. 

Since the stiff-axion-dominated phase occurs for very early epochs of the Universe, we may easily assume
that the potential-gradient term on the right-hand-side of \eqref{potgrad} can be ommitted when compared to the gravitational friction term $H \, \dot \sigma = H_{\rm stiff~axion} \, \dot \sigma \gg \partial \widetilde V/\partial \sigma$.
Such a condition also characterises the exit from the first hill-top inflationary phase.
In this case, to leading order, the solution of \eqref{potgrad} is an approximate constant {\it classical} gravitino condensate field $\sigma_{\rm cl}$:
\be\label{constcond}
\sigma_{\rm cl}  \simeq \vartheta = {\rm constant}
\ee
where $\vartheta$ is essentially arbitrary (see also discussion in \cite{ross}). We note that among the allowed constants $\vartheta$ are of course the vacuum-expectation-values (VEV) of the condensate, for which $\partial \widetilde V/\partial \sigma =0$~\cite{houston,ellis}.

This (approximate) constancy of the classical part of the condensate field, $\sigma_{\rm cl}$ persists right up to the first inflationary phase, as one goes backwards in cosmic time.
The formal necessity for having such an inflationary phase is explained below. 

To this end, we first notice that the full quantum condensate field, $\sigma $ can be written as an expansion about the classical $\sigma_{\rm cl}$ field:
\be\label{split}
\sigma (x) = \sigma_{\rm cl} + \sigma_q (x)  = \vartheta + \sigma_q (x)
\ee
where $\sigma_q(x)$ denotes the quantum fluctuations. The existence of an inflationary epoch allows
first of all to associate the quantum fluctuations $\sigma_q(x)$ with a {\it very-weakly inhomogeneous}
{\it semi-classical} scalar field, which can be represented as the sum of a zero mode, to be discussed below, and a
part which is decomposed into Fourier components  with wavelengths $\lambda$ satisfying the condition~\cite{ross}
\be\label{hmin1}
H_i^{-1} \le \lambda \le L
\ee
where $L$ is the radius of the Universe, and $H_i^{-1}$ is the radius of the event (Hubble) horizon at the end of the (first, hill-top) inflation. The condition \eqref{hmin1} implies that $H_i$ acts as an UV cutoff scale for the Fourier momentum scale $k$ of the modes.
For subsequent eras, {\it e.g.} the stiff-axion-dominated epoch in our model, only the modes with wavelength
\be\label{cutoff}
H(t)^{-1} \equiv \ell_c (t) \le \lambda \le L
\ee
i.e. inside the Hubble horizon of the corresponding era,
will remain frozen and thus constitute the components of the semi-classical field. 

The existence of an inflationary era allows us, according to standard analysis, to compute the two-point correlation function~\cite{ross}
\be\label{corr}
\xi (\ell) = <0| \sigma_q (x + \ell) \, \sigma_q(x)|0>
\ee
where $\ell$ is an arbitrary length. The reader should have noticed that  in \eqref{corr} the vacuum state $|0>$ used
is the Bunch-Davies-vacuum, which is appropriate for the inflationary space time, and which is characterised by translational invariance. This implies that the two-point function $\xi (\ell)$ is independent of the  space-time coordinate $x$. For
$\ell_c \ll \ell \ll L$, which is a range relevant for our purposes,  as we shall explain below, one has
\be\label{lcL}
\xi (\ell) \simeq = \frac{H_i}{4\,\pi^2} \, \ln \Big(\frac{L}{\ell} \Big).
\ee
This result is dominated by long wavelength modes with $\ell \le \lambda \le L$. 

On the other hand, by dividing the Universe into coarse-grained causally-independent regions of radius $\ell \ll L$,
one may consider random (Gaussian) fluctuations of the quantum gravitino condensate field inside each such region. As explained in \cite{ross}, the various spheres are correlated by the longer wavelength modes of the field, which they share in common. This leads to an approximately constant  ($x$-independent)
zero-mode  background:
\be\label{constS}
\Sigma_\ell = \sqrt{\xi (\ell)}~.
\ee
The random fluctuations of the
field within each causal sphere of radius $\ell$ are due to short wavelengths modes, satisfying $\ell_c (t) \le \lambda \le \ell$, which are characterised by the fluctuation parameter (variance)
\be\label{variance}
\Delta (\ell) \simeq \frac{H_i}{4\,\pi^2} \, \ln \Big(\frac{\ell}{\ell_c} \Big).
\ee
These random quantum  fluctuations of the gravitino field inside a causal coarse-grained region of the Universe, of radius $\ell$, centered at a point $x$ of space-time, at the exit of the first-inflation, is then given by the Gaussian distribution
\be\label{gaussprob}
\mathcal P(\mathcal F_\ell (x)) = \frac{1}{\sqrt{2\pi \, \Delta (\ell)}} \, \exp\Big(-\frac{\mathcal F_\ell (x)^2}{2\, \Delta (\ell)}\Big)
\ee
and the full quantum condensate field $\sigma (x, \ell)$ \eqref{split}, inside a coarse-grained region of radius
$\ell_c \ll \ell \ll L$,  appropriate for our situation described here, is then given by:
\be\label{split2}
\sigma (x,\ell) = \sigma_{\rm cl} + \sigma_q (x,\ell)  = \vartheta + \Sigma_\ell  + \mathcal F_\ell (x).
\ee
We next notice that, as the cosmic time elapses, the potential-gradient term in \eqref{potgrad} will become comparable to the gravitational friction term $H \, \dot \sigma$, and eventually dominate it. The system then, at each point in space $\vec x$ (for a given cosmic time $t$) must roll towards one of the two vacua ($\pm$) of the gravitino double-well one-loop effective potential~\cite{houston}, with probability for, say the ($+$) vacuum (with the definition $(+) ((-)) $ vacuum corresponding to positive (negative) VEV of the gravitino condensate):
\be\label{prop}
p^+ = \int_0^{+\infty} D\mathcal F_\ell \, \mathcal P(\mathcal F_\ell)  = \int_{\vartheta + \Sigma_\ell}^{+\infty}  \, d\sigma (x) \,
\frac{1}{\sqrt{2\pi \, \Delta (\ell)}} \, \exp\Big(-\frac{(\sigma (x) - \vartheta - \Sigma_\ell)^2}{2\, \Delta (\ell)}\Big)
\ee
while the probability for the system to roll towards the ($-$) vacuum is $p^-= 1-p^+ \ne p^+$ in general. A {\it biased} non-equilibrium phase transition then occurs. In passing from the middle to the last equality in \eqref{prop}, we made use of \eqref{split2}, taking into account that $\vartheta$ and $\Sigma_\ell$ are $x$-independent constants, as we have discussed above.
\begin{figure}
\begin{center}
\includegraphics[width=0.8\textwidth]{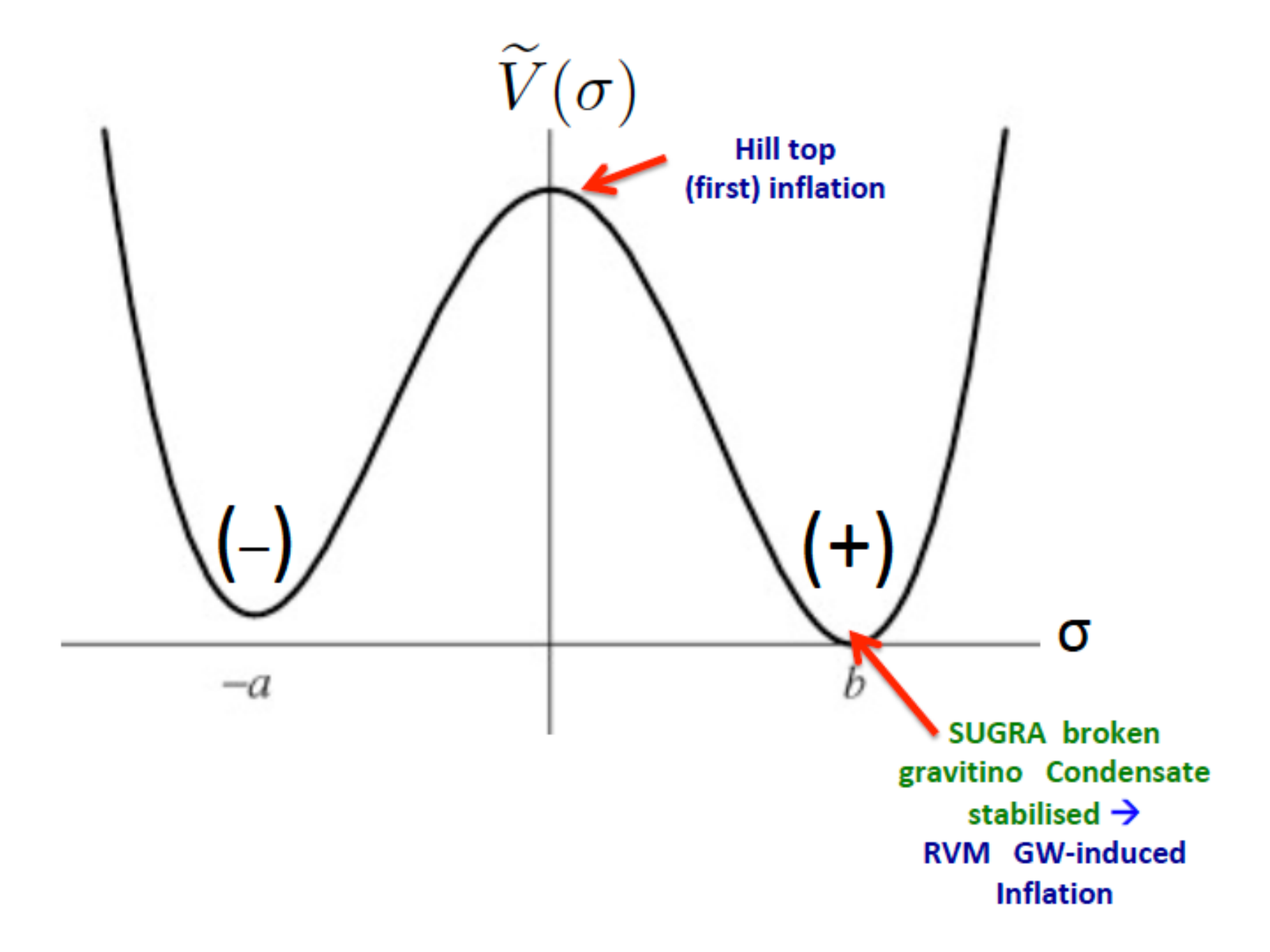}
\end{center}
\caption{{\it The (bias) double-wall potential corresponding to the gravitino condensate $\sigma$ in models with dynamical
breaking of (say $N=1$) supergravities~\cite{houston,ellis}, after percolation effects of the corresponding vacuum bubbles~\cite{ross} are taken into account. The near-zero-field-value region of the potential corresponds to
a `hill-top-inflationary phase, which in the context of our stringy RVM model~\cite{bms1} would correspond to a
first inflationary phase, preceding the GW-induced RVM inflation. This latter inflation would occur
at the broken-supergravity phase, in which the gravitino-condensate field has been stabilised to a constant {\it translationally-invariant} value $\sigma_0$ at the bottom of one of the two non-trivial vacua. The GW are generated by the collapse of unstable domain walls created due to the vacuum-bubble percolation effects in the phase preceding the RVM inflation.} }
\label{fig:ias}
\end{figure}

The reader should notice that the probability $p^+$ is arbitrary, since $\vartheta$ is arbitrary. It is only in thermal-equilibrium situations that $p^+=p^-=1/2$, but here, the gravitino condensation occurs as a result of only non-thermal gravitational interactions, as we have already mentioned~\cite{houston}. The formation of DW solitons interpolating between (+) and (-) vacua occurs, but as a result of the bias, such DW are unstable.
One may then discuss percolation properties of the system, as done in \cite{ross}, which we shall not pursue further here. For our purposes, the important point is that such percolating unstable DW systems, forming clusters of vacua of different sizes, lead to the formation of GW, as a result of non spherical collapse or annihilation of DW. Generic phenomenological studies of the formation of GW in such systems have been performed in \cite{gwwalls}, where generically a bias in, say, a double-well potential, including the statistical case discussed above, is described phenomenologically by assuming the existence of small  linear and cubic  correction terms in the field $\sigma$ in the respective effective double-well, $Z_2$-symmetric potential. Such perturbations break the $Z_2$ symmetry, resulting in unstable DW.

In our context, during the final stages of the evolution of the gravitino condensate towards the non-trivial vacua of its double-well potential, there is production of KR (or other stringy) axions, which dominate the phase as stiff ``matter'' ({\it cf.} \eqref{sea4}, or \eqref{seamulti} for the multi-string-axion case). Before the GW formation, the anomaly terms are irrelevant, since for a FLRW background, assumed to characterise also this early phase, these terms vanish. As depicted in fig.~\ref{fig:ias}, and discussed above, percolation of the bubbles corresponding to the two vacua~\cite{ross} would lead to an ``effective bias'' of the (gravitino) double-wall potential, in the sense of the situation being described phenomenologically by an ``effective shift'' in energy between the two vacua~\cite{gwwalls}. This would then lead to domain walls, whose collisions and annihilation would result in generation of GW.\footnote{The frequency of such GW is expected to be of the order of the gravitino mass, which  in the model of \cite{houston,ellis} is proportional to (and actually of the same order as)  the gravitino-condensate mass.}
Once GW perturbations are created, we assume that, as the time elapses, the conditions in  this early Universe become appropriate
for their condensation. The stringy axion matter couples to the GW condensates through gravitational anomalies, leading eventually to a RVM-like second dynamical inflation \cite{bms1,bms2}, as reviewed above. It is this second inflation that can be tested phenomenologically by means of cosmological data~\cite{Planck}. 

Although the existence of a first hill-top inflationary phase does not have any phenomenological sequences in this model, nevertheless
it allowes us to perform concrete computations for the percolation probability and thus understand better the evolution of the DW network in this early phase of this string-inspired Universe, characterised by gravitino condensation. Moreover the presence of this first hill-top inflationary era  implies that any spatial inhomogeneities of various fields
are washed away, thereby providing a microscopic explanation of the existence of cosmological (time-dependent, homogeneous to leading order) backgrounds, which have been used in \cite{bms1,bms2} to discuss the RVM-like dynamical inflation. On the other hand, the second RVM-inflationary phase, implies that
any remnant of massive gravitinos or domain walls from the early phase of the Universe, is washed out during the second inflationary period, at the end of which only KR (or other stringy) axion backgrounds remain, while chiral matter is generated~\cite{bms1}.

For completeness, we also mention at this point, that, as discussed in
\cite{sugraRVM}, this first inflation can also be described within the RVM unifying framework, which in this way can connect the Big-Bang to the present era of a string-inspired Cosmology. However, as mentioned in \cite{bms1}, and reviewed above, during the GW-condensate inflation, the $\nu$
coefficient of the $H^2$ terms of the energy density of the running vacuum \eqref{rLRVM}, turns out to be negative, in contrast to the stiff-axion-dominated (\eqref{stiffnu}) and post-second-RVM-inflationary (radiation, matter and present) eras, for which this coefficient is positive.  This feature is a unique feature of our gravitational-anomaly string-inspired RVM, which could perhaps be tested by cosmological data of the early Universe in the foreseeable future.

\section{Stringy RVM and the Swampland Criteria \label{sec:swamp}}

Last but not least, we would like to discuss briefly in this section the so-called swampland criteria~\cite{dSC,SC1,SC2a,SC2,branden} for embedding the (stringy) RVM framework in an UV complete quantum-gravity model, such as strings.
The swampland criteria refer to conditions on the potential $V$ of scalar fields used in inflationary or other models, which, if satisfied, guarantee that the model is embeddable in string theory, which is a consistent quantum-gravity framework. The criteria
require that one of the following two inequalities is satisfied,  either
\be\label{SC1a}
 \frac{|\nabla V|}{V} \gtrsim c_2 \, \kappa > 0
\ee
or
\be\label{SC2}
\frac{{\rm min}(\nabla_i \, \nabla_j V)}{V} \, \le \, -  c_3 \, \kappa^2 \, < 0
\ee
where $c_2,c_3$ are dimensionless (positive) constants of $\mathcal O(1)$.  The gradient $\nabla_i$ in field space refers to the multicomponent space of scalar fields $\phi_i$, $i,j =1,\dots N$ contained in the effective field theory. The second swampland conjecture \eqref{SC2}, refers to potentials that have a local maximum in field space, and the application of the criterion is near that maximum. It is evident that the swampland criteria \eqref{SC1a} and \eqref{SC2} disfavour slow-roll inflation.

Hence, they appear incompatible with the stringy RVM inflation, which is compatible with the slow-roll data. However, as explained in
\cite{msb}, there is no contradiction, because the GW-induced RVM inflation is due to a condensate of the anomaly terms due to GW condensed perturbations, and as such, its potential is still not fully known. Nonetheless, because of its composite nature, the swampland criteria are avoided altogether, which accounts for the phenomenological compatibility of this phase with the slow-roll-inflationary data.

\begin{figure}
\begin{center}
\includegraphics[width=0.7\textwidth]{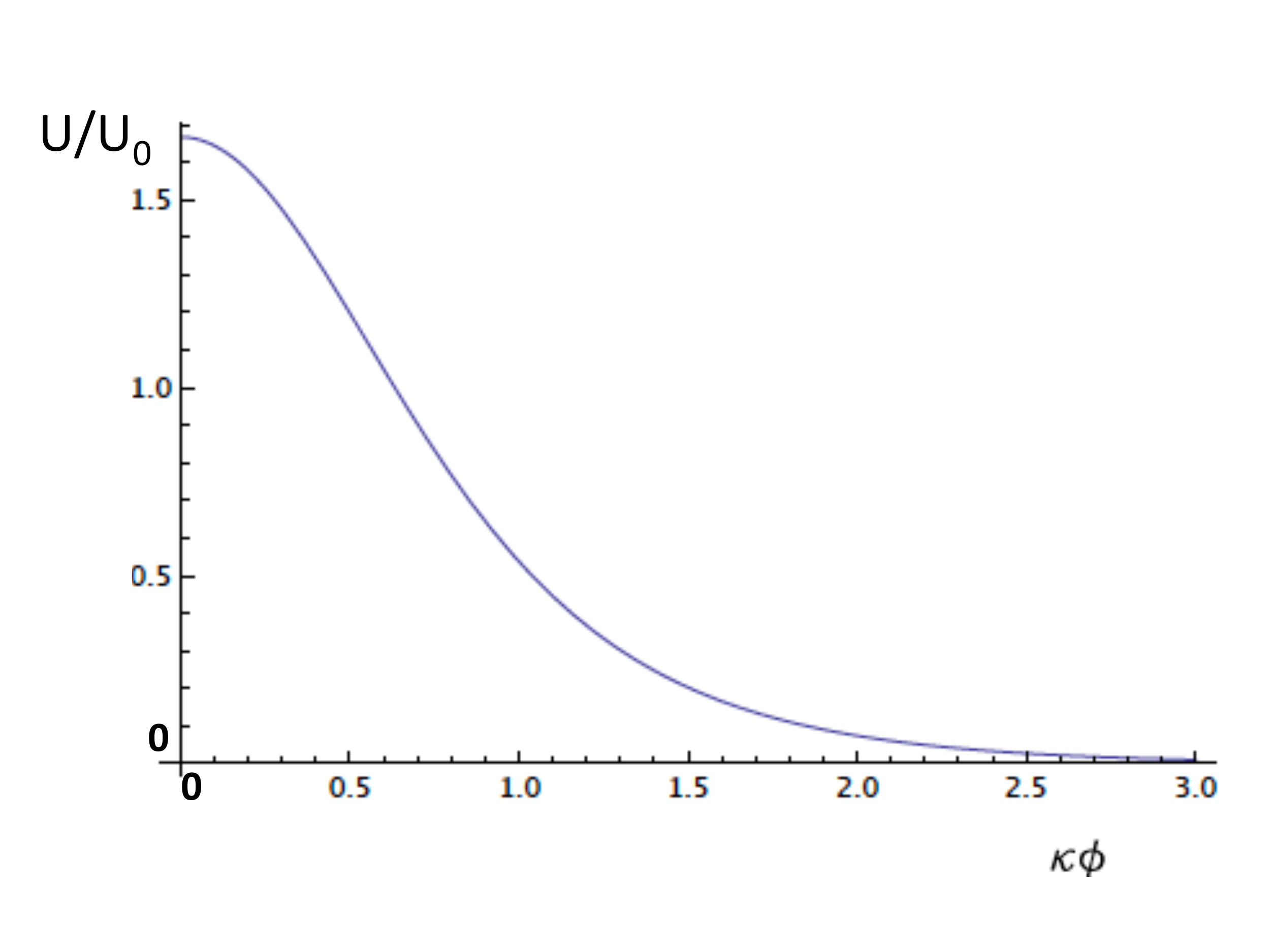}
\end{center}
\caption{{\it The ``vacuumon'' potential  \eqref{stringpot}, $U/U_0$, $U_0 \equiv  \frac{9H^{2}_{I}}{\alpha \kappa^{2}} $
for a classical scalar field representation of the early-Universe string-inspired RVM~\cite{bms1,bms2}, with
early-epoch massless `stiff' stringy (gravitational axion) ``matter'' present, with EoS $w_m =1$.
 The potential is defined for positive values of the vacuumon field $\phi > 0$. The potential leads to hill-top (not of slow-roll type) inflation, near the origin $\phi \simeq 0$, where it satisfies the second swampland conjecture \eqref{SC2}.
}}
\label{fig:swamppot}
\end{figure}

On the other hand, the so-called ``vacuumon  representation'' of the RVM~\cite{vacuumon} appears compatible with these criteria. According to
the vacuumon picture,  one can represent {\it classically} the total energy density and pressure of the generic RVM model (including its stringy version) in terms of a scalar (``vacuumon'') field $\phi$ by means of the correspondences~\cite{vacuumon,msb}:
\be\label{sfrvm}
\rho_{\rm tot}\equiv \rho_{\phi}={\dot \phi}^{2}/2+U(\phi), \quad
p_{\rm tot}\equiv p_{\phi}={\dot \phi}^{2}/2-U(\phi).
\ee
from which we obtain
\begin{align}
\dot{\phi}^{2} =-\frac{2}{\kappa^{2}}\dot{H} \quad \Rightarrow \quad
\phi= \pm \frac{\sqrt{2}}{\kappa}\int
\left(-\frac{H^{'}}{aH}\right)^{1/2}da\;,
\label{ff3}
\end{align}
and
\be\label{potH}
U =\frac{3H^{2}}{\kappa^{2}}\left(1+\frac{a}{6H^2}\,\frac{d H^2}{da}\right) \;.
\ee
Using \eqref{HS1rad} for the conventional RVM, or \eqref{stringHdS} for its stringy version, and \eqref{potH}, one could
construct the vacuumon potential directly, which can lead to hill-top inflation.  The potential has the form~\cite{vacuumon,msb}:
\begin{align}
\label{stringpot}
U(\phi) &= \frac{9H^{2}_{I}}{\alpha \kappa^{2}} \; \frac{\tfrac{2}{3}+{\rm
cosh}^{2}(\kappa \phi)} {{\rm cosh}^{4}(\kappa \phi)}~, \nonumber \\
\kappa\, \phi (a) &= \sqrt{\frac{2}{3}}\, {\rm sinh}^{-1} (\sqrt{D_{\rm string}} a^3) = \sqrt{\frac{2}{3}}\,\ln \Big(\sqrt{D_{\rm string}}\, a^3 + \sqrt{D_{\rm string} \, a^6 + 1}\Big).
\end{align}
where $H_I$ is the inflationary scale, $\alpha$ is the coefficient of the $H^4$ term in the RVM energy density \eqref{rLRVM}, and $a$ is the scale factor of the Universe. The potential is depicted in fig.~\ref{fig:swamppot}.

It can be seen from \eqref{stringpot} that the potential satisfies the second swampland conjecture
\eqref{SC2} for small values of the vacuumon field, near the origin, with $0 \le \kappa \, \phi \lesssim 0.4$~\cite{msb}, where the potential leads to hill-top inflation. For field values larger than $\kappa \phi \gtrsim 0.4$, the first swampland condition \eqref{SC1a} can be seen to be satisfied, since in that region $\kappa^{-1}\, |U'|/U >  1.04$. As also noted in \cite{msb}, for large $\kappa \phi > 1$, $|U'|/U $ asymptotes to the value 2,  which can be understood by the saturation of the entropy of the string-inspired RVM by the Bousso entropy bound~\cite{Bousso2002} (maximum Bekenstein-Gibbons-Hawking entropy~\cite{Bekenstein,Hawking71,GH}) during the exit from the early de Sitter phase.

Thus one could
naively think that they can describe the early $H^4$-dominated vacuum inflationary phase by means of the effective  potential $U(\phi)$ of the vacuumon which plays the role of the scalar degree of freedom associated with the RVM inflation.
However, given that the potentials $U(\phi)$ are compatible with the
swampland criteria, as shown in \cite{msb} and mentioned above, which disfavour slow-roll inflation, while the dynamical RVM inflation is compatible with the slow-roll data, such a description cannot be extended to the full quantum RVM.

At this point we note for completeness that the swampland criteria appear to be consistent with the thermodynamical properties of the RVM, discussed in section \ref{sec:ThermoAspects}. Indeed, as discussed in \cite{msb}, most of the entropy of the stringy RVM is produced near the exit phase of the RVM inflation, which occurs for large values of the vacuumon field $\kappa \phi > 1$, due to towers of stringy states becoming light, thus contaminating the effective field theory approach. Imposing the Bousso holographic entropy bound~\cite{Bousso2002}, which pertains to field theories, and is equivalent to the Bekenstein-Gibbons-Hawking entropy in a cosmological setting~\cite{Bekenstein,Hawking71,GH} \eqref{eq;SAentropy} ({\it cf.} section \ref{sec:ThermoAspects}), one obtains actually the second swampland conjecture \eqref{SC2}~\cite{SC2a}.
However, although such a compatibility implies that the vacuumon representation is embeddable in principle in
an UV complete quantum-gravity framework, it also implies that the vacuumon model fails to provide a faithful representation of the fully quantum RVM, given the agreement of the latter with the slow-roll inflationary data, as mentioned above. Hence the vacuumon is {\it not} the fully quantum scalar degree of freedom that underlies the condensate-induced inflation in the stringy RVM framework~\cite{msb}.

We do notice, however, that  within the context of a supergravity-breaking first inflation model, the double-well potential of the gravitino condensate, viewed as a hill-top-inflation-inducing potential, with the condensate playing the role of the inflaton field, can be made compatible, with the second swampland criterion \eqref{SC2}, near the origin of the potential. This would imply that the first inflation would not be a slow-roll one, which will have no phenomenological consequences, as mentioned previously, given that this first inflationary phase cannot be detected by the current data.
This compatibility though, would allow the early-Universe model to be embedded in a UV complete theory, such as strings, consistent with the fact that the underlying supergravity theory can be obtained as a low-energy limit of string theory. 

\section{Conclusions and Outlook \label{sec:concl}}

In this review we discussed a stringy version of the Running-Vacuum-Model (RVM) (``stringy RVM''), which provides an effective  description for the evolution of a string-inspired cosmological model from the Big-Bang to the present era.
For completeness. we have first recalled some of the basic properties of the RVM of the Universe, including its thermodynamical behaviour,
and stressed its important differences, as a model leading to dynamical-inflation without external inflatons, from other such frameworks, for instance Starobinsky inflation. This will hopefully assist the reader in their quest towards a better understanding of the physics underlying the embedding of the RVM framework in a string-inspired low energy approach, and an appreciation of its distinctive features as compared to the conventional field-theoretic RVM.

We have argued that an RVM-behaviour characterises the early phases of the string Universe, in which only gravitational degrees of freedom from the massless bosonic ground-state string multiplet (i.e. dilatons, graviton and antisymmetric (KR) axion fields) are included. Other stringy axions, arising from compactification might also be present.
For constant dilatons, one obtains consistent cosmological solutions, and this is the case we restrict ourselves on in this review and in \cite{bms1}, which this work is based upon.

The model is in general characterised by gravitational anomalies,
which become non trivial in the presence of primordial gravitational waves (GW).
In the review we have discussed potential scenarios on the origin of such waves. One of the simplest scenarios is that of
dynamical breaking of a supergravity model, which could be viewed as a low-energy limit of our string theory.
Condensation of gravitino fields in such models, at a very early epoch near the Big Bang, may break supergravity dynamically, and lead to percolation effects of the bubbles associated with any of the two vacua corresponding to the non-trivial minima of the double-wall
gravitino-condensate potential. The percolation phenomenon, in turn, results in formation of unstable domain walls, whose collapse produces GW. The reader should notice that in such extended scenarios, still only gravitational degrees of freedom are assumed to be present in the early Universe, given that gravitinos are the (local)supersymmetric partners of gravitons.\footnote{Of course, one may also assume more complicated gauge supergravity models, in which gauge degrees of freedom appear in hidden sectors of the model, thus still maintaining the spirit of only gravitational degrees of freedom being present in the observable sectors of the early Universe.}
\begin{figure}
\includegraphics[width=0.9\textwidth]{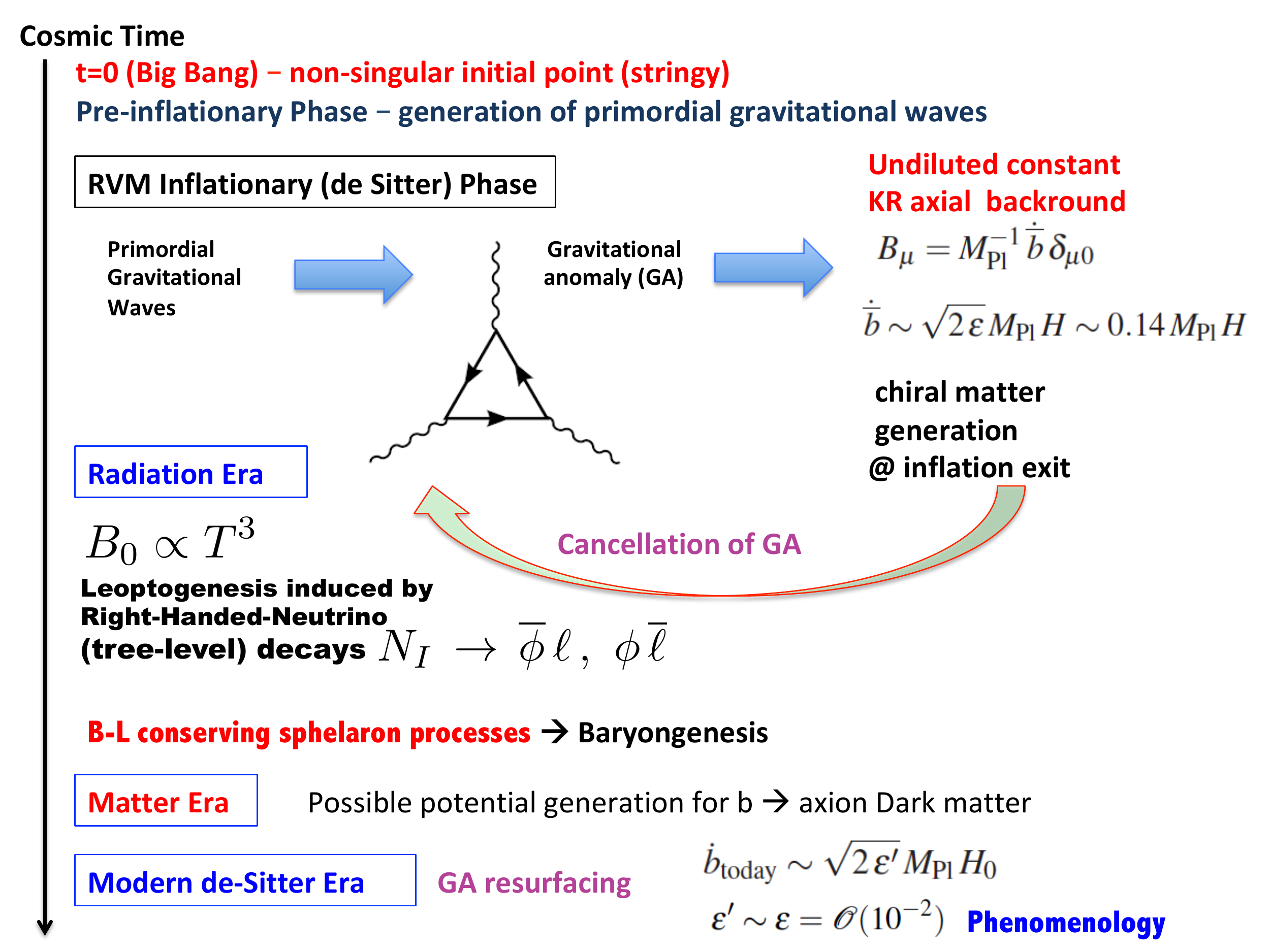}
\caption{{\it The Cosmic Evolution of the stringy Running Vacuum Model (RVM) of the Universe, from an early era after the Big Bang, till the present epoch~\cite{bms1,bms2}. The main eras examined in detail in our studies are depicted inside boxes. The main characteristics of these epochs are independent of the underlying microscopic string model. On the other hand the (non-boxed) initial and pre-inflatiomnary eras, do depend on details of the underlying microscopic string theory model.
At the Big-Bang era, corresponding to the origin of time $t=0$, all (infinite-order) higher-curvature terms in the gravitational action play a role, and thus the initial singularity is expected to be smoothened out. This is compatible with an effective RVM description.
The main RVM-inflationary phase is due to a condensate of Gravitational Waves (GW). The GW could be produced in a pre-RVM-inflationary phase, which could include a first inflation in some string-inspired supergravity models, which our framework can be embedded to. A microscopic-model-independent feature of this epoch is that after the first inflation, in the phase where supergravity is dynamical broken, there is a ``stiff''-KR-axion dominance.}}
\label{fig:stringrvm}
\end{figure}

A schematic evolution of the stringy RVM is given in fig.~\ref{fig:stringrvm}, where the various epochs, that have been studied in detail in \cite{bms1,bms2}, are depicted inside boxes. The main features of these epochs are largely independent of the underlying microscopic string theory.  Their main characteristic is the existence of (string-model independent)  KR axions, which in the early inflationary era, in the presence of GW, couple to gravitational anomalies via CP-violating anomalous couplings, responsible for inducing inflation. Additional stringy axions that depend on the details of compactification may be present, and couple to the gravitational anomalies in the same way, but the KR axion is always present.
In each of these eras the energy density of the cosmic fluid
assumes an RVM form \eqref{rLRVM}, but the coefficients of the $H^2$ and $H^4$ may differ from era to era, due to phase transitions in the stringy Universe. In this respect, we mention that the $\nu$ coefficient of the $H^2$ term during the GW-induced RVM-inflationary era is negative, due to the dominant contributions of the gravitational anomaly terms. This should be contrasted with the positive signature of the corresponding $\nu$ coefficients in the post- or pre-inflationary epochs, for which gravitational anomalies are absent. The reader should bear in mind that ordinary QFT effects associated with matter and radiation fields are responsible for the generation of positive $\nu > 0$ coefficients of the $H^2$ terms during the radiation- and matter-dominated epochs of the post inflationary Universe, as per the study of \cite{Cristian2020}, reviewed here, in particular in section \ref{Sec:RVMQFT} ({\it cf.} \eqref{eq:nueff} for the case of a scalar matter field non-minimally coupled to gravity. Qualitatively similar positive contributions are made by other matter and/or radiation fields of various spins, as we discussed above).
This difference in sign between the $\nu$ coefficients in the RVM-inflationary and post-inflationary eras is one of the most important, phenomenologically relevant, features of the stringy RVM, which might be, in principle, falsifiable, provided that sufficiently accurate data from  the inflationary era become available, leaving sufficiently significant imprints on CMB spectra~\cite{Planck}. At present this is an open issue.

Our stringy RVM framework, make the prediction that the primordial KR and other stringy axions could constitute a significant (or a dominant, depending on the model) dark-matter (DM) component. Given the `torsion' interpretation of the
KR axion, one obtains a `geometric origin' for DM in this case. Moreover, the undiluted KR axion background at the end of the RVM-inflation, whose Lorentz-violating form is due to the formation of GW-induced, CP-violating, gravitational-anomaly condensates, implies CPT- and CP-violating leptogenesis during the radiation era, in models including right-handed-neutrino matter in their spectra~\cite{decesare} (the latter are produced together with the rest of chiral matter, at the exit of the GW-induced RVM inflation~\cite{bms1}). Baryon-minus-Lepton (B-L)-number conserving sphaleron processes in the standard-model matter sector, then, are responsible for producing baryogenesis. In this sense, the observed matter-antimatter asymmetry in the Universe is attributed to gravitational anomalies during the GW-induced RVM-inflationary era. This provides an affirmative answer to the question `do we come from an anomaly?'~\cite{bms3}.

During the post-inflationary radiation- and matter-dominated eras of the Universe, gravitational anomalies are cancelled~\cite{bms1} by the chiral matter generated at the exit phase of the GW-induced RVM-inflation, but chiral anomalies in general remain, which are then responsible, through non-perturbative effects in the gluon (Quantum Chromodynamics (QCD)) sector of the model, for the generation of potentials for the axion field, which thus behave as dark matter in modern eras~\cite{bms2}.

In the present epoch, the plethora of cosmological data~\cite{Planck} suggest that the Universe re-enters a de Sitter phase. One may assume several scenarios for the onset of this second de Sitter phase in the history of our stringy RVM. The simplest is to assume that there is an underlying cosmological constant $c_0$ (which is allowed by the RG evolution of the RVM ({\it cf.} \eqref{runningmu},  \eqref{rLRVM} ). In terms of microscopic string or brane models, underlying the stringy-RVM, there is a plethora of reasons for the appearance of such constant vacuum terms, ranging form brane-Universe-tension contributions, to condensates of effectively point-like brane defects~\cite{dparticle}, which can be fine-tuned to de dominant only in the current era. We shall not go through them in this review.

What we shall assume instead, for the purposes of this work, is that the conditions in the Universe just before modern-era ``cosmological-constant'' dominance, are similar to the ones before the RVM-induced inflationary phase, which favour GW condensation.
Today, matter started to be a subleading contribution to the Universe energy budget compared to the vacuum energy, and thus it is possible that gravitational
anomalies resurface. The latter are much weaker of course than their ancient counterparts, but still they may lead to a new inflationary de Sitter era, corresponding though to
the present-era Hubble parameter, $H_0 \sim 10^{-60} M_{\rm Pl}$, which is much smaller than the RVM-inflationary epoch Hubble parameter \eqref{HIscale}. As we discussed in \cite{bms1},
during the modern era one can only phenomenologically impose a slow-running with the cosmic time of the KR axion,
\be\label{today}
\dot b_{\rm today} \sim \sqrt{2\epsilon^\prime} \, H_0 \, M_{\rm Pl}~,
\ee
where $\epsilon^\prime \sim 10^{-1}$ to fit the data, {\it i.e.} we require it to be of the same order as the $\epsilon$ of the early-Universe RVM-inflationary era \eqref{dotearly}.\footnote{Some microscopic models for $\epsilon^\prime$, associated with cosmic magnetic fields present in the current era have been presented in \cite{bms1}, but we have no way at present of estimating the magnitude of such fields, and thus verifying the assumption \eqref{today}.}

In view of this, when the anomaly condensate forms, the condition \eqref{cutoff1} for its constancy during the entirety of the new inflationary period ({\it cf.} \eqref{k02}) cannot be satisfied, unless unnaturally large transplanckian values of the UV cutoff $\mu$ for the respective GW modes are assumed. Hence the Physics of this new inflation, although formally resembling RVM, will be very different from the early inflationary epoch.
It is not possible at present to predict the future of our Universe within the stringy RVM effective formalism. This would require a detailed microscopic knowledge of the underlying string/brane theory, which at present falls way beyond the scope of our discussion.

We close by emphasizing that the scenario we have described here is not merely a theoretical proposal for new physics in the very early universe that  is able to connect the physics of primordial gravitational waves with the physics of inflation. If only for this, we believe it would be of significant value. However,  the new ``stringy RVM'' indeed  has two main phenomenological implications for the current universe, to wit: i)   there is an almost constant vacuum energy density (a bulk cosmological constant term as in the $\CC$CDM) accompanied  by a residual  --  kind of fossil\,\cite{Fossil2008} --  dynamical  DE, which is  reminiscent of extremely vigorous events that occurred at early times (which generated most of the entropy we now see) and presently showing up in the manner of a mild time-evolving DE component $\sim\nu H^2 \, (0<\nu\ll1$) which mimics quintessence; and ii)  the KR axion from the bosonic part of the original gravitational supermultiplet becomes the Dark-Matter axion, which through an appropriate mass generation mechanism might be responsible for part or the whole DM that is needed in our universe to explain structure formation.  The first implication is common with the original RVM, whereas the second is characteristic of the stringy version.   Both  fossil remnants of the very early universe may well be   living testimonies of the potential truth behind this fascinating story, which might provide a consistent overarching view of the cosmological evolution.

\section*{Acknowledgements}

The work  of NEM is supported in part by the UK Science and Technology Facilities  research Council (STFC) under the research grant ST/T000759/1. The work of JS has
been partially supported by projects  PID2019-105614GB-C21 and FPA2016-76005-C2-1-P (MINECO, Spain), 2017-SGR-929 (Generalitat de Catalunya) and MDM-2014-0369 (ICCUB). The authors also acknowledge participation in the COST Association Action CA18108 ``{\it Quantum Gravity Phenomenology in the Multimessenger Approach (QG-MM)}''.
NEM acknowledges a scientific associateship (``\emph{Doctor Vinculado}'') at IFIC-CSIC-Valencia University, Valencia, Spain.

\end{document}